\documentclass[twocolumn,showpacs,amsmath,amssymb,prd,10pt,nofootinbib]{revtex4-1}

\usepackage{graphicx}
\usepackage{dcolumn}
\usepackage{bm}
\usepackage{hyperref}
\usepackage{color}

\def\l#1{\left#1}
\def\r#1{\right#1}
\def\ds{\displaystyle}
\def\eref#1{(\ref{#1})}

\def\simlt{\lower.5ex\hbox{$\; \buildrel < \over \sim \;$}}
\def\simgt{\lower.5ex\hbox{$\; \buildrel > \over \sim \;$}}

\newcommand{\Ylm}[3]{{}_{#1}Y_{#2 #3}}
\newcommand{\Ylmcc}[3]{{}_{#1}Y^\star_{#2 #3}}

\begin{document}
\title{CMB {\it EB} and {\it TB} cross-spectrum estimation via pseudospectrum techniques}

\author{J. Grain}
 \email{julien.grain@ias.u-psud.fr}
 \affiliation{%
 Universit\'e Paris-Sud 11, Institut d'Astrophysique Spatiale, UMR8617, Orsay, France, F-91405}
\affiliation{%
CNRS, Orsay, France, F-91405}

\author{M. Tristram}%
 \email{tristram@lal.in2p3.fr}
\affiliation{CNRS, Laboratoire de l'Acc\'el\'erateur Lin\'eaire, Universit\'e Paris-Sud 11\\ B\^atiment 200, 91898 Orsay Cedex, France}

 \author{R.Stompor}%
 \email{radek@apc.univ-paris-diderot.fr}
\affiliation{%
AstroParticule et Cosmologie, Universit\'e Paris Diderot, CNRS/IN2P3, CEA/Irfu, Obs. de Paris, Sorbonne Paris Cit\'e, France 
}

\begin{abstract}
We discuss methods for estimating  $EB$ and $TB$ spectra of the cosmic microwave background anisotropy maps covering limited sky area.  Such odd-parity correlations are expected to vanish whenever parity is not broken.  As this is indeed the case in the standard cosmologies, any evidence to the contrary would have a profound impact on our theories of the early Universe. 
Such correlations  could also become a sensitive diagnostic of some particularly insidious instrumental systematics.
In this work we introduce three different unbiased estimators based on the so-called standard and pure pseudo-spectrum techniques and later assess  their performance by means of extensive Monte Carlo simulations performed for different experimental configurations. We find that a hybrid approach combining a pure estimate of $B$-mode multipoles with a standard one for $E$-mode (or $T$)  multipoles, leads to the smallest error bars for both $EB$ (or $TB$ respectively) spectra as well as for the three other polarization-related angular power spectra ({\it i.e.} $EE$, $BB$, and $TE$). However, if both $E$ and $B$ multipoles are estimated using the pure technique the loss of precision for the $EB$ spectrum is not larger than $\sim30$\%. Moreover, for the  experimental configurations considered here, the statistical uncertainties --due to sampling variance and instrumental noise-- of the pseudo-spectrum estimates is at most a factor $\sim1.4$ for $TT$, $EE$, and $TE$ spectra and a factor $\sim2$ for $BB$, $TB$, and $EB$ spectra, higher than the most optimistic Fisher estimate of the variance.
\end{abstract}


\maketitle

\section{Introduction}
\label{sec:intro}

A reliable and complete characterization of the polarized cosmic microwave background (CMB) anisotropies is one of the main challenges in observational cosmology, targeted by a large set of ongoing \cite{bicep_website,planck_website}, being-deployed
\cite{polarbear_website,quiet_website,ebex_website,spider_website} or planned \cite{brain_website, bpol_website} experiments. CMB (linear) polarization is completely described by two Stokes parameters, $Q$ and $U$, which are mapped over the celestial sphere by CMB experiments. In the harmonic domain, the polarization field can be described either with help of spin-2 and spin-$(-2)$  or  gradient, $E$, and curl, $B$, multipoles. From the physics point of view, the gradient/curl decomposition is more natural as it is directly linked to the cosmological perturbations produced in the primordial universe. Indeed, for symmetry reasons, scalar perturbations can produce solely $E$-mode-like polarization patterns and therefore the $B$ component of the polarization field can be viewed as a direct tracer of the primordial gravity waves \cite{zaldarriaga_seljak_1997,kamionkowski_etal_1997} and thus as a window onto the primordial Universe. Due to the presence of a secondary $B$-mode contribution generated by the gravitational lensing of the CMB photons by the Universe's large scale structure~\cite{zaldarriaga_seljak_1998},  such a picture is not fully correct, however  it is hoped that the latter signal can be accurately subtracted permitting a recovery of the primordial $B$-modes.

In standard cosmology, the physics governing photon propagation is parity invariant resulting in vanishing odd parity correlations, $\left<TB\right>$ and $\left<EB\right>$. Nevertheless in the context of the next generation CMB $B$-mode experiments, the $EB$ and $TB$ spectra will play an important role, both for the data characterization and their scientific interpretation. This is because on the one hand, these odd-parity cross-spectra are comprehensive, end-to-end, null tests of the presence of instrumental and/or astrophysical systematic effects still present in the data (see {\it e.g.} \cite{hu_etal_2003,yadav_etal_2010}). On the other hand, as some non-standard cosmological mechanisms could produce nonvanishing odd-parity cross-spectra, their detection
could become a smoking gun of such effects with potentially far-reaching consequences for our understanding of the Universe. Examples of such mechanisms include a primordial stochastic magnetic field,
 which generates $TB$ and $EB$ correlations, if this magnetic field possesses a helical component \cite{pogosian_etal_2002,caprini_etal_2004,kahniashvili_etal_2005}. Similar effects can be obtained due to a rotation of the plane of linear polarization of the  CMB photons traveling from the last scattering surface to our detectors. This could result from either the Faraday rotation induced by interaction with background magnetic fields \cite{kosowsky_loeb_1996,kosowsky_etal_2005,campanelli_etal_2004,scoccola_etal_2004} or interactions with pseudoscalar fields \cite{carroll_1998,lue_etal_1999}.
 In all these circumstances a robust and reliable $EB$ and $TB$ spectrum estimator could be needed to constrain, and potentially correct for, such effects.

The goal of this paper is to investigate some of the possible extensions of the existing power spectrum estimation methods to incorporate the odd-parity power spectra. Hereafter we work within a framework of  the pseudospectrum techniques. In this context the polarized CMB multipoles can be calculated with the help of either the  standard or  pure estimators, with the latter specifically designed to suppress (or to nearly suppress)  $E$-to-$B$ and $B$-to-$E$ leakages. In this work we use a specific prescription for calculating those as introduced in \cite{smith_2006} and later elaborated on in \cite{smith_zaldarriaga_2007, grain_etal_2009}, for some alternatives see, e.g.,   \cite{zhao_baskaran_2010,kim_2010,bowyer_etal_2011}. 
The relative merit of using either  the standard or the pure estimates of the pseudo-multipoles has been discussed in the context of the $B$-mode power spectra showing that the pure approach leads to smaller error bars \cite{smith_2006,smith_zaldarriaga_2007,grain_etal_2009}.  
This can be understood as follows. The pure estimator by construction removes some spatial modes, referred to hereafter as ambiguous modes, which are the source of the leakage between the two polarization components. This  would unavoidably lead to some information loss, reflected in an increase of the final variance of the estimated quantity, were not for a gain incurred as a result of removal
the $E$-mode power leaked to the $B$-spectrum and residing in the ambiguous modes.
 The latter effect offsets the former one at least as long as $E\gg B$ as indeed is expected for the CMB signal. Consequently, a significant gain in precision is usually observed as compared with the standard estimator,
which retains all the modes together with the leaked power.
By the same token,  for estimating the $E$-mode power spectrum the standard approach is expected to be more efficient as this time
extra variance due to the power of the $B$ modes leaked into $E$ is by far subdominant to that resulting from the removal of the ambiguous modes.

The question, whether to include or not the ambiguous modes, and whether to do it consistently or only partially, is less straightforward to answer in the context of the two odd-parity power spectra estimation and has not been  addressed to date on either a qualitative or quantitative level. This is the problem that is at the focus of this work. To address it
hereafter we construct and investigate performance of three alternative $EB$  pseudospectrum estimators\footnote{In principle, a fourth alternative is possible: using standard estimation for the $B$-modes and the pure estimation for the $E$-modes. However, such an alternative is a priori disfavored as a significant amount of information on the $E$-modes is lost and the $B$-modes still suffer from $E$-to-$B$ leakage and therefore will not be considered in this work.}: \\
$\bullet$~ {\it a standard approach} \cite{hivon_etal_2002,tristram_etal_2005} involving the standard pseudoharmonic coefficients for both the signal maps and no explicit correction for 
the $E$-to-$B$ leakage; \\
$\bullet$~ {\it a pure approach}, when both sets of the multipoles are pure; \cite{smith_2006,smith_zaldarriaga_2007,grain_etal_2009}; and \\
$\bullet$~ {\it a hybrid approach}, where the $E$ multipoles are obtained from the standard estimator and the $B$ signal from the pure estimator \cite{smith_2006,smith_zaldarriaga_2007}. \\

Temperature multipoles are always estimated using the standard pseudo-multipole approach, it may therefore appear that there are only two possible estimators for the $TB$ spectra 
corresponding to $B$ multipoles being either standard or pure. In fact, the situation is somewhat more complex due to the fact that, as discussed in Sec.~\ref{sec:FormalismTB}., an unbiased estimation of the $TB$ spectrum requires treating it together with the corresponding $TE$ spectrum.
Consequently, we arrive again at three types of  estimators of  possible interest, which mirror the case of the $EB$ estimators, and are therefore defined by the type of multipoles used for $E$ and $B$ signals. 
We will refer to them as {\em standard}, {\em pure}, and {\em hybrid} estimators.
  
Our paper is organized as follows. In Sec. \ref{sec:Conv}, we present our conventions and notation for describing polarization fields on the sphere. Section \ref{sec:Formalism} is devoted to the definition of the different pseudo-$C_\ell$ estimators of the $EB$ (Sec. \ref{sec:FormalismEB}) \& $TB$ spectrum (Sec. \ref{sec:FormalismTB}), with a particular emphasis on the computation of the mode-mode coupling matrices. The influence of the sky apodization and pixelization scheme is discussed in Sec. \ref{sec:Skyap}. Finally, the performances of each estimator is discussed in Sec. \ref{sec:Application} by quantifying the uncertainties of the reconstructed $EB$ and $TB$ spectra considered as null tests.
 In that section, we assume a typical sky-coverage and noise of suborbital experiments,  as inspired by the ongoing {\sc ebex} experiment \cite{britt_etal_2010}, and consider a series of mock observations with progressively more realistic features.
A simple Fisher analysis estimate of the power spectra uncertainties, which provides a lower limit of the variance of the estimators, is used as a theoretical benchmark. Our conclusions are presented in the last section (Sec. \ref{sec:Conclusion}) of this paper. 

Most of the technical details are provided in the appendices of this paper. Appendix \ref{app:Gaunt} summarizes the main properties of Gaunt integrals and Wigner-$3j$ symbols used in our derivation of the different mode-mode coupling matrices,  while the details  of their computation and  explicit formulas~are provided in Appendix \ref{app:mixtotal}.
  In Appendix \ref{app:Noise}, we provide the noise contribution expected to bias the pseudo-$C_\ell$'s for the three formalisms.

In the following of this paper, the term {\it mask} will systematically denote the binary, pixelized maps assuming two values, 0 or 1, corresponding to the observed (or kept in the analysis) or unobserved pixels. The pixel weights which can assume any nonzero value will be referred to {\it sky apodization} or {\it window function}.

\section{Convention and notation}
\label{sec:Conv}
\subsection{Spin fields on the sphere}
Any spin-$s$ field on the sphere satisfying $\Phi^\star_s=\Phi_{-s}$ (in the following, $s$ is always assumed to be $\ge 0$) can be decomposed using an $E$ and $B$ basis,
\begin{eqnarray}
\begin{array}{l c l}
	\Phi_s(\vec{n})&=&-\displaystyle\sum_{\ell m}\left(E_{\ell m}+iB_{\ell m}\right)\Ylm{s}{\ell}{m}, \\
	\Phi_{-s}(\vec{n})&=&(-1)^{s+1}\displaystyle\sum_{\ell m}\left(E_{\ell m}-iB_{\ell m}\right)\Ylm{-s}{\ell}{m},
\end{array}
\label{eqn:phiHarmRep}
\end{eqnarray}
where $\Ylm{\pm s}{\ell}{m}$ stands for the spin-weighted spherical harmonics obtained by applying the spin-raising and spin-lowering operators ($\partial$ and $\bar\partial$) on the standard (spin-0) spherical harmonics.
\begin{eqnarray}
\ds{}_{s}Y_{\ell m}=\frac{1}{N_{\ell,s}}\,\partial^sY_{\ell m}~~\mathrm{and}~~\ds{}_{-s}Y_{\ell m}=\frac{(-1)^s}{N_{\ell,s}}\,\bar\partial^sY_{\ell m},
\label{eqn:sharmDef}
\end{eqnarray}
with 
$$
N_{\ell,s}=\sqrt{\frac{(\ell+s)!}{(\ell-s)!}}.
$$
We remind the reader that $\Ylmcc{s}{\ell}{m}=(-1)^{s+m}\Ylm{-s}{\ell}{(-m)}$. For the special case of a spin-0 and real valued field (like the CMB temperature field), the $B$ component is vanishing. Moreover, this also applies to any spin-$s$ field built from a spin-0, real-valued field by use of the spin-raising operator  (i.e. $\Phi_s=\partial^s\Phi_0$), as can be directly seen by applying the spin-raising or lowering operators on 
the harmonic representation of the field as in Eq.~\eref{eqn:phiHarmRep} and using Eq.~\eref{eqn:sharmDef}.

CMB anisotropies are described by three fields: a spin-0 field, corresponding to temperature anisotropies $T(\vec{n})$, and spin-$(\pm2)$ fields describing the polarization denoted $_{\pm2}P$. In terms of the Stokes parameters, the two polarization fields are given by $_{\pm2}P\equiv(Q\pm iU)$. The multipole decompositions of such anisotropies are given by,
\begin{eqnarray}
	\ds T(\vec{n}) & = & \ds\sum_{\ell m}a^T_{\ell m}~Y_{\ell m}(\vec{n}), \\
	\ds _{\pm2}P(\vec{n}) & = & -\displaystyle\sum_{\ell m}\left(a^E_{\ell m}\pm ia^B_{\ell m}\right)~{}_{\pm2}Y_{\ell m}(\vec{n}).
\end{eqnarray}
Conversely, at least as long as CMB is mapped over the full sky, the $T$, $E$, and $B$ multipoles can be reconstructed from the Stokes parameters by applying the following scalar product,
\begin{eqnarray}
	a^{T}_{\ell m}&\equiv&\ds\int_{4\pi}\,T(\vec{n})\,Y^\star_{\ell m}(\vec{n})\,d\vec{n}, \nonumber \\
	a^E_{\ell m}&\equiv&-\frac{1}{2}\displaystyle\int_{4\pi}\left[_2P \, _2Y_{\ell m}^\star+_{-2}P  _{-2}Y_{\ell m}^\star\right]d\vec{n}, 
	\label{eqn:cmbMultFullSky}
	\\
	a^B_{\ell m}&\equiv&\frac{i}{2}\displaystyle\int_{4\pi} \left[_2P\, _2Y_{\ell m}^\star-_{-2}P _{-2}Y_{\ell m}^\star\right]d\vec{n}. \nonumber
\end{eqnarray}


Hereafter we will use $a^X_{\ell m}$, ($X=T, E, B$) to denote, the  full-sky multipoles as  in Eq.~\eref{eqn:cmbMultFullSky} and referred to them as
the CMB multipoles  and use $X_{\ell m}$ to denote partial-sky, pseudo-multipoles defined in the next section.

\section{pseudospectrum estimators for the {\it EB} \& {\it TB} cross-correlation}
\label{sec:Formalism}
\subsection{Standard and pure pseudo-multipoles}

\subsubsection{Temperature case} 
For temperature anisotropies, the pseudo-multipoles are defined as
\begin{equation}
	T_{\ell m}=\displaystyle\int_{4\pi}\left(\frac{\Delta{T}}{T}\right)\,W^{(T)}\,Y^\star_{\ell m}d\vec{n},
	\label{eqn:pseudoMultTemp}
\end{equation}
where $W^{(T)}$ is a window function applied to the temperature map. Such pseudo-multipoles are given by a convolution of the CMB temperature multipoles $a^T_{\ell m}$ with multipoles of the window function,
\begin{equation}
	T_{\ell m}=\displaystyle\sum_{\ell'm'}K^{(T)}_{\ell m;\ell'm'}a^T_{\ell'm'}.
\end{equation}
An exact expression for the convolution kernel $K^{(T)}_{\ell m;\ell'm'}$ has been derived in \cite{hauser_peebles_1973,hivon_etal_2002,hinshaw_etal_2003,tristram_etal_2005} and is recalled in Eq. (\ref{equ:TempMulti}).

\subsubsection{Polarized case: Standard pseudo-multipoles} 
In the case of the polarization, pseudo-multipoles can be introduced most straightforwardly by directly adapting the definition from the temperature case, Eq.~\eref{eqn:pseudoMultTemp}, 
\begin{eqnarray}
	E^{(std)}_{\ell m}&\equiv&-\frac{1}{2}\displaystyle\int_{4\pi}W\left[ _2P \, _2Y_{\ell m}^\star
	+\,_{-2}P\,_{-2}Y_{\ell m}^\star\right]d\vec{n},  \\
	B^{(std)}_{\ell m}&\equiv&\frac{i}{2}\displaystyle\int_{4\pi}W\left[ _2P\,_2Y_{\ell m}^\star \, -\,  _{-2}P\,_{-2}Y_{\ell m}^\star\right]d\vec{n}.
\end{eqnarray}
We will refer to an approach based on these definitions as the {\it standard estimator}. The resulting standard pseudo-multipoles are then related to the CMB polarization multipoles by,
\begin{eqnarray}
	E^{(std)}_{\ell m}&=&\ds\sum_{\ell'm'}\left[K^{diag}_{\ell m,\ell'm'}a^{E}_{\ell'm'}+iK^{off}_{\ell m,\ell'm'}a^{B}_{\ell'm'}\right], 
	\label{pseudostdE}
	\\
	B^{(std)}_{\ell m}&=&\ds\sum_{\ell'm'}\left[K^{diag}_{\ell m,\ell'm'}a^{B}_{\ell'm'}-iK^{off}_{\ell m,\ell'm'}a^{E}_{\ell'm'}\right].
	\label{pseudostdB}
\end{eqnarray}	
The convolution kernels, $K^{diag}$ and $K^{off}$, are functions of the Gaunt integrals and their explicit  expression are given in Eq. \eref{Kll-std} of App. \ref{app:mixingAlm}.

\subsubsection{Polarized  case: Pure pseudo-multipoles}
As emphasized in \cite{bunn_etal_2003}, the pseudo-multipoles as defined above suffer from $E$-to-$B$ leakage. Following \cite{bunn_etal_2003}, the polarization field on a partial sky can be split into three subspaces: the pure $E$-modes, the pure $B$-modes and the ambiguous modes, which contain simultaneously information about $E$ and $B$ modes.
 These are these last modes, which are the source of the $E$-to-$B$ leakage due to impartial sky coverage. Their presence is due to the fact that on the cut-sky the spin harmonics,  $_{\pm2}Y_{\ell m}$, 
 are not anymore orthogonal and any alternative basis constructed out of their linear combinations will include some which will be neither of the $B$- or $E$- type.
 
   Because, for the CMB the power of the $E$ signal is much bigger than that of $B$, this leakage is a dominant factor for the $B$-modes estimation giving rise to a so called  $E$-to-$B$ leakage problem. Nevertheless, by the same token a certain amount of $B$-mode power is also present in the $E$ pseudo-multipoles, which occasionally may also become problematic.

The $E$-to-$B$ leakage problem can be resolved if the ambiguous modes are identified and excluded from the spectrum estimation procedure or by using some (incomplete) set of basis functions, which by construction are orthogonal to such modes.  Following ~\cite{smith_2006} such a basis can be constructed with help of the spherical harmonics and appropriate window functions $W$ and $W_1=\partial W$ both vanishing at the boundaries of the observed sky. Denoting these basis functions as,
\begin{eqnarray}
\ds{}_{2}Y^{\l(W\r)}_{\ell m}&\equiv&\frac{1}{N_{\ell,2}}\,\partial^2\l(W\,Y_{\ell m}\r)\\
\ds{}_{-2}Y^{\l(W\r)}_{\ell m}&\equiv&\frac{(-1)^2}{N_{\ell,2}}\,\bar\partial^2\l(W\,Y_{\ell m}\r),
\end{eqnarray}
we can write now expressions for pseudo-multipoles, which are free of any $E/B$ leakage as,
\begin{eqnarray}
	E^{(pure)}_{\ell m}\equiv-\frac{1}{2}{\displaystyle\int_{4\pi}}\left[_2P \, _2{Y^{\l(W\r)}_{\ell m}}^\star
	+\,_{-2}P\,_{-2}{Y^{\l(W\r)}_{\ell m}}^\star\right]d\vec{n},  &&
	\label{eqn:pureMultE}
	\\
	B^{(pure)}_{\ell m}\equiv\phantom{-}\frac{i}{2}\displaystyle\int_{4\pi}\left[ _2P\,{_2Y^{\l(W\r)}_{\ell m}}^\star - _{-2}P\,{_{-2}Y^{\l(W\r)}_{\ell m}}^\star\right]d\vec{n}.&&
		\label{eqn:pureMultB}
\end{eqnarray}
As in the standard case these multipoles can be related to the actual CMB multipoles as,
\begin{eqnarray}
	E^{(pure)}_{\ell m}&=&\ds\sum_{\ell'm'}\left[H^{diag}_{\ell m,\ell'm'}a^{E}_{\ell'm'}+iH^{off}_{\ell m,\ell'm'}a^{B}_{\ell'm'}\right],
	\label{pseudopureE}
	\\
	B^{(pure)}_{\ell m}&=&\ds\sum_{\ell'm'}\left[H^{diag}_{\ell m,\ell'm'}a^{B}_{\ell'm'}-iH^{off}_{\ell m,\ell'm'}a^{E}_{\ell'm'}\right]. 
	\label{pseudopureB}
\end{eqnarray}	
We note that in principle by construction $H^{(off)}$ should be identically zero. This would be however only  the case were the cut-sky a sole source of the leakage. In practice, other sources of residual leakage exist, most notably due to sky pixelization effects,\footnote{More particularly, the Dirichlet and Neumann boundary conditions can never be completely fulfilled by $W$ because of pixelization.} and it is prudent and useful to correct for such effects at least on average, what can be achieved with help of an appropriately calculated off-diagonal block. Such off-diagonal blocks, though not precisely zero, have been shown to be typically much smaller than the diagonal blocks for the pure estimates, i.e., $H^{off}_{\ell m,\ell'm'}\ll H^{diag}_{\ell m,\ell'm'}$ and, more importantly, than the off-diagonal blocks of the standard approach, i.e., $H^{off}_{\ell m,\ell'm'}\ll K^{off}_{\ell m,\ell'm'}$ \cite{smith_2006,grain_etal_2009}. This is this latter condition, which ensures that the pure estimator typically results in significant improvements over the standard one at least as far
as $B$-mode power spectra are concerned. This emphasizes the fact that an estimator does not have to be strictly pure, i.e., $H^{off} \ne 0$, to provide good estimates of the $B$ modes. This indeed turns
out to be the case with so-called optimized apodizations~\cite{smith_zaldarriaga_2007}, which lead to the smallest overall uncertainty of the estimated $BB$ spectra even if not strictly fulfilling the required boundary conditions.

The pure pseudo-multipoles, Eqs.~\eref{eqn:pureMultE}~\&~\eref{eqn:pureMultB}, can be related to the standard pseudo-multipoles as follows,
\begin{eqnarray}
	E^{(pure)}_{\ell{m}}&=&{E}_{2,\ell{m}}+2\frac{N_{\ell,1}}{N_{\ell,2}}{E}_{1,\ell{m}}+\frac{1}{N_{\ell,2}}{E}_{0,\ell{m}}, 
	\label{aeapp}
	\\
	B^{(pure)}_{\ell{m}}&=&{B}_{2,\ell{m}}+2\frac{N_{\ell,1}}{N_{\ell,2}}{B}_{1,\ell{m}}+\frac{1}{N_{\ell,2}}{B}_{0,\ell{m}}.
	\label{abapp}
\end{eqnarray}
Here, $E_{s, \ell m} (B_{s,\ell m}),\ s =0, 1, 2$  is a spin-$s$ multipole of type $E$ ($B$) of a field, $W^\star_{2-s}\, _2P$ , where $W_s \equiv \partial^s\, W$ are derivatives of the scalar window, $W$.
It is easy to show  that $E_{2,\ell m}=E^{(std)}_{\ell m}$ and $B_{2,\ell m}=B^{(std)}_{\ell m}$, while the two remaining terms, called the counterterms,  are there to cancel the ambiguous modes present in the standard term, what they do prefectly whenever the window fulfills the required conditions.
Each of the spin-$s$ pseudo-multipoles is related to the CMB multipoles via spin-dependent, mode-mode coupling matrices as in Eqs. (\ref{pseudostdE}) and (\ref{pseudostdB}). Combined together, those lead to expressions for the convolution kernels, $H^{diag}$ and $H^{off}$,
introduced in Eqs.~\eref{pseudopureE}~\&~\eref{pseudopureB}. The explicit forms for all those matrices can be found in App. \ref{app:mixingAlm}, Eqs. \eref{Hll-pure} and \eref{Kll-pure}.

\subsubsection{Coupling between $a_{\ell m}$ and pseudo-$a_{\ell m}$}
It is instructive to briefly comment on how the pseudo-multipoles are coupled to the CMB and the window functions multipoles. The mode-mode coupling corresponds to the coupling of three angular momenta: the momentum $(\ell,m)$ of either the $E$ or $B$ pseudo-multipoles, the momentum $(\ell',m')$ of either the $E$ or $B$ CMB multipoles and the momentum $(\ell'',m'')$ of either the $E$ or $B$ components of each spin-$s$ window functions. The total number of such {\it angular momentum couplings} is eight as any angular momentum can be of either $E$- or $B$-type. Each of these couplings can be classified according to its parity: even or odd. For example, the coupling between $E_{\ell m}$, $a^{E}_{\ell'm'}$ and $w^{(E)}_{s,\ell''m''}$  does not vanish only for even values of $(\ell+\ell'+\ell'')$,
as it is made of the $G^{+}$ quantities (see App. \ref{app:Gaunt}). We call such a coupling {\it even}. In contrast, the coupling between $E_{\ell m}$, $a^{B}_{\ell'm'}$ and $w^{(E)}_{s,\ell''m''}$, made of the $G^{-}$ quantities, is an {\it odd coupling} as it is nonzero only for odd values of $(\ell+\ell'+\ell'')$. From this perspective, the pseudo-multipoles can be written using the following matrix notation
\begin{eqnarray}
	E_{\ell m}&=&\ds\sum_{\ell'm'}\sum_{\ell''m''}\sum_{s=0}^2\left(\begin{array}{cc}
			a^{E}_{\ell'm'} & a^{B}_{\ell'm'}
		\end{array}\right)\left(\begin{array}{cc}
			g^{even}_s & -ig^{odd}_s \\
			ig^{odd}_s & g^{even}_s
		\end{array}\right) \nonumber\\
		&\times&\left(\begin{array}{c}
			w^{(E)}_{s,\ell''m''} \\
			w^{(B)}_{s,\ell''m''}
		\end{array}\right), \\
	B _{\ell m}&=&\ds\sum_{\ell'm'}\sum_{\ell''m''}\sum_{s=0}^2\left(\begin{array}{cc}a^{E}_{\ell'm'} & a^{B}_{\ell'm'}
		\end{array}\right)\left(\begin{array}{cc}
			-ig^{odd}_s & -g^{even}_s \\
			g^{even}_s & -ig^{odd}_s
		\end{array}\right) \nonumber\\
		&\times&\left(\begin{array}{c}
			w^{(E)}_{s,\ell''m''} \\
			w^{(B)}_{s,\ell''m''}
		\end{array}\right),
\end{eqnarray}
where both $g^{even}_s$ and $g^{odd}_s$ depend on $(\ell,m)$, $(\ell',m')$, and $(\ell'',m'')$, and $g^{even}_s\times g^{odd}_s=0$ if they both have the same arguments.  This classification is helpful during the computation of the pseudo-power spectra as any term involving the product of an odd coupling times an even coupling will vanish. The parity nature of each coupling is summarized in Table \ref{tab:Coupling}. 
\begin{table}[ht]
\begin{center}
	\begin{tabular}{c|c|c|c|c} \hline\hline
		    & $a^{E}_{\ell'm'}w^{(E)}_{s,\ell''m''}$ & $a^{E}_{\ell'm'}w^{(B)}_{s,\ell''m''}$ & $a^{B}_{\ell'm'}w^{(E)}_{s,\ell''m''}$ & $a^{B}_{\ell'm'}w^{(B)}_{s,\ell''m''}$ \\ \hline
		$E_{\ell m}$ & even & odd & odd & even \\
		$B_{\ell m}$ & odd & even & even & odd \\ \hline\hline
	\end{tabular}
	\caption{Parity classification of the coupling between pseudo-multipoles, CMB multipoles and window function multipoles.}
	\label{tab:Coupling}
\end{center}
\end{table}

\subsection{Pseudo-power spectra estimators: {\it EB} cross-correlations}
\label{sec:FormalismEB}
The two types of the pseudo-multipoles defined in the previous section permit us to introduce three different pseudospectrum estimators.  These include {\it a standard estimator} in which both $E$ and $B$ pseudo-multipoles used to construct pseudo-$C_\ell$'s are standard; {\it  a pure estimator} in which the $E$ and $B$ pseudo-multipoles are pure; and {\it a hybrid estimator} which combines a standard estimate of the $E$ pseudo-multipoles with a pure estimate of the $B$ pseudo-multipoles.

For each of the three alternatives, the pseudo-power spectra, $\mathcal{C}^{PP'}_\ell$, are defined as follows
\begin{eqnarray}
	\mathcal{C}^{EE}_\ell&\equiv&\frac{1}{2\ell+1}\ds\sum_mE^{(x)}_{\ell m}E^{(x)\star}_{\ell m}, \\
	\mathcal{C}^{BB}_\ell&\equiv&\frac{1}{2\ell+1}\ds\sum_mB^{(y)}_{\ell m}B^{(y)\star}_{\ell m}, \\
	\mathcal{C}^{EB}_\ell&\equiv&\frac{1}{2(2\ell+1)}\ds\sum_m\left[E^{(x)}_{\ell m}B^{(y)\star}_{\ell m}+E^{(x)\star}_{\ell m}B^{(y)}_{\ell m}\right], 
\end{eqnarray}
where $(x)$ and $(y)$ stands for the standard and the pure pseudo-multipoles respectively. Because of the limited sky-coverage and pixel dependent weights, these estimators are biased as their average over CMB realizations involves a mixing between different $\ell$-modes and between different polarization type. Those two classes of mixing are encoded in the mode-mode coupling matrices $M^{PP',P''P'''}_{\ell\ell'}$ and the unbiased estimator, $C^{P''P'''}_\ell$, is computed by solving the following linear system
\begin{equation}
\begin{array}{r}
	\left(\begin{array}{ccc}
		M^{EE,EE}_{\ell\ell'} & M^{EE,BB}_{\ell\ell'} & M^{EE,EB}_{\ell\ell'} \\
		M^{BB,EE}_{\ell\ell'} & M^{BB,BB}_{\ell\ell'} & M^{BB,EB}_{\ell\ell'} \\
		M^{EB,EE}_{\ell\ell'} & M^{EB,BB}_{\ell\ell'} & M^{EB,EB}_{\ell\ell'}
	\end{array}\right)\left(\begin{array}{c}
		C^{EE}_{\ell'} \\
		C^{BB}_{\ell'} \\
		C^{EB}_{\ell'}
	\end{array}\right)= \\
	\left(\begin{array}{c}
		\mathcal{C}^{EE}_{\ell}-\mathcal{N}^{EE}_\ell \\
		\mathcal{C}^{BB}_{\ell}-\mathcal{N}^{BB}_\ell \\
		\mathcal{C}^{EB}_{\ell}-\mathcal{N}^{EB}_\ell
	\end{array}\right),
\end{array} \nonumber
\end{equation}
where $\mathcal{N}^{PP'}_\ell$ stands for the noise contribution to the pseudo-power spectra, {\it a priori} considered as nonzero. Expressions for the different noise biases under the assumption of white noise as well as those for the mode-mode coupling matrix depend on a choice of a specific formalism. They are discussed in Appendices~\ref{app:Noise} and~\ref{app:MixingEB}, respectively, and below we only briefly summarize main properties of the coupling matrices referring the reader for all the details there.

\subsubsection{Standard formalism} 
In this formalism, assuming the same window function is used for estimating the $E$ and the $B$ pseudo-multipoles, the different blocks are related to each other as follows,
\begin{equation}
\begin{array}{l}
	\ds M^{EE,EE}_{\ell\ell'}=M^{BB,BB}_{\ell\ell'} \equiv M^{diag}_{\ell\ell'}, \\
	\ds M^{EE,BB}_{\ell\ell'}=M^{BB,EE}_{\ell\ell'} \equiv M^{off}_{\ell\ell'}, \\
	\ds M^{EB,EB}_{\ell\ell'}=M^{diag}_{\ell\ell'}-M^{off}_{\ell\ell'},
\end{array}
\label{block-link1}
\end{equation}
and
\begin{equation}
\begin{array}{ll}
	\ds M^{EE,EB}_{\ell\ell'}&=-M^{BB,EB}_{\ell\ell'}=-2M^{EB,EE}_{\ell\ell'} \\
	&=2M^{EB,BB}_{\ell\ell'} \equiv M^{cross}_{\ell\ell'}.
\end{array}
\label{block-link2}
\end{equation}
Their explicit expressions are summarized in Eqs. \eref{Mll-std1} and \eref{Mll-std2}. 

Though $M^{off}_{\ell\ell'}$ is smaller than $M^{diag}_{\ell\ell'}$, it cannot generally be neglected as it describes the leakage from $E$-modes to $B$-modes, which has to be corrected for, in particular if the maps to be analyzed cover a tiny amount of the celestial sphere (see for example Fig. 9 of \cite{grain_etal_2009}). As shown in App. \ref{app:MixingEB}, the $M^{cross}_{\ell\ell'}$ matrix is equal to zero, emphasizing the fact that the $EB$ cross-spectrum is not coupled to the $EE$ and $BB$ auto-spectra. The main reason for this is that whenever the window functions contains only a $E$-component, as is precisely the case in the standard formalism, $E$ pseudo-multipoles are evenly coupled to $E$ multipoles and oddly coupled to $B$ multipoles while the $B$ pseudo-multipoles are oddly coupled to $E$ multipoles and evenly coupled to $B$ multipoles. As a consequence, the $M^{cross}_{\ell\ell'}$ block involves product of the even and odd coupling and thus vanishes. This demonstrates that  the $EB$ cross-spectrum can therefore be treated independently of the $EE$ and $BB$ spectra in this formalism.

\subsubsection{Pure formalism}
In the pure formalism, the different mode-mode coupling blocks satisfy the same relations as given in Eqs. \eref{block-link1} and \eref{block-link2}. Nevertheless their respective expressions  are qualitatively and quantitatively different in both these cases, see Eqs. \eref{Mll-pure1} and \eref{Mll-pure2}.  First of all, the pure estimation ensures the $M^{off}_{\ell\ell'}$ block be small enough to substantially reduce  amount of $E$-to-$B$ leakage even in the cases of small-sky coverage (e.g., see Fig. 9 of \cite{grain_etal_2009}). Second, unlike the standard case, the four blocks $M^{cross}_{\ell\ell'}$, which couple the $EB$ cross-spectrum to the $EE$ and $BB$ autospectra, may not vanish. These blocks involve two types of couplings: i) odd-even couplings corresponding to the $EE$ and $BB$ power spectra of the window functions and ii) even-even and odd-odd couplings via the cross-correlation of the $E$-component of the window functions with their $B$-component. Though the first type of couplings is identically equal to zero, the second class of couplings does not vanish if the spin-1 and spin-2 windows have a nonvanishing $B$-component. 

\subsubsection{Hybrid formalism}
The case of the hybrid formalism is more intricate as it involves the cross-correlation of the pure pseudo-multipoles with the standard pseudo-multipoles. 
Denoting by $X,$ and $X'$ either $E$ or $B$, we find that the $(EE,XX')$ blocks are equal to the $(EE,XX')$ blocks as computed in the standard formalism while the $(BB,XX')$ blocks are given by the $(BB,XX')$ blocks as derived in the pure formalism. The expression for the $(EB,XX')$ blocks is given by Eq. \eref{Mll-hybrid} and involves a cross-product of the  matrices $K^{diag/off}_{\ell\ell'}$ and $H^{diag/off}_{\ell\ell'}$. As in the case of the pure formalism, the $(EB,XX')$ blocks generally do not vanish as they contain even-even and odd-odd couplings corresponding to the cross-correlation of the $E$-component of the spin-0 window with the $B$-components of the spin-1 and spin-2 window functions.

\subsection{Pseudo-power spectra estimators: {\it TB} cross-correlations}
\label{sec:FormalismTB}
The formalism developed in the previous section is easily adaptable to the second odd-parity power spectrum, $C^{TB}_\ell$. Because of the $E$-to-$B$ (and $B$-to-$E$) leakage, the $TB$ spectrum has to be analyzed with the $TE$ in a unified framework. However, cut-sky effects do not introduce any leakages from temperature to $Q$ or $U$ Stokes parameter. As a consequence, there is no leakage from $TT$, $EE$, $BB$, and $EB$ spectra into the $TB$ and $TE$ pseudospectra and the $TB$  spectrum can be treated independently of the $EE,BB,EB$.

The two pseudospectra are defined as follows
\begin{eqnarray}
	\ds\mathcal{C}^{TE}_\ell&\equiv&\ds\frac{1}{2(2\ell+1)}\ds\sum_m\left[T_{\ell m}E^{(x)\star}_{\ell m}+T^{\star}_{\ell m}E^{(x)}_{\ell m}\right], \\
	\ds\mathcal{C}^{TB}_\ell&\equiv&\ds\frac{1}{2(2\ell+1)}\ds\sum_m\left[T_{\ell m}B^{(y)\star}_{\ell m}+T^{\star}_{\ell m}B^{(y)}_{\ell m}\right],
\end{eqnarray}
from which an unbiased estimator is built by inverting the usual mode-mode coupling matrix, i.e.
\begin{equation}
	\left(\begin{array}{cc}
		M^{TE,TE}_{\ell\ell'} & M^{TE,TB}_{\ell\ell'} \\
		M^{TB,TE}_{\ell\ell'} & M^{TB,TB}_{\ell\ell'} 
	\end{array}\right)\left(\begin{array}{c}
		C^{TE}_{\ell'} \\
		C^{TB}_{\ell'} 
	\end{array}\right)=\left(\begin{array}{c}
		\mathcal{C}^{TE}_{\ell}-\mathcal{N}^{TE}_\ell \\
		\mathcal{C}^{TB}_{\ell}-\mathcal{N}^{TB}_\ell
	\end{array}\right). \nonumber
\end{equation}
The explicit expressions of the mode-mode coupling matrix for the three formalisms are provided in App. \ref{app:MixingTB}. 
We observe that in the standard and pure formalisms, $M^{TE,TE}_{\ell\ell'}=M^{TB,TB}_{\ell\ell'}$ and $M^{TE,TB}_{\ell\ell'}=-M^{TB,TE}_{\ell\ell'}$ as long as the same window functions are used to compute the $E$ and $B$ multipoles. However, the off-diagonal blocks, $M^{TE,TB}_{\ell\ell'}$ are equal to zero in the standard formalism but may not vanish in the pure one if the spin-1 and spin-2 windows have a nonzero $B$-component. The fundamental reason is that pseudo-$a^T_{\ell m}$ are coupled only to the product of the $E$-component of the window functions applied to the maps times the $a^T_{\ell m}$ via an even coupling while in the standard formalism, $E(B)$-pseudo-multipoles are coupled to $B(E)$-multipoles via an odd coupling. However, the pure formalism introduces an additional even coupling of the $E(B)$-pseudo-multipoles with the $B(E)$-multipoles via the $B$-component of the nonzero-spin window functions, which translates into nonvanishing off-diagonal blocks. In the hybrid formalism, the $M^{TE,TE}_{\ell\ell'}$ and $M^{TE,TB}_{\ell\ell'}$ blocks are formally equal to the ones computed in the standard formalism while the $M^{TB,TB}_{\ell\ell'}$ and $M^{TB,TE}_{\ell\ell'}$ blocks are equal to the ones as computed in the pure formalism.

\section{Sky apodization and pixelization effects}
\label{sec:Skyap}
\subsection{Sky apodization}
Applicability of the pure pseudospectrum estimator depends on our ability to compute sky apodization with the appropriate boundary conditions. Different functions have been proposed. They range from ones derived as a result of an optimization procedure to ones based on some analytic expressions. Their relative merit has been extensively discussed in \cite{grain_etal_2009} to show that the pixel-domain optimization scheme proposed in \cite{smith_zaldarriaga_2007} leads to the lowest variance on the power spectrum estimation. 

The underlying strategy to derive such an optimized sky apodization is to search for  window functions, $W$, which make the pure pseudo-$C_\ell$ as close as possible to the optimal, quadratic power-spectrum estimator (see Sec. V of \cite{smith_zaldarriaga_2007}). An optimized weighting scheme therefore involves a specific sky apodization for each $\ell$-band where the power spectrum is to be estimated, according to the signal and noise power aliasing in each band. Such an optimization procedure assumes that noise and signal are known. For the noise and  $E$-modes sufficiently precise assumptions  are usually available, e.g., $E$-mode power spectrum can be recovered precisely enough without any optimization.  Though this is not the case with the $B$-modes, it can be shown \cite{grain_etal_2009} that the 
assumptions about the $B$-modes power spectrum very weakly affect the results of the power-spectrum estimation using the optimized sky apodization. This is due to the fact that  the optimization process foremost attempts to reduce the $E$-to-$B$ leakage and then the noise variance and thus no precise $B$-mode knowledge is necessary.
We note that in this approach as  the derivative relationship is relaxed during the computation, the $B$-component of the spin-1 and spin-2 windows becomes nonzero. Therefore,  the  $(EB,XX)$ and $(XX,EB)$ blocks of the mode-mode coupling will not vanish. Nevertheless, numerical studies have shown that the $B$-component of their spin-1 and spin-2 windows is very small.
 As shown in Appendix \ref{app:MixingEB}, only the {\it cross-spectra} of the $E$-component and $B$-component of the spin-weighted window functions are involved in the $(EB,XX)$ and $(XX,EB)$ while the {\it autospectra} of both $E$- and $B$-component of those window functions enter in the other blocks. As a consequence, we expect the $(EB,XX)$ and $(XX,EB)$ blocks of the mode-mode coupling matrix to be much smaller than the other blocks. 
 
Other options for the apodizations considered to date \cite{grain_etal_2009} include  i) analytic window functions, and ii) harmonic-based optimized window functions.  In these cases, the spin-1 and spin-2 windows are computed by taking numerically the derivative of the spin-0 window. As a consequence, none of the spin-$s$ sky apodizations have a $B$-component and the $(EB,XX)$ and $(XX,EB)$ blocks of the mode-mode coupling are expected to vanish within numerical precision in both the pure and the hybrid formalism. Hence,  in these cases the $EB$ spectrum is  effectively decoupled from $EE$ and $BB$, as in the standard formalism, but in a contrast with the pixel-domain apodization cases.

Similar arguments apply to the case of mode-mode coupling for the $TB$ cross-spectrum. In the pure and hybrid formalism, the potential cancellation of the off-diagonal blocks  depends on the sky apodizations used in the numerical computation of the pure pseudo-multipoles. Those blocks are expected to be zero for either the analytic or the harmonic-based optimized window functions but may become nonzero for
 the pixel-based optimized sky apodizations. From a quantitative point of view, the off-diagonal blocks share common features with the $(XX,EB)$ and $(EB,XX)$ blocks. They in particular depend only on the {\it cross-spectra} of the $E$-component and $B$-component of window functions and we therefore expect those off-diagonal blocks to be much smaller than the diagonal ones if pixel-based optimized sky apodizations are used.
\begin{figure}
\begin{center}
	\includegraphics[scale=0.375]{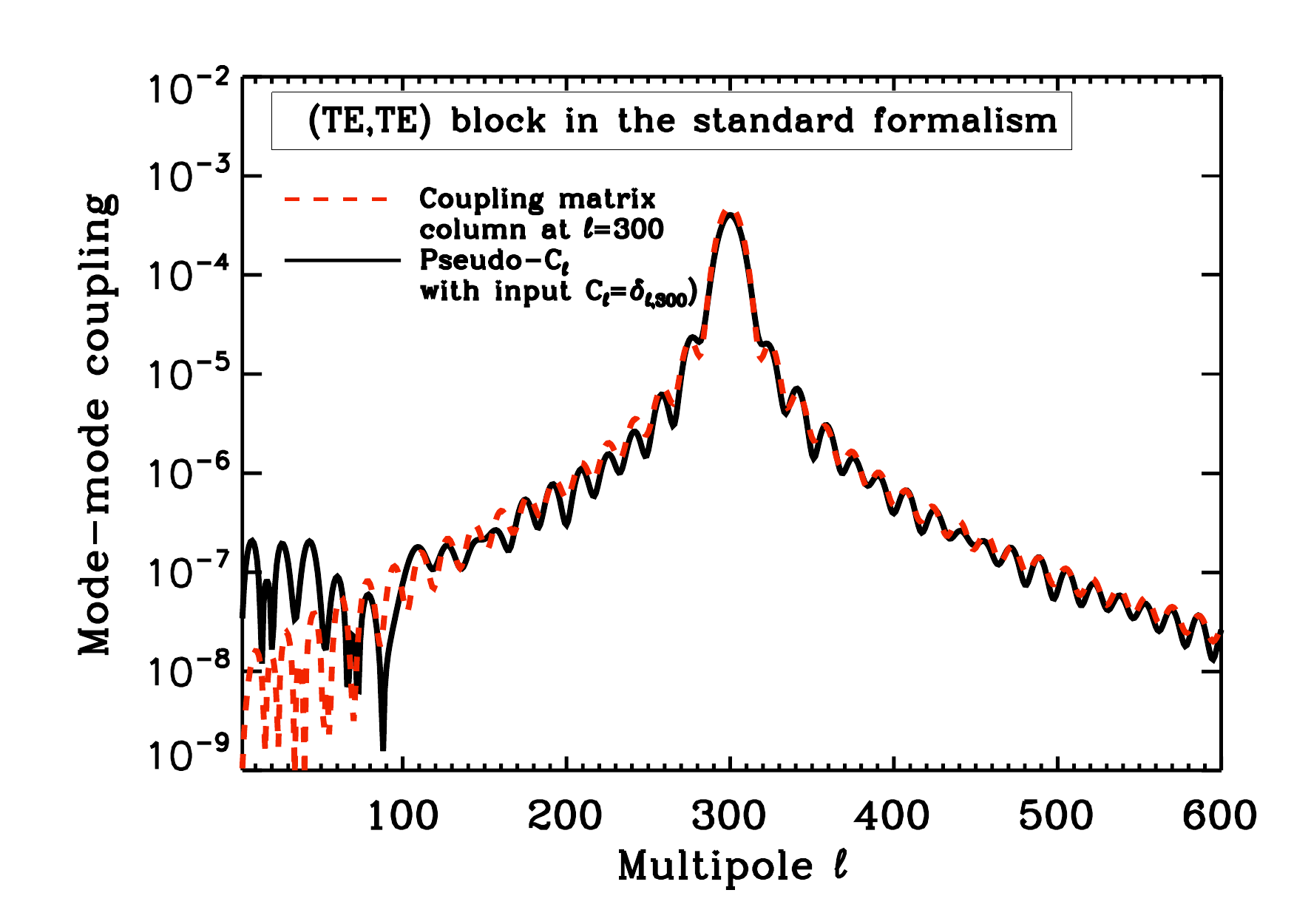} \includegraphics[scale=0.375]{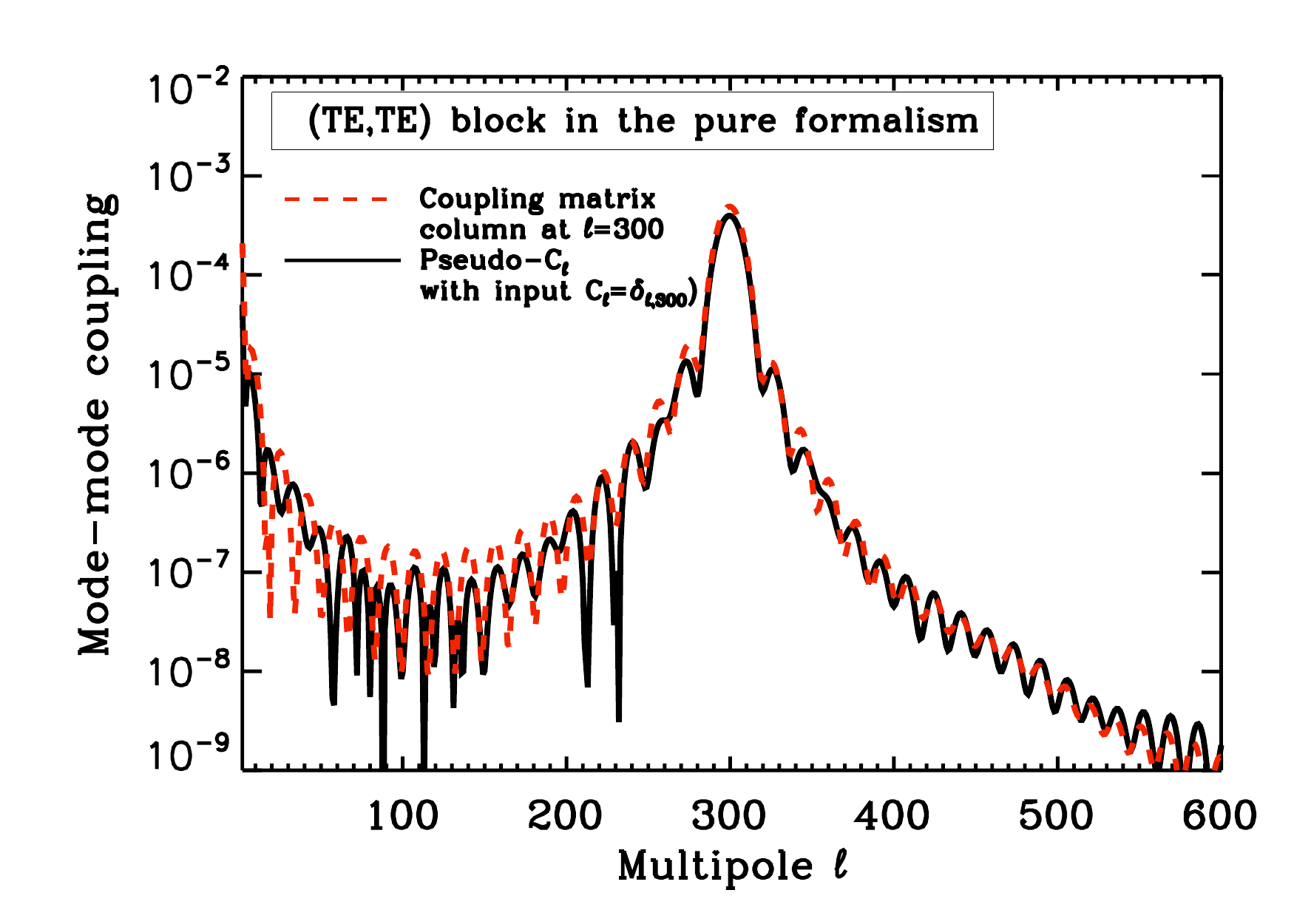} 
	\caption{The red-dashed curves correspond to the $\ell=300$ column of the  $(TE,TE)$ coupling kernels computed in the standard (left) and pure (right) formalisms. The standard $(TE,TE)$ block equals the hybrid $(TE,TE)$ block and standard $(TB,TB)$ block. The overlapping black curve shows the $TE$ obtained from a map containing $TE$ and $TB$ correlations at $\ell=300$ only. The agreement between the red and black curves underlines the correctness of the numerical computation of the different the mode-mode coupling matrices.}
	\label{fig:kernel1}
\end{center}
\end{figure}	

\begin{figure*}
\begin{center}
	\includegraphics[scale=0.35]{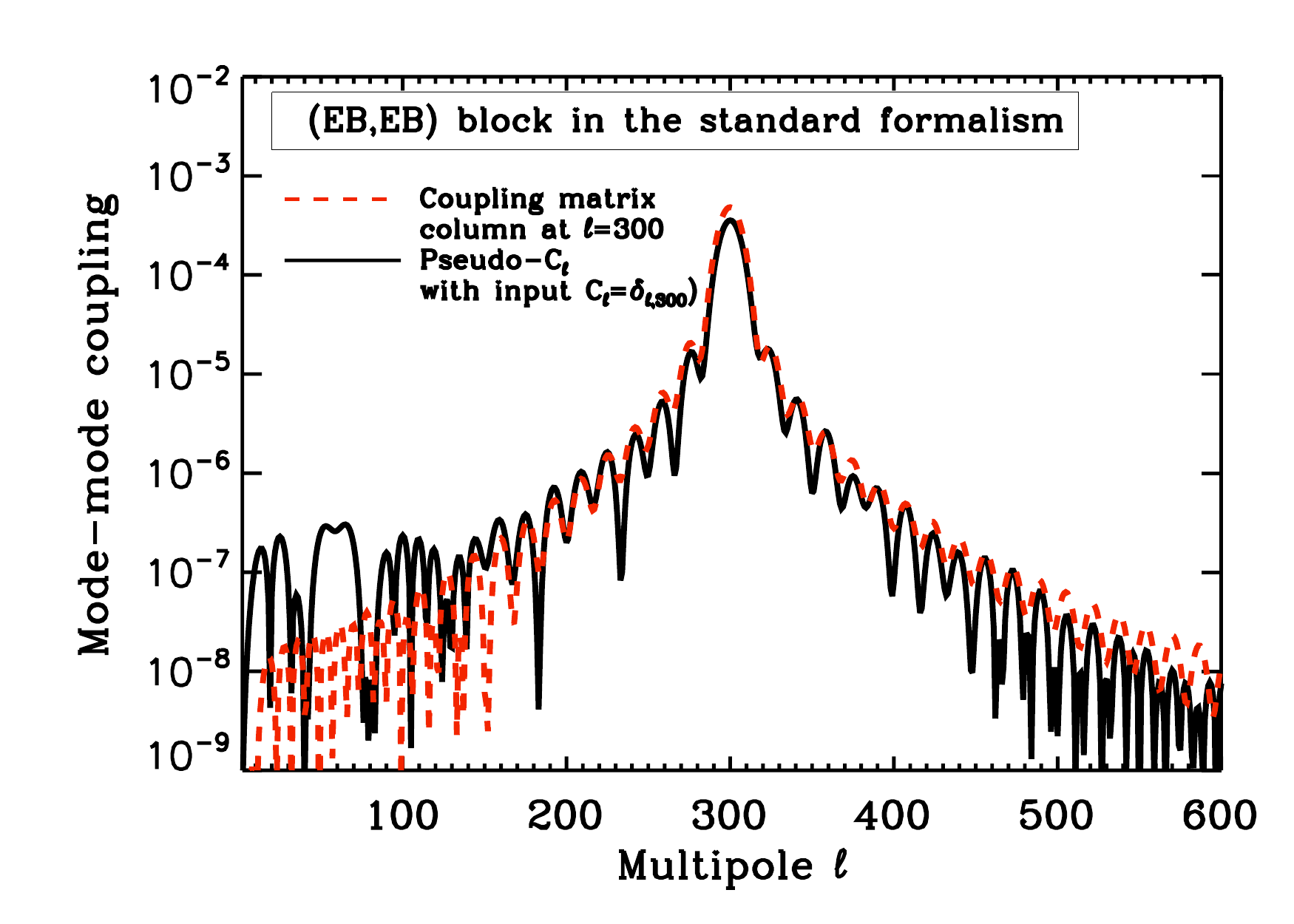}\includegraphics[scale=0.35]{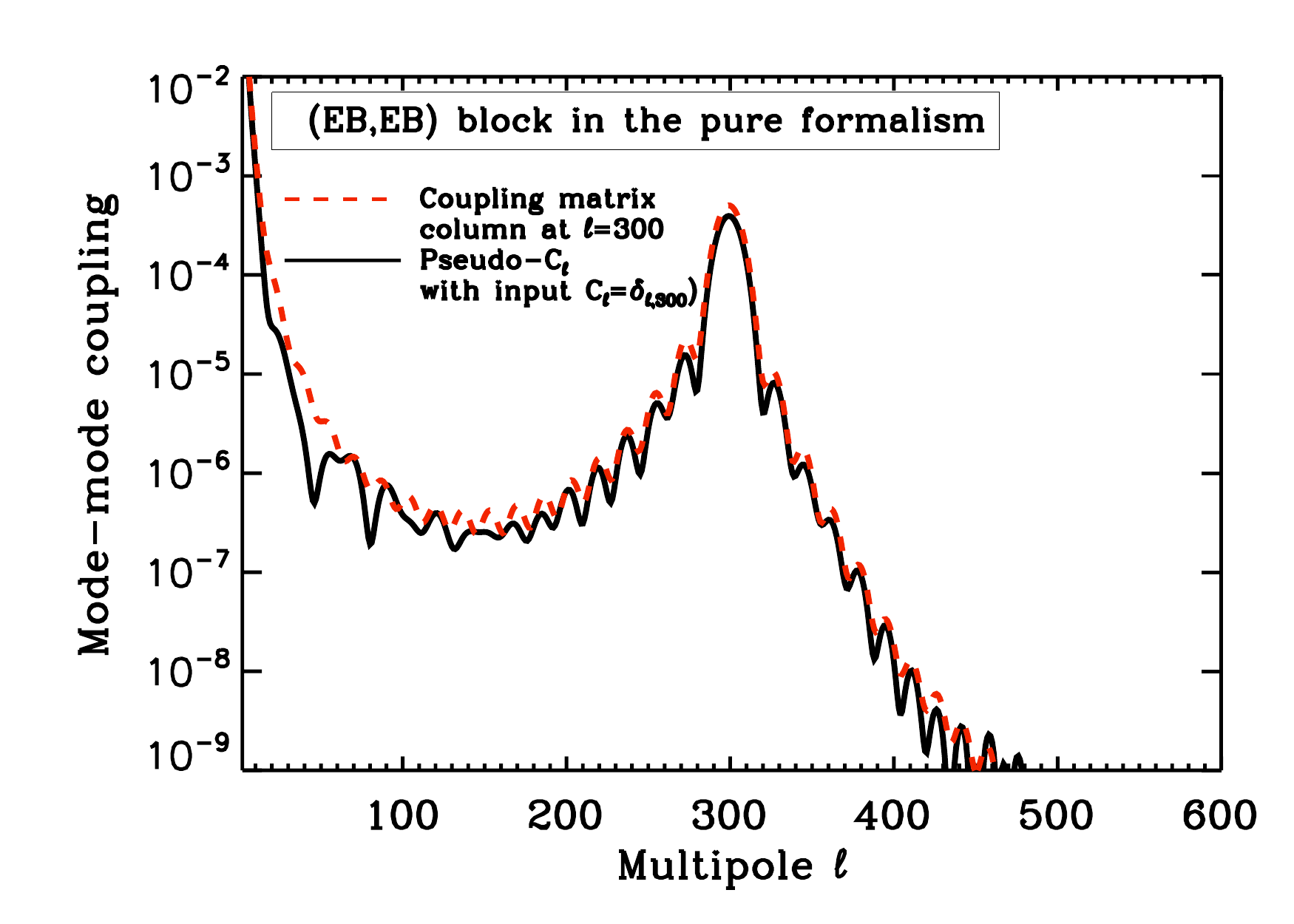}\includegraphics[scale=0.35]{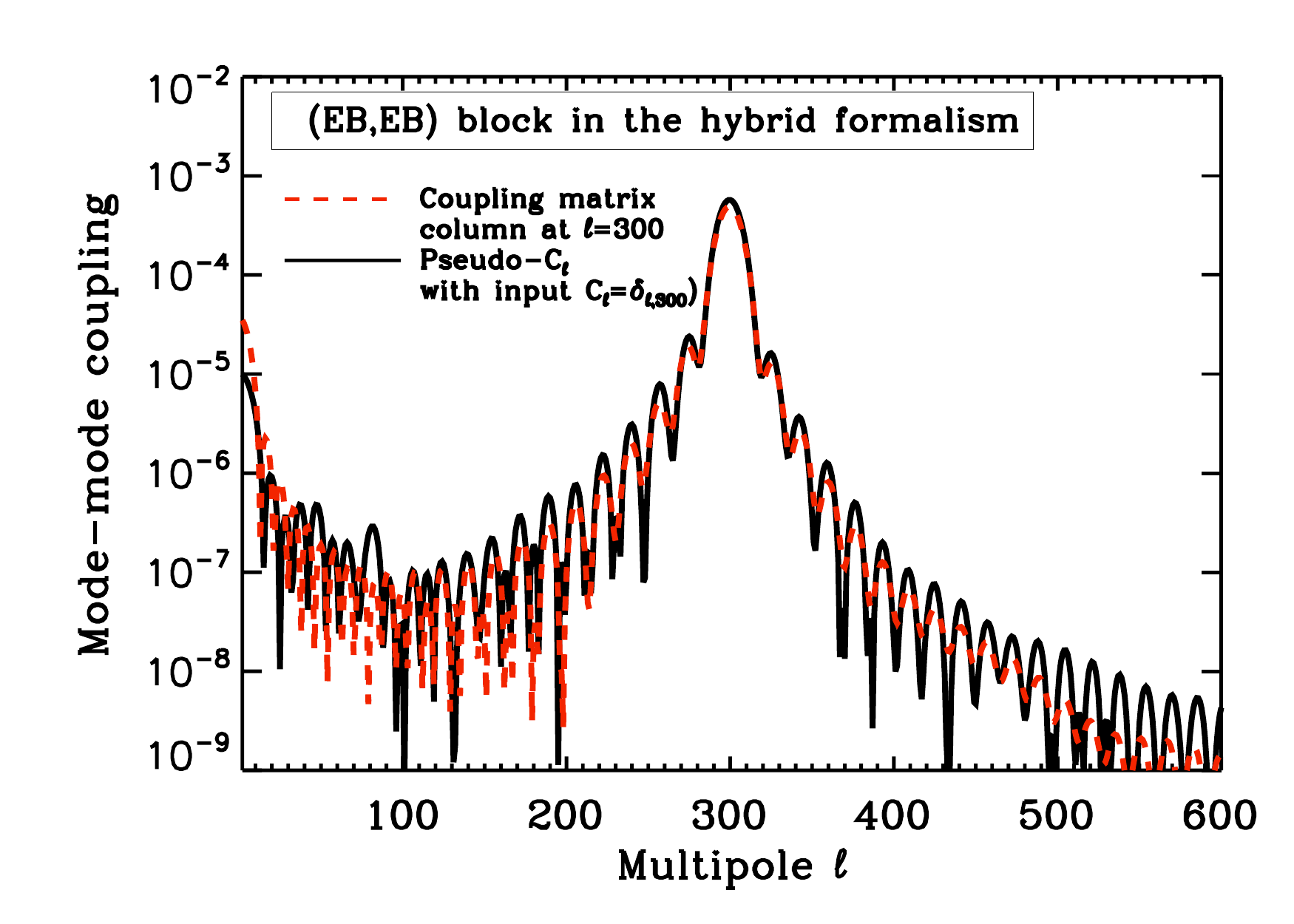}
	\caption{The red-dashed curves corresponds to the $\ell=300$ column of the $(EB,EB)$ coupling kernel for the three formalisms: standard, pure, and hybrid formalisms (from left to right). The overlapping black curve shows the $EB$ pseudospectrum obtained from a map containing $EB$ correlation at $\ell=300$ only. The agreement between the red and black curves underlines the correctness of the numerical computation of the different the mode-mode coupling matrices.}
	\label{fig:kernel2}
\end{center}
\end{figure*}	

\subsection{Pixelization impact on the mode-mode coupling matrices}
To test our implementation of the mode-mode coupling matrices, we compare it to the pseudospectrum obtained for an initial power spectrum with nonzero power only at a single angular frequency, i.e., 
$\propto \delta_{\ell,\ell_0}$. For example, the $TE$ pseudospectrum derived from an initial power spectrum such as $C^{TE}_{\ell}=\delta_{\ell,\ell_0}$ and $C^{TB}_{\ell}=0$ is equal to the elements of the $\ell_0$ th column of $M^{TE,TE}_{\ell,\ell'}$. Similarly, the $TB$ pseudospectrum for such a process is given by the $\ell_0$ th column of the $M^{TB,TE}_{\ell,\ell'}$ block. Such a calculation has also an additional advantage as the resulting pseudospectra include other effects, which could potentially lead to nonzero couplings, for instance, such as  possible pixelization effects, and which are not taken into account in the computation of the mode-mode coupling matrices. This test has already been applied in \cite{grain_etal_2009} to the $(EE,EE)$, $(BB,BB)$, and $(EE,BB)$ blocks to show that most of the pixelization effects are carefully taken into account and we will only present our results concerning the remaining blocks of the full mode-mode coupling kernels.

\begin{figure}
\begin{center}
	\includegraphics[scale=0.245]{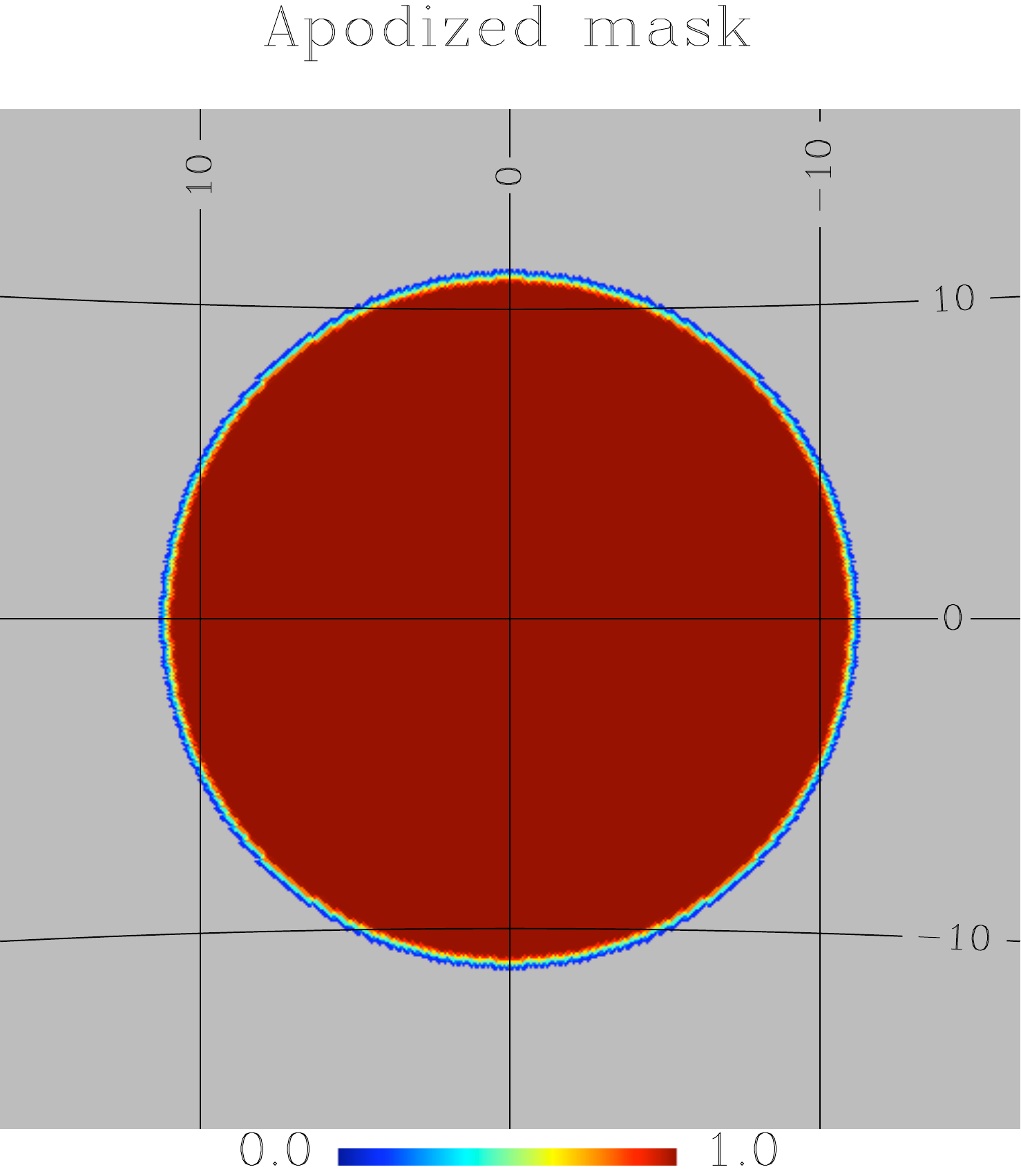} \includegraphics[scale=0.245]{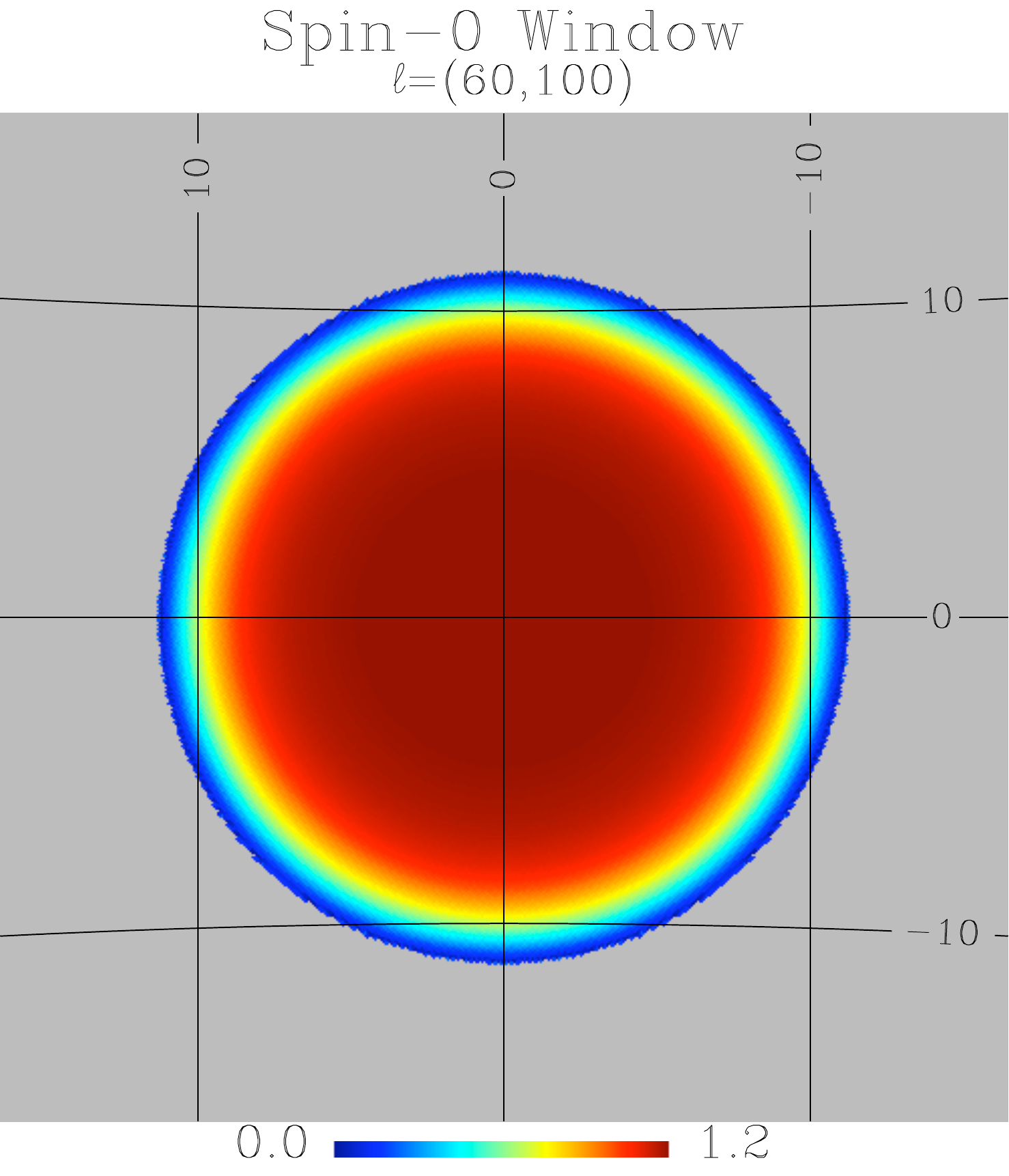} \includegraphics[scale=0.245]{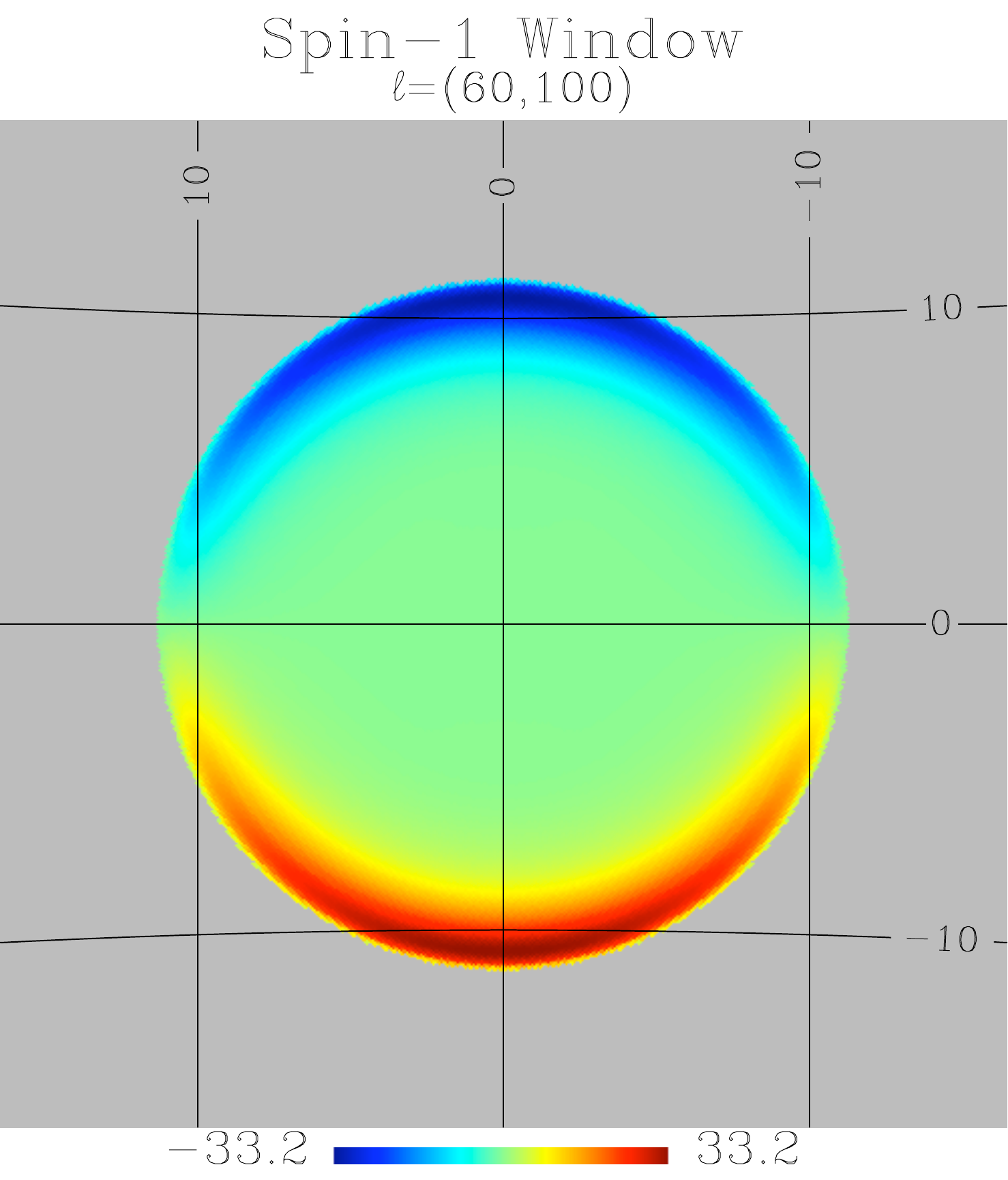} \includegraphics[scale=0.245]{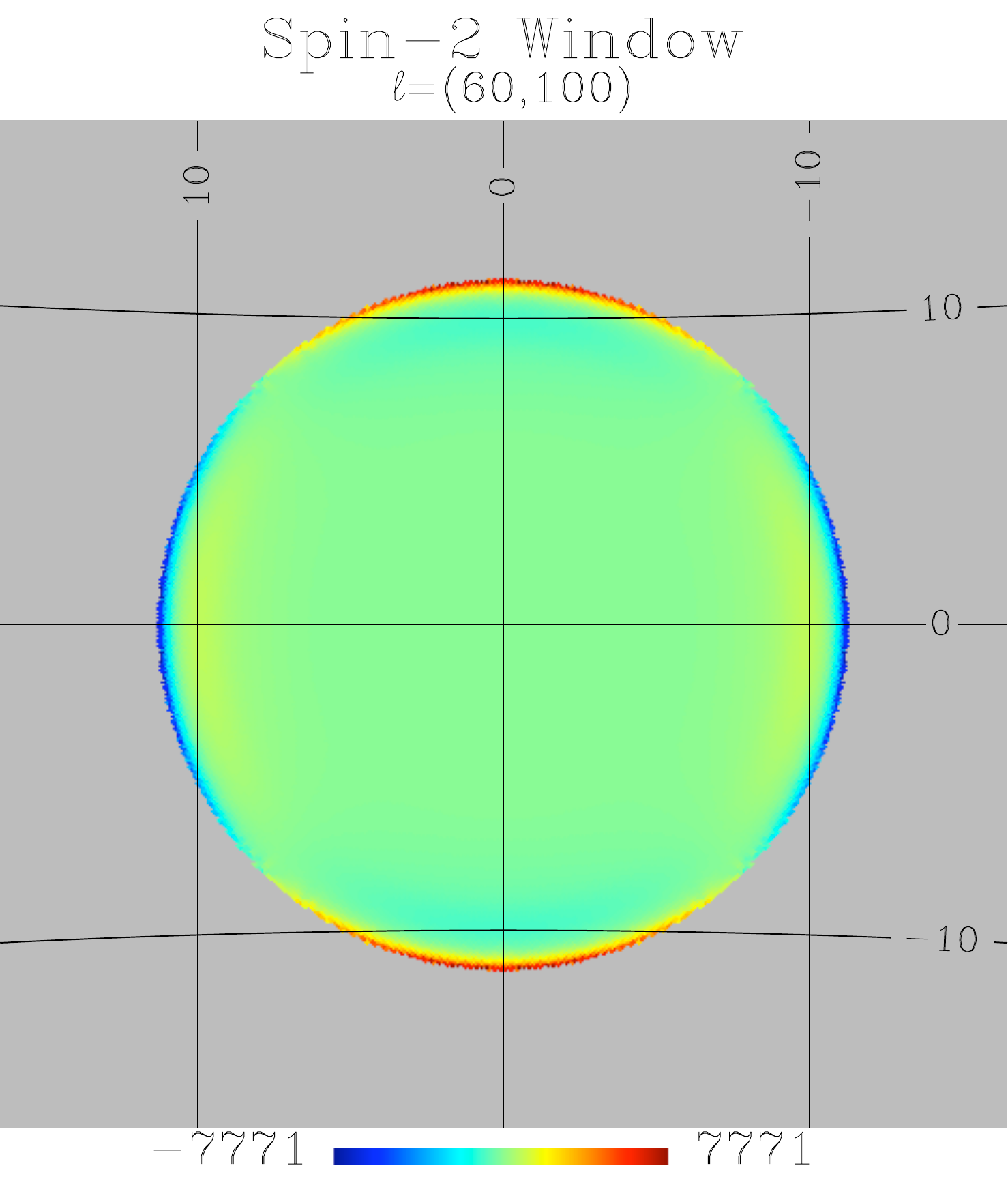}
	\caption{Analytic sky apodization (to be applied to intensity map and polarization map for standard pseudo-$a_{\ell m}$'s estimate) and spin-0, spin-1, and spin-2 sky apodizations, optimized for the second bin, $\ell\in(60,100)$ (to be applied to the polarization maps for pure pseudo-$a_{\ell m}$'s estimate), in the case of the idealized mock survey. (Only the real parts of the spin-1 and spin-2 windows are shown.)}
	\label{fig:SkyapIdeal}
\end{center}
\end{figure}

An example of such tests is shown on Fig. \ref{fig:kernel1} for the case of the $(TE,TE)$, $(TB,TB)$ and in Fig. \ref{fig:kernel2} for $(EB,EB)$ blocks in the three proposed formalism (standard, pure and hybrid formalisms). The sky coverage adopted here is a spherical cap with a radius of $\sim$11.3 degrees. The sky apodizations used are the analytic windows with a 30 arcminutes apodization length to be applied to temperature and polarized {\it standard} pseudo-$a_{\ell m}$'s (see \cite{grain_etal_2009} for details) and the pixel-based optimized windows for polarized pure pseudo-$a_{\ell m}$'s. (Those peculiar sky apodizations are shown on Fig. \ref{fig:SkyapIdeal}.) In this peculiar case, a first sky apodization is identically applied to compute standard $E$-and $B$-pseudo-multipoles and a second one is identically applied to compute pure $E$-and $B$-pseudo-multipoles. As already mentioned in the previous section, with such a weighting, the $(TE,TE)$ block in the standard and hybrid formalisms equals the $(TB,TB)$ block in the standard formalism, and  the $(TE,TE)$ block in the pure formalism equals the $(TB,TB)$ block in the pure  and hybrid formalisms. Therefore, only the $(TE,TE)$ block in the standard approach and in the pure approach is depicted in Fig. \ref{fig:kernel1}. However, the $(EB,EB)$ block systematically differs from one formalism to another and such a block is displayed in Fig. \ref{fig:kernel2} for the three techniques. In such figures, the red-dashed curves correspond to the $\ell=300$ column of the computed mode-mode coupling while the overlapping black curves  stand for $TE(TB)$ and $EB$ pseudospectra as derived  from a map containing correlated $T$-, $B$-, and $E$-modes at $\ell=300$ only. The overall agreement of red-dashed and black curves pins down the right computation of the mode-mode coupling matrices. The agreement is nevertheless not perfect for two reasons. First, the equality between the column of the mode-mode coupling and the pseudospectrum from a single mode process is exact on {\it average}. However, the black curves displayed on Figs. \ref{fig:kernel1} and  \ref{fig:kernel2} are obtained from one realization only, explaining the scattering of the black curves around the red ones. Second, we cannot exclude the influence of residual pixel effects.

As already explained, the $(TB,TE)$ and $(EB,PP)$ blocks are nonvanishing (though expected to be small) in the peculiar case of the pure and hybrid formalisms {\it if and only if} the spin-1 and spin-2 sky apodizations have a nonzero $B$-component. This is the case of pixel-based optimized sky apodizations. However, the above described tests shows that within the HEALPix scheme, and irrespectively of any formalisms, {\it unmodeled} pixel effects induce some nonvanishing $(TB,TE)$ and $(EB,PP)$ blocks with an amplitude {\it higher} than the cut-sky effects expected in the the pure and hybrid formalisms\footnote{As is already shown in \cite{grain_etal_2009} (Sec. IV.A.5) for the specific case of $EE$ and $BB$ spectra , such unmodeled pixelization effects are significantly reduced by using ECP-like pixelization schemes (for which the number of azimuthal grid points remains unchanged for each constant zenithal ring and therefore providing a nearly exact quadrature over the polar angle).}. However, the pixel-induced nonvanishing $(TB,TE)$ and $(EB,PP)$ blocks in the HEALPix scheme remains much smaller than the cut-sky (and modeled) $(TE,TE)$ and $(EB,EB)$ blocks. We will therefore neglect such effects and systematically set the $(TB,TE)$ and $(EB,PP)$ blocks to zero in the following. 

As the pixel-induced leakages are not corrected, the final estimate of {\it all} the angular power spectra but the $TT$ one is {\it a priori} biased as long as the $TB$ and $EB$ correlations do not vanish. Fortunately, such cross-correlations are expected to be zero in the standard cosmological paradigm. Moreover, we have tested that with nonvanishing $TB$ and $EB$ input power spectra at a level of ${\ell(\ell+1)}C^{TB}_\ell/(2\pi)\sim0.2~\mu K^2$ and ${\ell(\ell+1)}C^{EB}_\ell/(2\pi)\sim0.004~\mu K^2$ at $\ell=2-1000$
(as motivated by the potential impact of an helical, primordial magnetic field \cite{caprini_etal_2004,kahniashvili_etal_2005}),  the $EE$, $BB$, $TE$, $TB$, and $EB$ angular power spectra estimated on 1\% of the sky with inhomogeneous noise (see Fig. \ref{fig:Nhit} for the noise distribution) are indeed not biased though residual $EB\leftrightarrow EE$, $EB\leftrightarrow BB$, and $TB\leftrightarrow TE$, leakages are not corrected for \cite{ferte_etal_2011}.

\section{Applications to small-sky experiments: power spectra uncertainty}
\label{sec:Application} 
In this section we discuss the performance and applicability of the above defined odd-parity power spectra estimators  in the context of vanishing true $TB$ and $EB$ cross-correlations, i.e., considered as a null test, potentially used for search of astrophysical and/or instrumental systematic effects. The uncertainty of each estimator is quantified by computing the power spectrum variance thanks to Monte-Carlo simulations. Our fiducial model for the input $T$ and $E$ CMB signal is given by the cosmological parameters as constrained by the WMAP 7-year data. The input $B$-mode is composed of a primordial component with a tensor-to-scalar ratio $r=0.05$ and a secondary component induced by lensing\footnote{Our convention for $r$ follows the WMAP convention: $r=\mathcal{P}_\mathrm{T}(k_0)/\mathcal{P}_\mathrm{S}(k_0)$ with $\mathcal{P}_\mathrm{S(T)}$, the {\it primordial} scalar(tensor) power spectrum and $k_0=0.002$~Mpc$^{-1}$ the pivot scale.}. The two input odd-parity power spectra are set to zero (as is predicted in a parity invariant cosmological scenario).

Our starting point is an idealized mock survey considered below covering a circular patch of  $\sim$1\% of the sky area with a level of homogeneous noise set to $5.75$ $\mu$K-arcmin for the three Stokes parameters. In order to get closer to realistic sky observations, two possible extensions of this ideal mock survey are implemented. We first consider the effect of more intricate boundaries using a square patch with holes due to {\it e.g.} point-sources removal, and later an implementation of an inhomogeneous sky survey as expected for balloon-borne experiment. The sky coverage is always taken to be $\sim$1\%  and the RMS of the noise in the inhomogeneous case is set equal to $5.75~\mu$K-arcmin. 
The values of both these parameters are motivated by the {\sc ebex} experiment \cite{britt_etal_2010}.
In addition, in all cases we assume that noise is uncorrelated from pixel to pixel and from one Stokes parameter to another, i.e., 
\begin{eqnarray}
\left<N_{S}(i)N_{S'}(j)\right>=\sigma^2_{S}(i)\delta_{i,j}\delta_{S,S'}.
\label{eqn:noiseMapDef}
\end{eqnarray}
As a consequence, the $TE$, $TB$ and $EB$ noise biases vanish (see Appendix \ref{app:Noise}).  

Hereafter, the variance is estimated as the standard deviation of $500$ MC simulations. The input \{CMB + noise\} maps are built using the HEALPix pixelization scheme at a resolution of $N_\mathrm{side}=512$ corresponding to a pixel size of $\sim7$~arcmin. The estimated power spectra are binned with a band width of $\Delta\ell=40$ and with the lowest bin starting at $\ell=20$. The pseudo-$a_{\ell m}$ for temperature are estimated applying the arc cosine apodization~\cite{grain_etal_2009} with an apodization length of $30$ arcminutes. The same apodization is used while calculating the standard pseudo-multipoles for polarization from the $Q$ and $U$ maps. For the pure-pseudo-multipoles of $E$- and $B$-modes computation we use a pixel-based, optimized sky apodization to the $Q$, $U$ maps.

As a reference against which to compare the variances obtained from the MC simulations we use a theoretical --and optimistic-- Fisher estimate of the variance, which for a cross-spectrum of two sets of $(I,Q,U)$ maps, labelled by $(A)$ and $(B)$, reads\footnote{We point out that in \cite{grain_etal_2009}, a factor $1/2$ is missing in front of the last term (pure noise term) of their $f_{\mathrm{sky}}$-formulas. Their equation (32) for cross-spectrum should read
$$
\Delta\mathcal{C}^B_\ell=\sqrt{\frac{2f^{-1}_\mathrm{sky}}{(2\ell+1)}\left(C^{B~2}_{\ell}+C^B_\ell\frac{\sigma^2}{B^2_\ell}+\frac{1}{2}\frac{\sigma^4}{B^4_\ell}\right)}.
$$
}

\begin{widetext}
\begin{eqnarray}
	\Delta^{X_{(A)}Y_{(B)}}_\ell&=&\frac{1}{(2\ell+1)f^{X_{(A)}Y_{(B)}}_\mathrm{sky}}\left[\left(C^{XY}_\ell+\frac{4\pi}{N_\mathrm{pix}}\sigma^2_{X_{(A)}Y_{(B)}}B^{-1}_{\ell,(A)}B^{-1}_{\ell,(B)}\right)^2\right. 
	\label{eq:fishvar}
	\\
	&+&\left.\left(C^{XX}_\ell+\frac{4\pi}{N_\mathrm{pix}}\sigma^2_{X_{(A)}X_{(A)}}B^{-2}_{\ell,(A)}\right)\left(C^{YY}_\ell+\frac{4\pi}{N_\mathrm{pix}}\sigma^2_{Y_{(B)}Y_{(B)}}B^{-2}_{\ell,(B)}\right)\right]. \nonumber
\end{eqnarray}
\end{widetext}
 Here, $\Delta^{X_{(A)}Y_{(B)}}_\ell\equiv\mathrm{Var}\left[C^{X_{(A)}Y_{(B)}}_\ell\right]$ denotes the variance of a cross-spectrum of $X_{(A)}$ and $Y_{(B)}$ where $X$ and $Y$ stand for either $T,~E$, or $B$ as derived from a map either $A$ or $B$, respectively, $N_\mathrm{pix}=12N^2_\mathrm{side}$ is the total number of pixels, and $B_{\ell,(A)}$ stands for a resolution of the map $A$. Given our assumptions about the noise, Eq.~\eref{eqn:noiseMapDef},  $\sigma^2_{X_{(A)}Y_{(B)}}=\sigma^2_{X_{(A)}X_{(A)}}\times\delta_{X,Y}\times\delta_{A,B}$, where  $\sigma^2_{X_{(A)}X_{(A)}}$ is related to the noise per pixel $\sigma^2_\mathrm{pix}(i)$ via $\sigma^2_{X_{(A)}X_{(A)}}=N^{-1}_\mathrm{obs}\sum_i\sigma^2_\mathrm{pix}(i)$. The effective sky fraction, $f^{X_{(A)}Y_{(B)}}_\mathrm{sky}$, is defined as
\begin{equation}
f^{X_{(A)}Y_{(B)}}_\mathrm{sky}=\frac{\left(\displaystyle\sum_{i=1}^{N_\mathrm{pix}}\mathcal{W}^{X_{(A)}Y_{(B)}}_{i}\right)^2}{N_\mathrm{pix}\times\displaystyle\sum_{i=1}^{N_\mathrm{pix}}\left(\mathcal{W}^{X_{(A)}Y_{(B)}}_{i}\right)^2}
\end{equation}
where $\mathcal{W}^{X_{(A)}Y_{(B)}}_{i}=W^{X_{(A)}}_{i}\times W^{Y_{(B)}}_{i}$ is a cross-product of the sky apodizations applied to the maps $A$ and $B$ from which the $X$- and $Y$-modes are estimated. As the sky fraction depends on the chosen sky apodizations, which in general will depend on the adopted formalism, on the map itself, and  the considered $\ell$-bin, the resulting Fisher estimate 
of the error bars defined in Eq. (\ref{eq:fishvar}), will also defined on all these factors. As a unique --formalism independent-- benchmark, we will therefore use the Fisher variance computed for the maximal value of 
$f_\mathrm{sky}$ for each considered map corresponding to setting the sky apodization equal to the binary mask. These theoretical uncertainties therefore correspond to the {\it lowest} Fisher estimate one 
may expect given a peculiar sky fraction, not taking into account specific apodization used in the calculations of the corresponding pseudospectra.

The entire set of operations needed to compute the estimates, i.e., (1) the mode-mode coupling matrices, (2) the six pseudo-cross-spectra, and (3) the six angular cross-power spectra estimates from the noise-debiased pseudo-$C_\ell$, using CMB maps and sky apodizations as inputs, are numerically implemented in the {\sc X}$^2${\sc pure} code. The code is an extension of the {\sc Xspect} and {\sc Xpol} codes \cite{tristram_etal_2005} and is based on the ({\sc {pure}}){\sc s$^2$hat} library --an efficient massively parallel implementation of spin-weighted spherical harmonic transforms \cite{s2hat,pures2hat,hupca_etal_2010,szydlarski_etal_2011}. This code can also be used in an MC setting allowing the user to simulate, and subsequently analyze, random realizations of CMB intensity and polarization maps from the {\it six} angular power spectra as an input model.

\subsection{Results for an ideal mock survey}
\label{subsec:ideal}
\begin{figure*}
\begin{center}
	\includegraphics[scale=0.35]{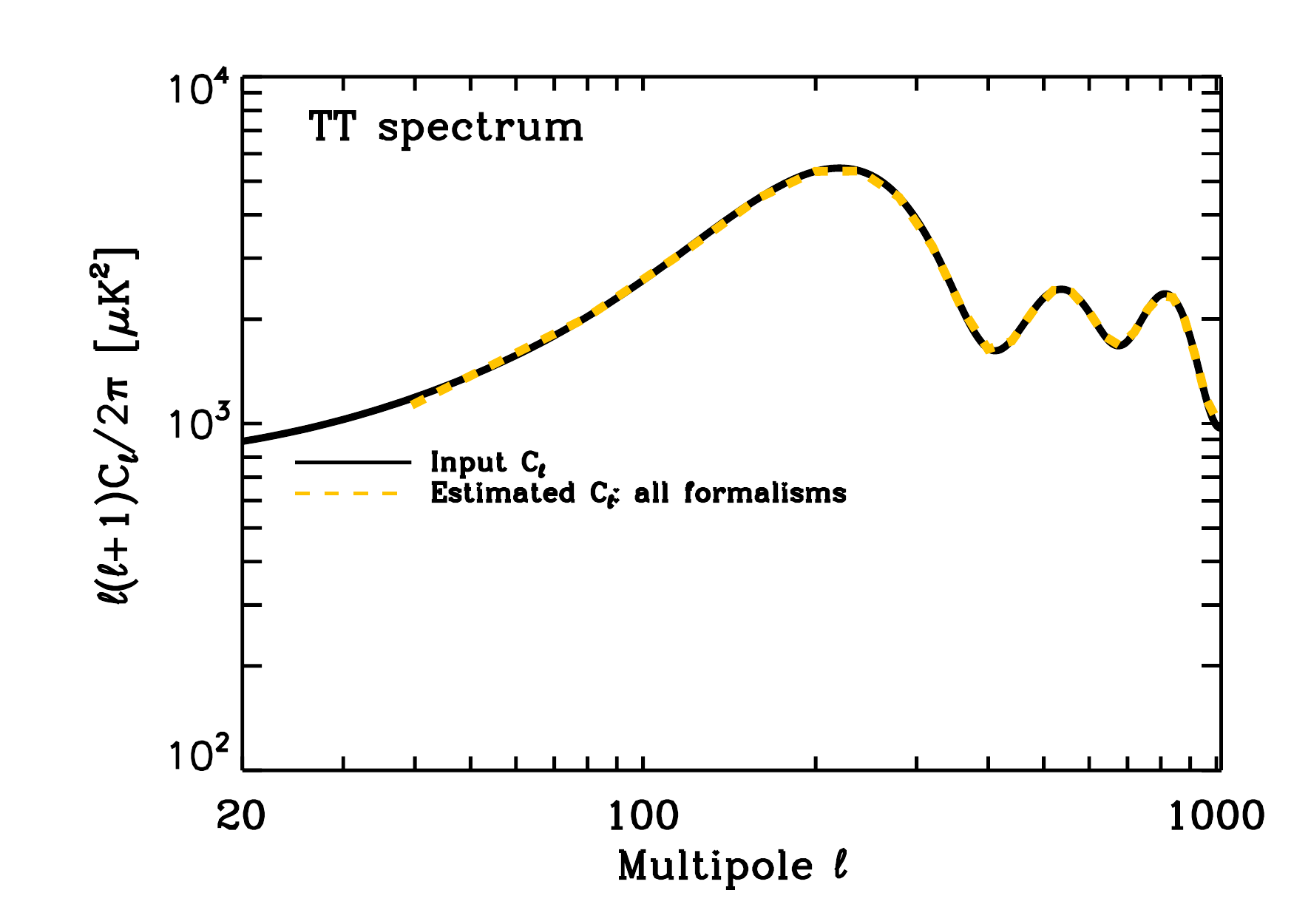} \includegraphics[scale=0.35]{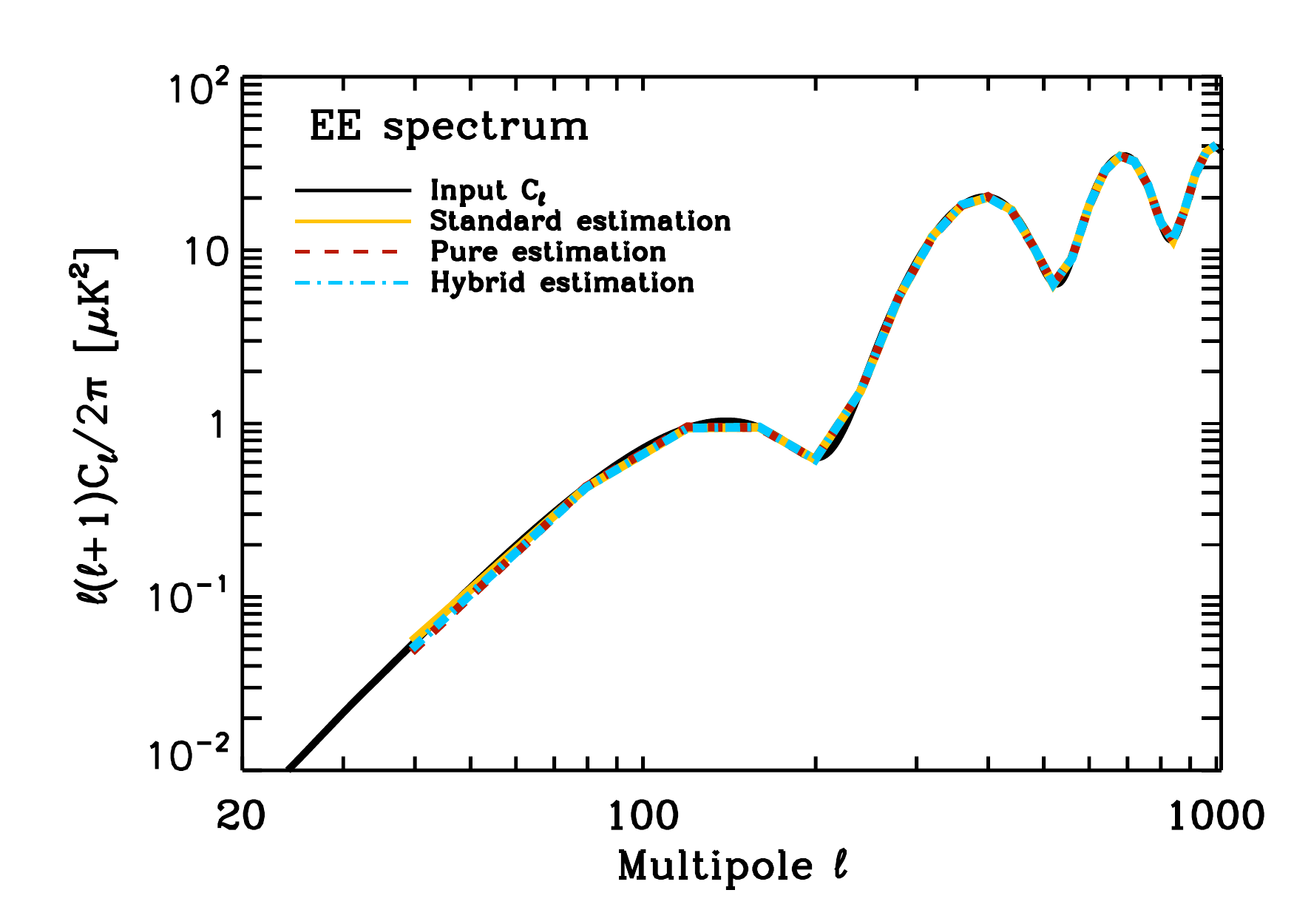}\includegraphics[scale=0.35]{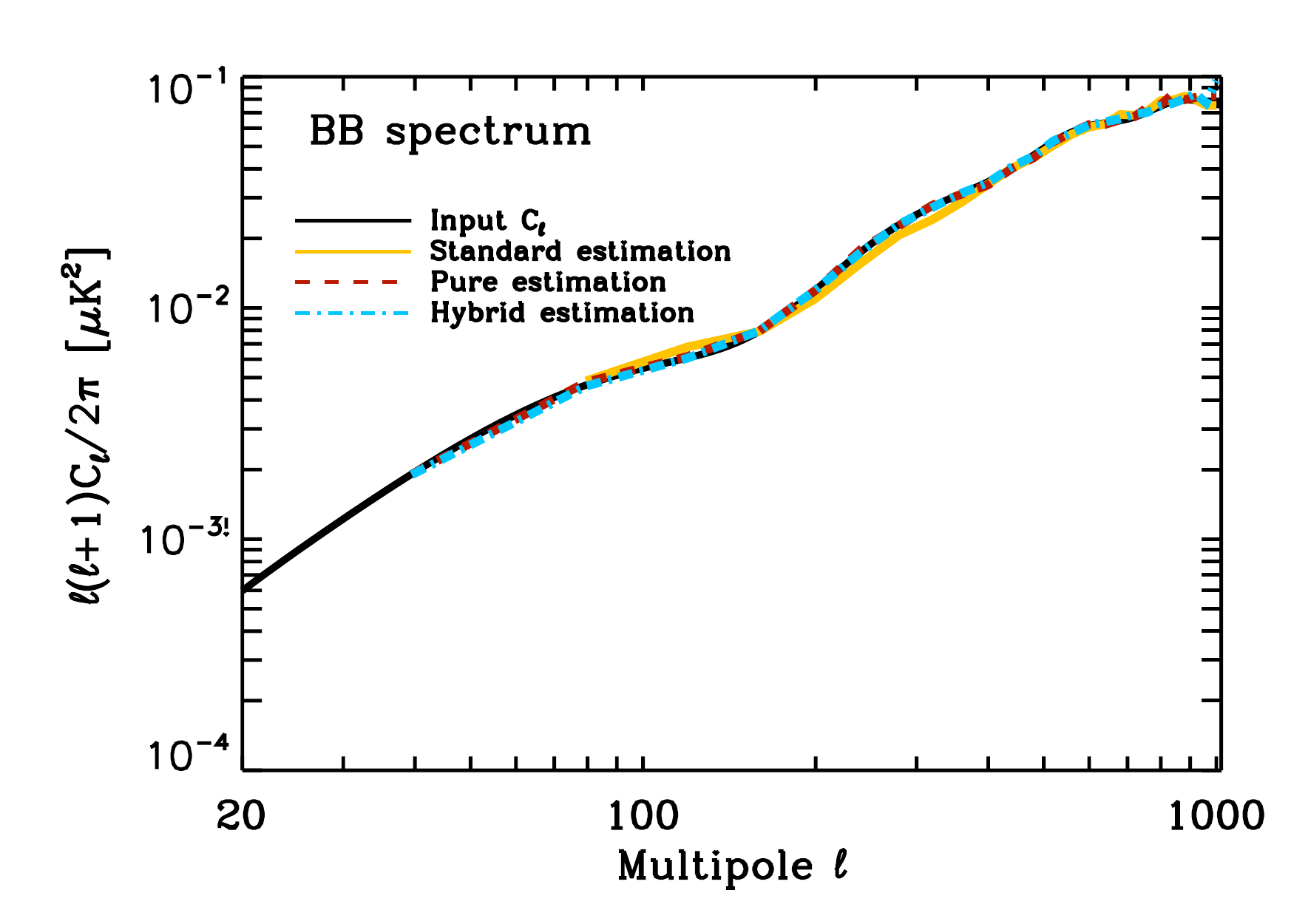}\\
	\includegraphics[scale=0.35]{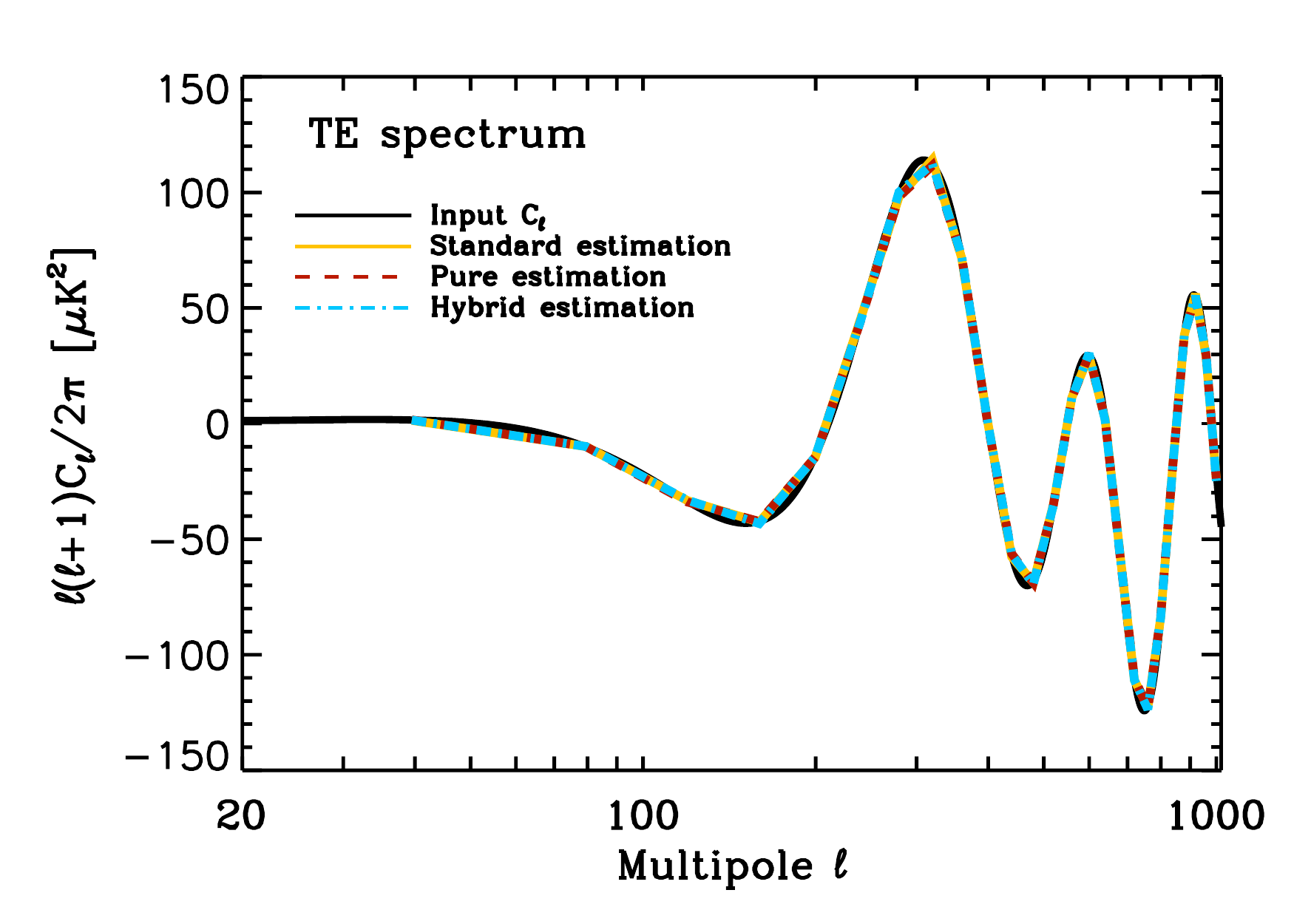} \includegraphics[scale=0.35]{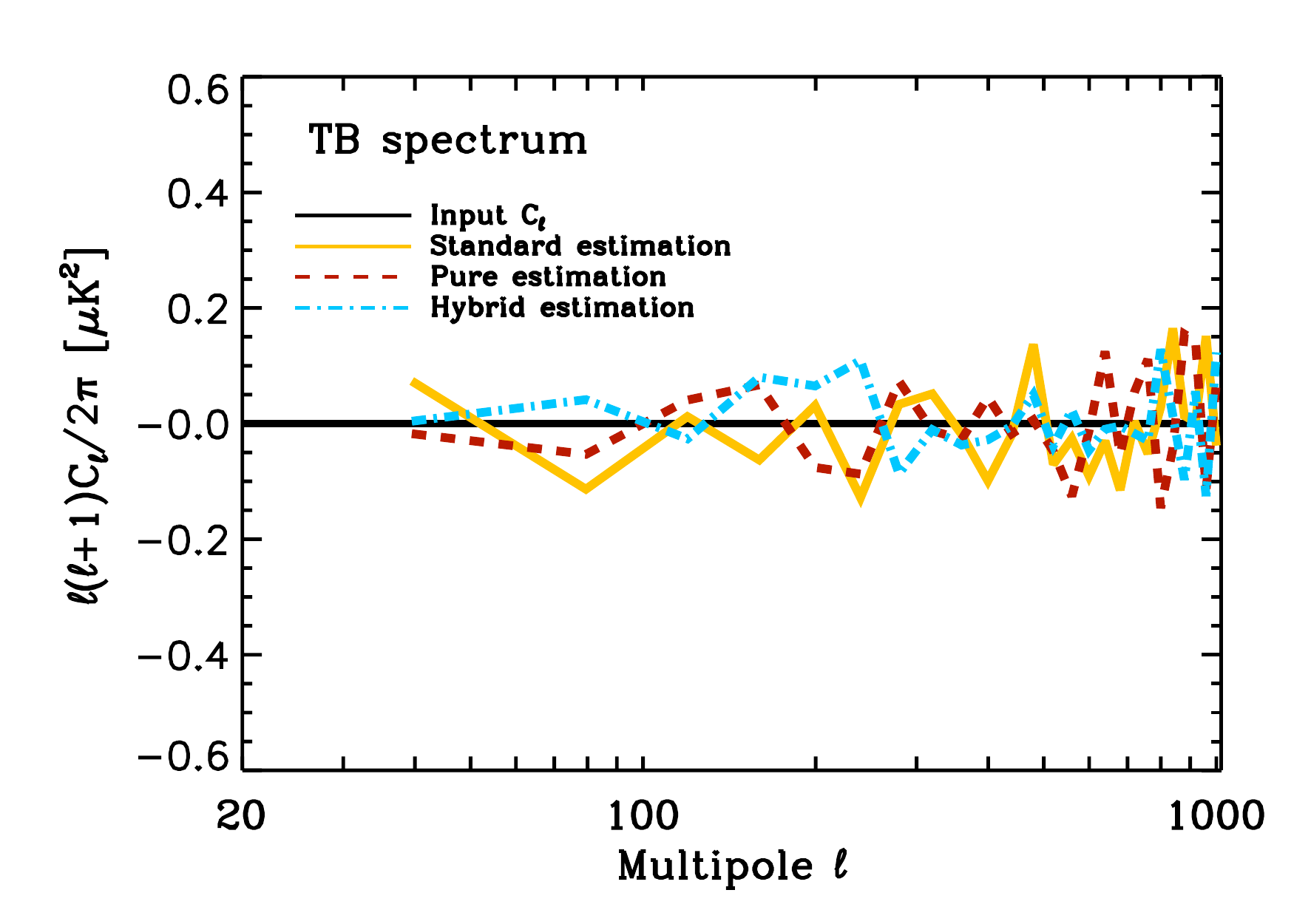}\includegraphics[scale=0.35]{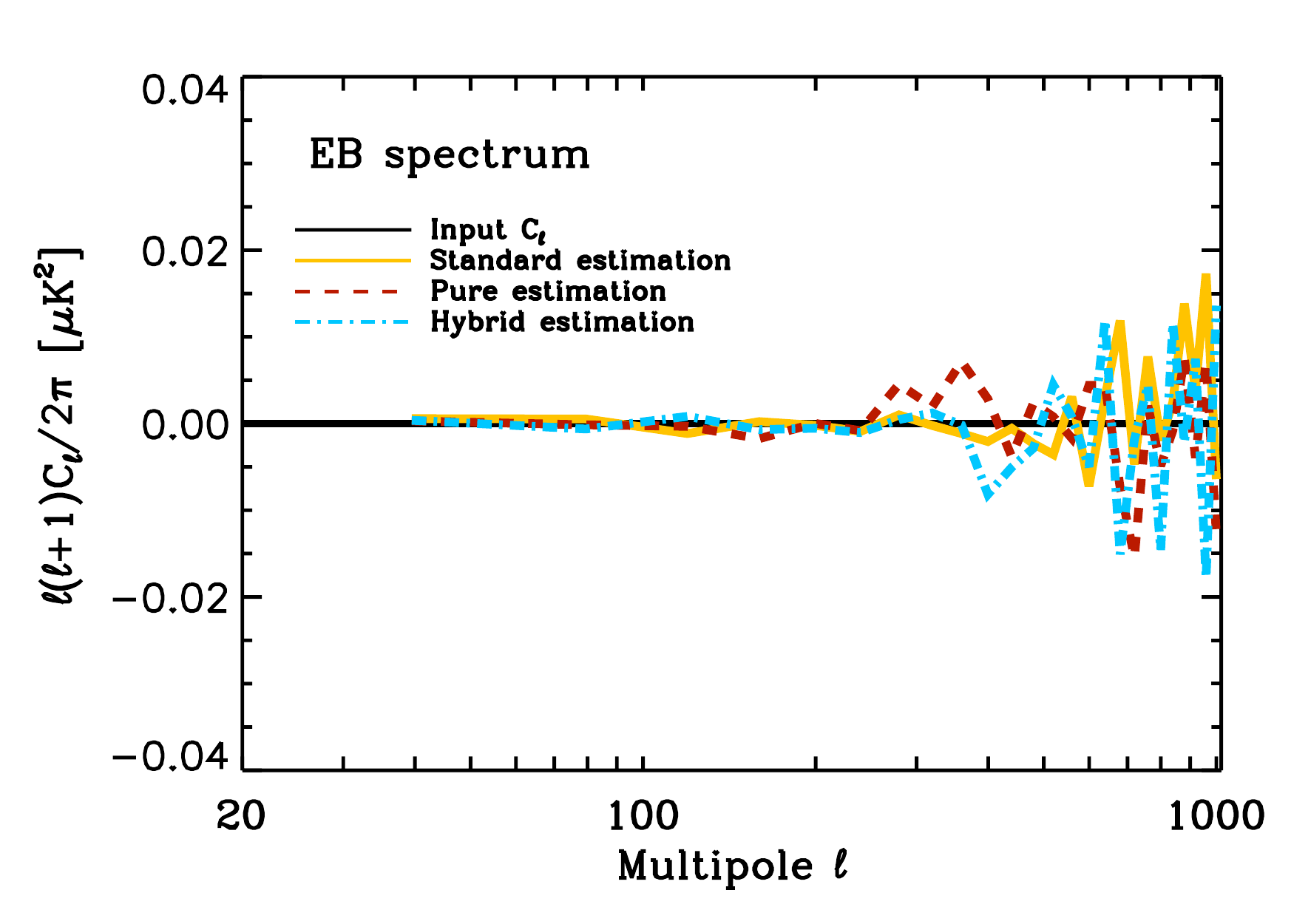}
	\caption{MC averaged estimation of the six angular power spectra (colored curves) alongside the theoretical input $C_\ell$'s (black curves) for the case of an ideal mock survey consisting of a spherical cap with homogeneous noise (see Fig. \ref{fig:SkyapIdeal}). The perfect agreement between the estimated and the theoretical $C_\ell$'s proves that each of the three formalisms leads to an unbiased estimation.}
	\label{fig:cellcap}
\end{center}
\end{figure*}
The idealized mock survey consists of a circular patch with a radius of $\sim$11.3~deg., centered at the equator to reduce pixel effects (see Sec. IV.A.5 of \cite{grain_etal_2009}), and with homogeneous noise at a level of $5.75~\mu$K-arcmin. Such a survey is ideal due to both the simplifying assumption of noise homogeneity and its simple shape. Indeed, a circular patch exhibits the shortest perimeter for a fixed value of the sky coverage. As $\ell$-to-$\ell'$ and $E$-to-$B$ mixing appears as the result of boundary effects, such shape permits us to minimize different mode-mode couplings for a given sky coverage. In addition, we do not allow here for any internal boundaries, e.g., due to masked point sources. The analytic sky apodization --to be applied to intensity map and polarization maps for standard pseudo-$a_{\ell m}$'s--, and the three spin-weighted sky apodization, optimized for the second bin, $\ell\in(60,100)$ --and to be applied to the polarization maps for pure pseudo-$a_{\ell m}$'s--, are displayed on Fig. \ref{fig:SkyapIdeal}.

For this ideal mock survey, the fraction of the sky is $f_\mathrm{sky}\simeq0.96$\%. For the $TT$ power spectrum (all formalisms),  the $EE$ and $TE$ power spectra (standard and hybrid formalisms), and the $BB$, $TB$, and $EB$ (standard formalism), the effective sky coverage is constant and equal to 0.92\%. For the $BB$ power spectrum (pure and hybrid formalism), and the $EE$ and $EB$ power spectra (pure formalism), it varies from 0.64\% at large angular scales ($20\leq\ell<60$) up to 0.95\% at small angular scales ($980\leq\ell<1020$). Finally, for the $TB$ power spectrum (pure and hybrid formalisms), the $TE$ power spectrum (pure formalism), and the $EB$ power spectrum (hybrid formalism), the sky coverage ranges from 0.78\% at large angular scales ($20\leq\ell<60$) up to 0.93\% at small angular scales ($980\leq\ell<1020$).

The MC averaged estimate of the six power spectra ({\it i.e.} $TT,~EE,~BB,~TE,~TB$, and $EB$ spectra), estimated using the three types of formalism, are displayed on Fig. \ref{fig:cellcap} alongside the theoretical input $C_\ell$'s. The perfect agreement between the estimated and the theoretical input shows that each of the formalism indeed leads to an {\it unbiased} estimate of the angular power spectra. As such estimators are built to be unbiased, this essentially confirms the correct numerical implementation of the three formalisms in the {\sc X}$^2${\sc pure} code.

\begin{figure*}
\begin{center}
	\includegraphics[scale=0.35]{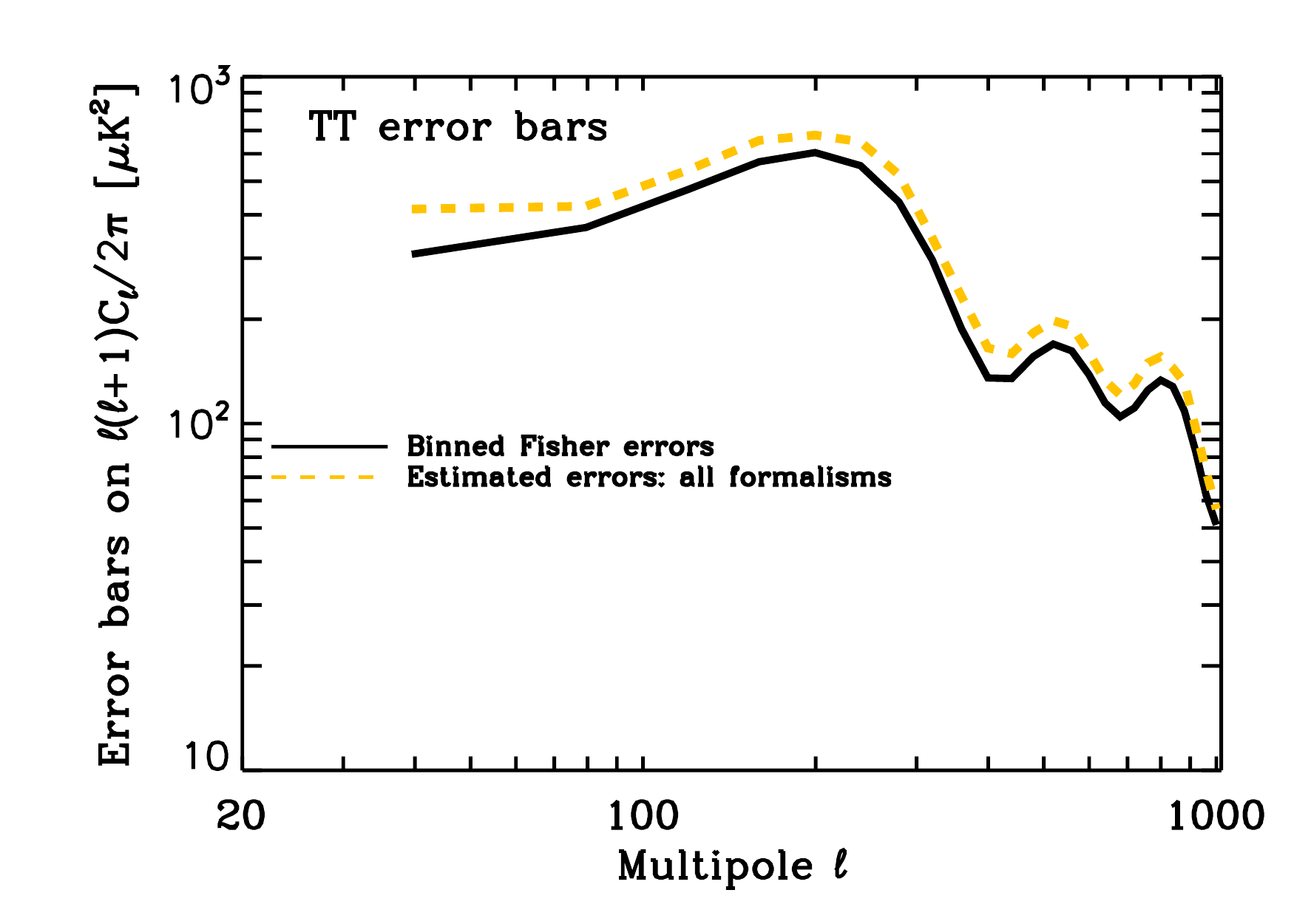}\includegraphics[scale=0.35]{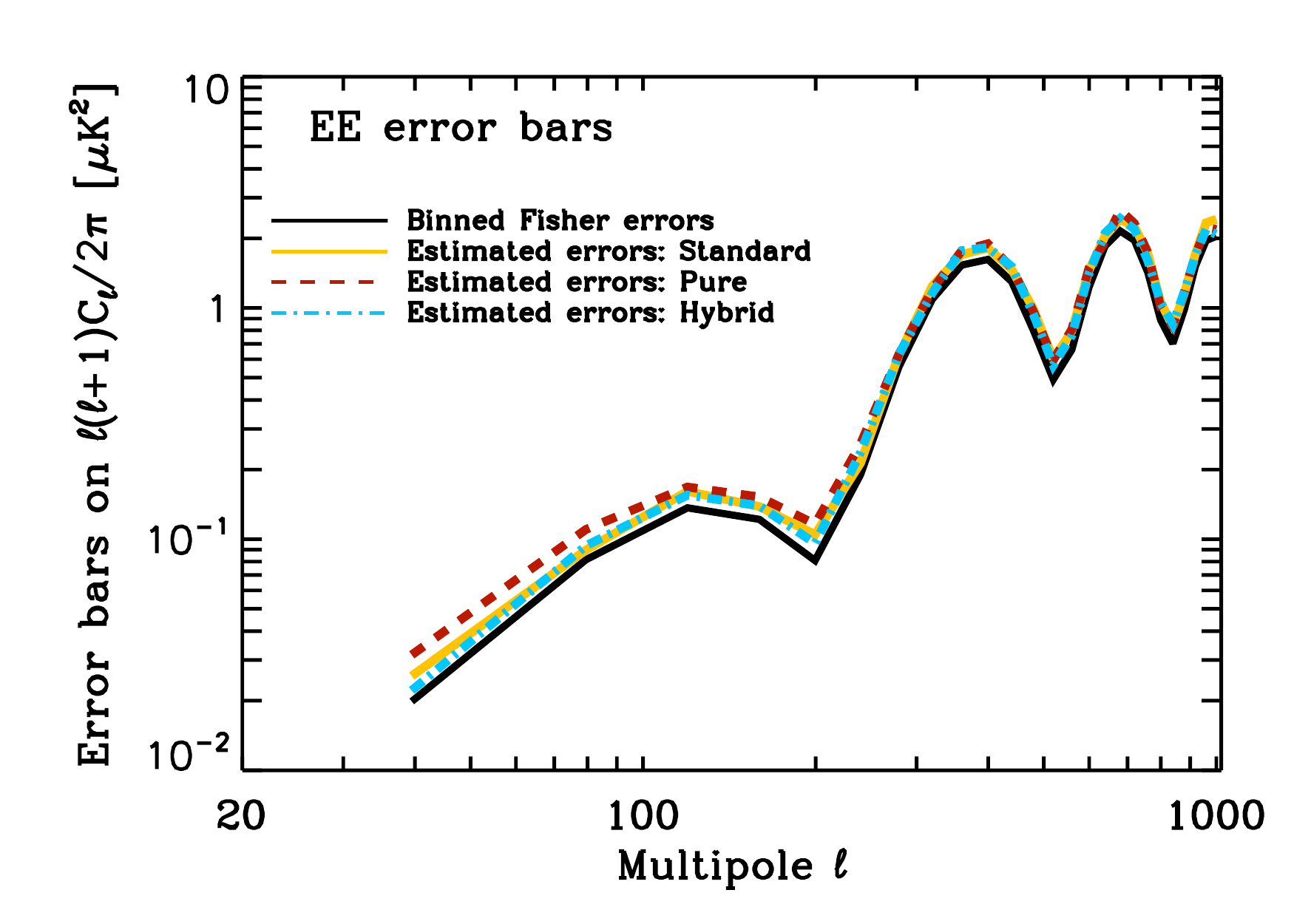}\includegraphics[scale=0.35]{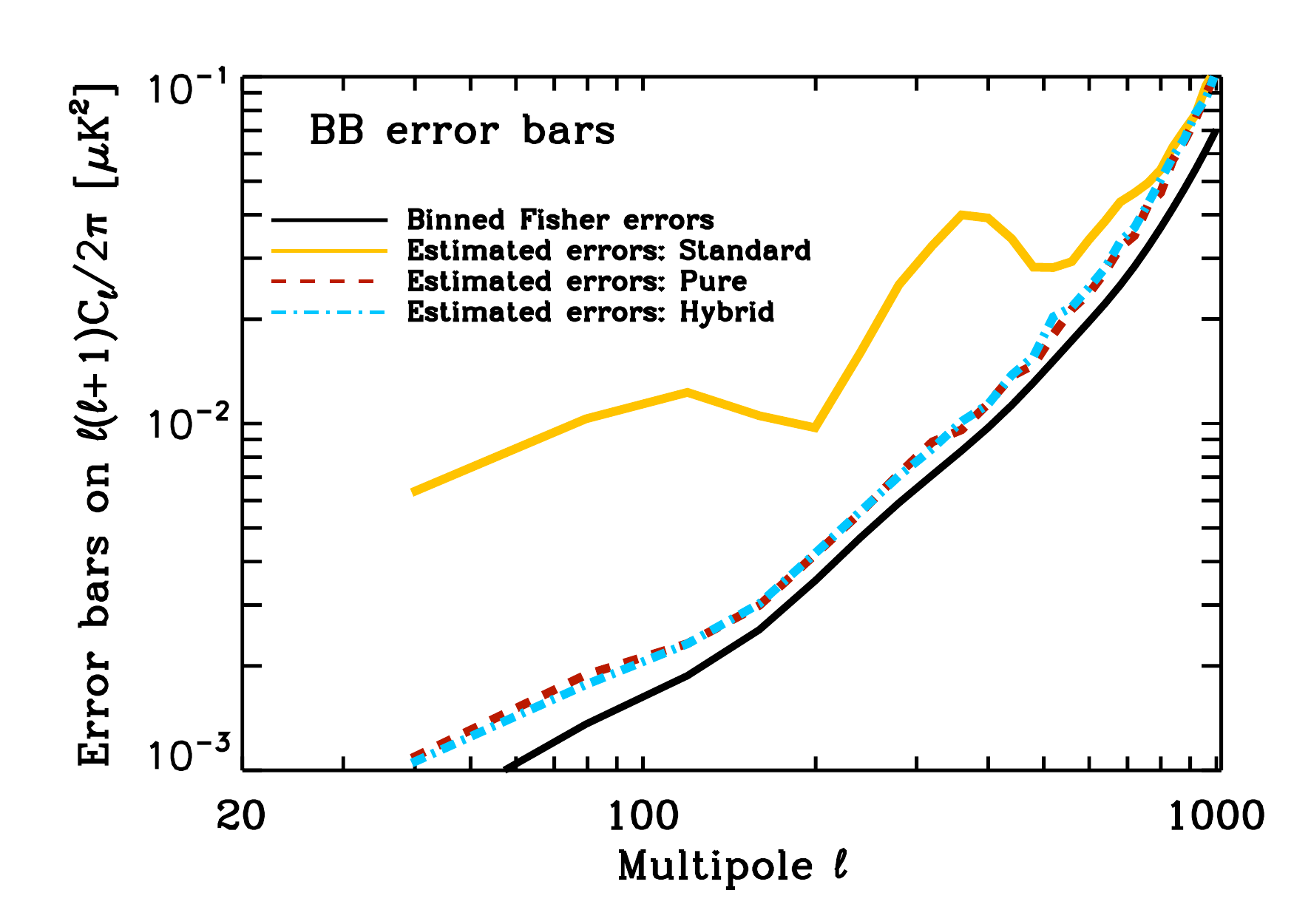}\\ 
	\includegraphics[scale=0.35]{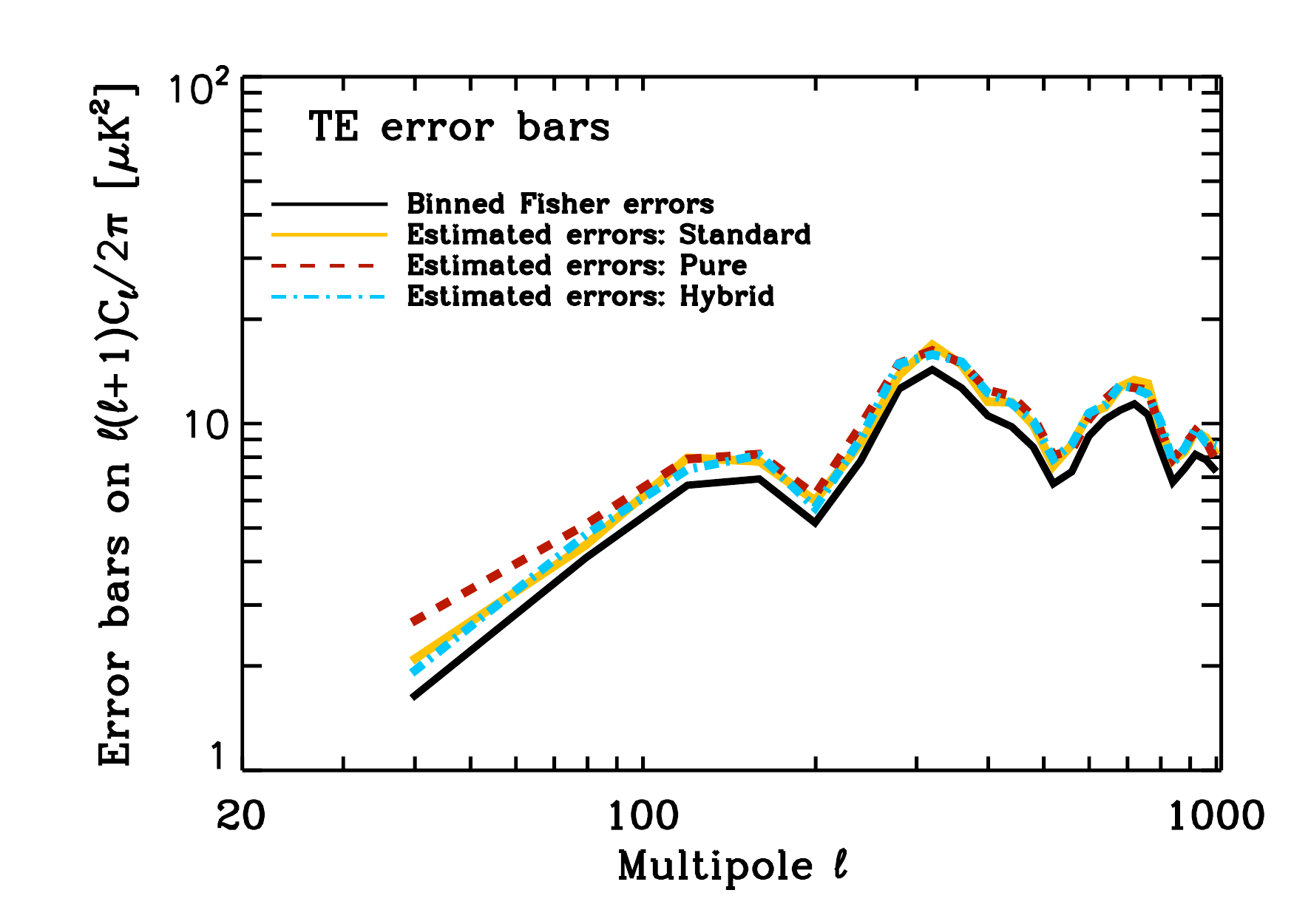}\includegraphics[scale=0.35]{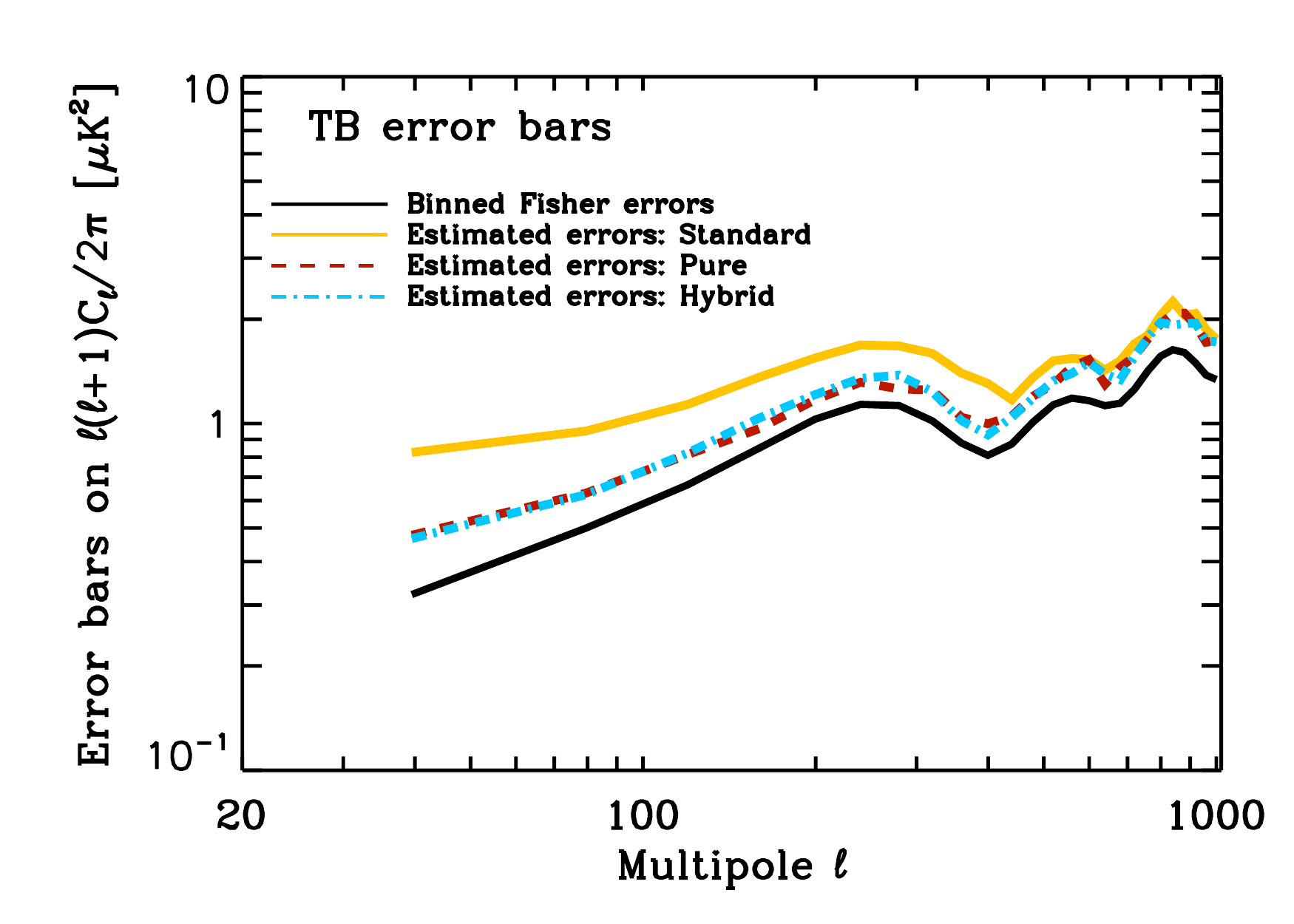}\includegraphics[scale=0.35]{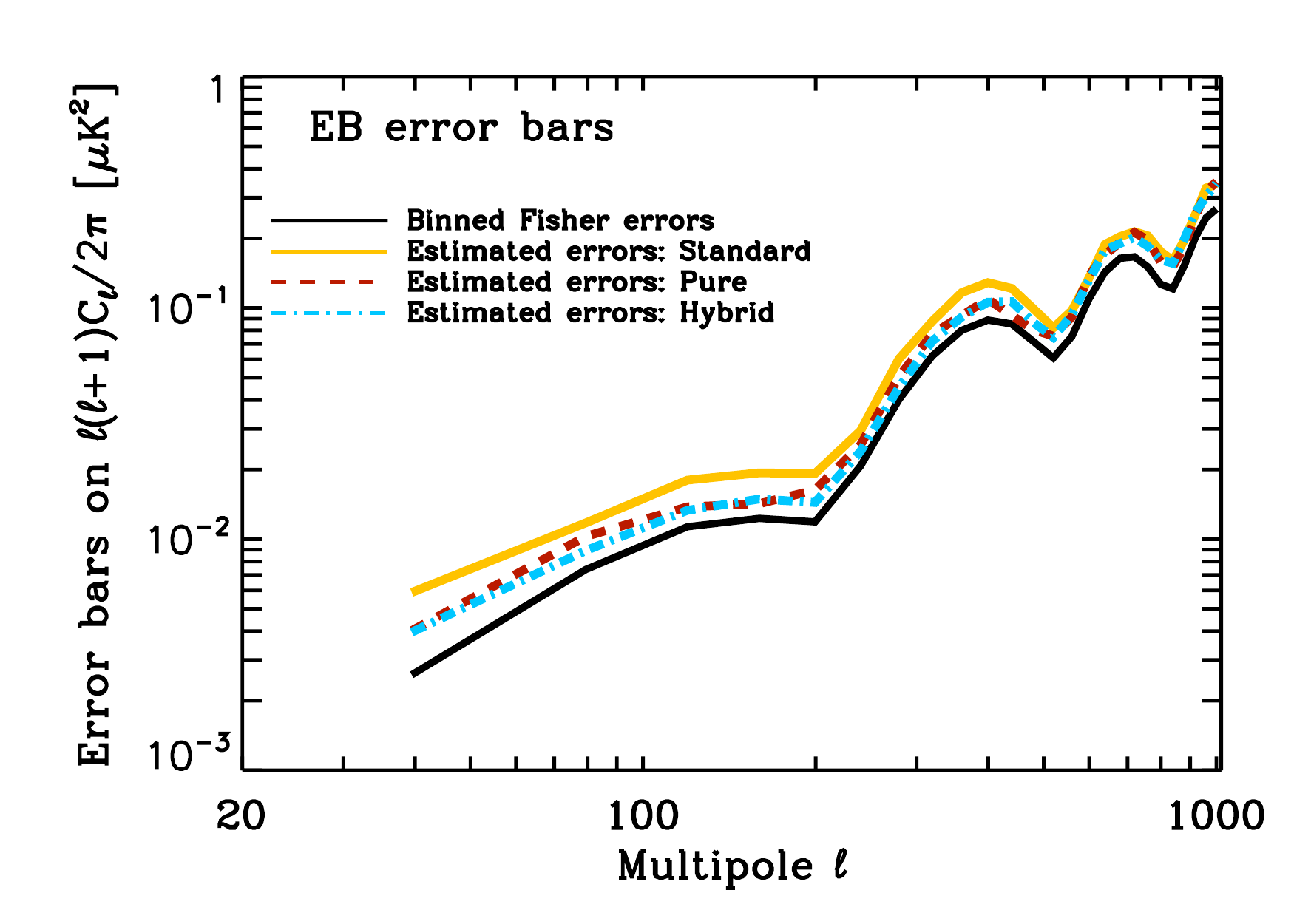}
	\caption{Error bars on the reconstructed angular power spectra for each of the three formalisms (colored curves) alongside the naive (binned) Fisher estimate of such uncertainties. The assumed sky survey is the ideal case consisting of a circular patch with homogeneous noise (see Fig. \ref{fig:SkyapIdeal}). The error bars are obtained as the standard deviation over 500 MC simulations.}
	\label{fig:varcap}
\end{center}
\end{figure*}

The statistical uncertainties of the estimated angular power spectra obtained via MC simulations are displayed in Fig. \ref{fig:varcap} alongside the Fisher estimates of the variance computed as explained in the previous section, Eq. (\ref{eq:fishvar}). 
We emphasize that except for $TT$, all the studied formalisms are theoretically different from each other, as for instance they lead to different  mode-mode coupling matrices, and thus could result
in different levels of the estimates uncertainties.
 Nevertheless, as the $(TE,TB)$ block is typically negligible, we can expect that the $TE$ estimate will be identical in the standard and hybrid formalisms, and that the $TB$ estimates in the pure and hybrid formalism will coincide. For the $EE$, $BB$, and $EB$ spectra, the mode-mode coupling matrix changes from one formalism to another, even if $(EE,EB)$ and $(BB,EB)$ blocks are neglected. We consequently expect differences --though the hybrid and standard estimates should be close in the case of the $EE$ spectrum and the hybrid and the pure estimates of the $BB$ spectrum should also be rather similar as well.  Some other expectations are as follows:\\
$\bullet$~{\it TT-spectrum:} For such an angular power spectrum, the three formalism are equivalent and we expect to reach the same level of uncertainties for all the formalism. This is clearly confirmed by the results exhibited in the top-left panel of Fig. \ref{fig:varcap}. \\
$\bullet$~{\it EE-spectrum:} As already suggested in \cite{smith_2006} and illustrated in \cite{stivoli_etal_2010}, the standard estimator leads to slightly smaller error bars than the pure estimator. The underlying reason is that, though $B$-mode leaks into $E$-mode in the standard approach (thus increasing the sampling variance), this effect is small and the information lost by removing ambiguous mode in the pure approach leads to a higher increase of the sampling variance. In addition, the hybrid estimator, being based on the standard pseudo-multipoles for $E$-modes, performs roughly the same than the standard estimators. The gain obtained by using the standard or hybrid estimators is clearly seen on the top-middle panel of Fig. \ref{fig:varcap} at large angular scales. \\ 
$\bullet$~{\it BB-spectrum:} The case of $BB$ power spectrum is precisely the opposite of the $EE$ one. The main source of uncertainties is the extra-sampling variance due to $E$-modes leaking into $B$-modes. As clearly illustrated on the top-right panel of Fig. \ref{fig:varcap}, both the pure and hybrid estimators perform the same, leading to uncertainties on par of the most optimistic Fisher estimate, while the standard estimator leads to uncertainties {\it greater} than $C^{BB}_\ell$ for $\ell$ smaller than $\sim300$. \\
$\bullet$~{\it TE-spectrum:} For the very same reason that standard and hybrid estimator leads to smaller uncertainties than pure estimator for $EE$ spectrum, one expects those two approaches to also perform the best for $TE$ cross-spectrum. Results displayed on Fig. \ref{fig:varcap} (bottom-left panel) show significant improvement by using the standard or hybrid estimator as compared to the pure ones especially at large angular scales. \\
$\bullet$~{\it TB-spectrum:} This case is conceptually identical to $BB$ power spectrum. The lowest uncertainties are achieved by removing $E$-modes leaking into $B$: pure or hybrid formalism are then preferred to get an estimate of $C^{TB}_\ell$ as accurate as possible with respect to statistical error bars, as exhibited on the bottom-middle panel of Fig. \ref{fig:varcap}. \\
$\bullet$~{\it EB-spectrum:} Standard estimators, pure estimators and hybrid estimators perform the best respectively for $E$-pseudo-multipoles only, for $B$-pseudo-multipoles only and for both $E$- and $B$-pseudo-multipoles respectively. Our results show that the gain in using pure $B$-pseudo-mulipoles is more significant than the gain in using standard $E$-pseudo-multipoles. One then expects the following hierarchy for the error bars: $\Delta^{EB}_{\ell,~\mathrm{standard}}>\Delta^{EB}_{\ell,~\mathrm{pure}}>\Delta^{EB}_{\ell,~\mathrm{hybrid}}$. Though uncertainties from the standard estimators are clearly higher (see bottom-right panel of Fig. \ref{fig:varcap}), it appears that the pure and hybrid estimators perform the same. \\

From these results, we could conclude that the hybrid estimator systematically leads to the lowest uncertainties for the six angular power spectra; an expected conclusion as the hybrid formalism is based on the best pseudo-multipoles computation for both $E$- and $B$-modes.

\begin{figure}
\begin{center}
	\includegraphics[scale=0.245]{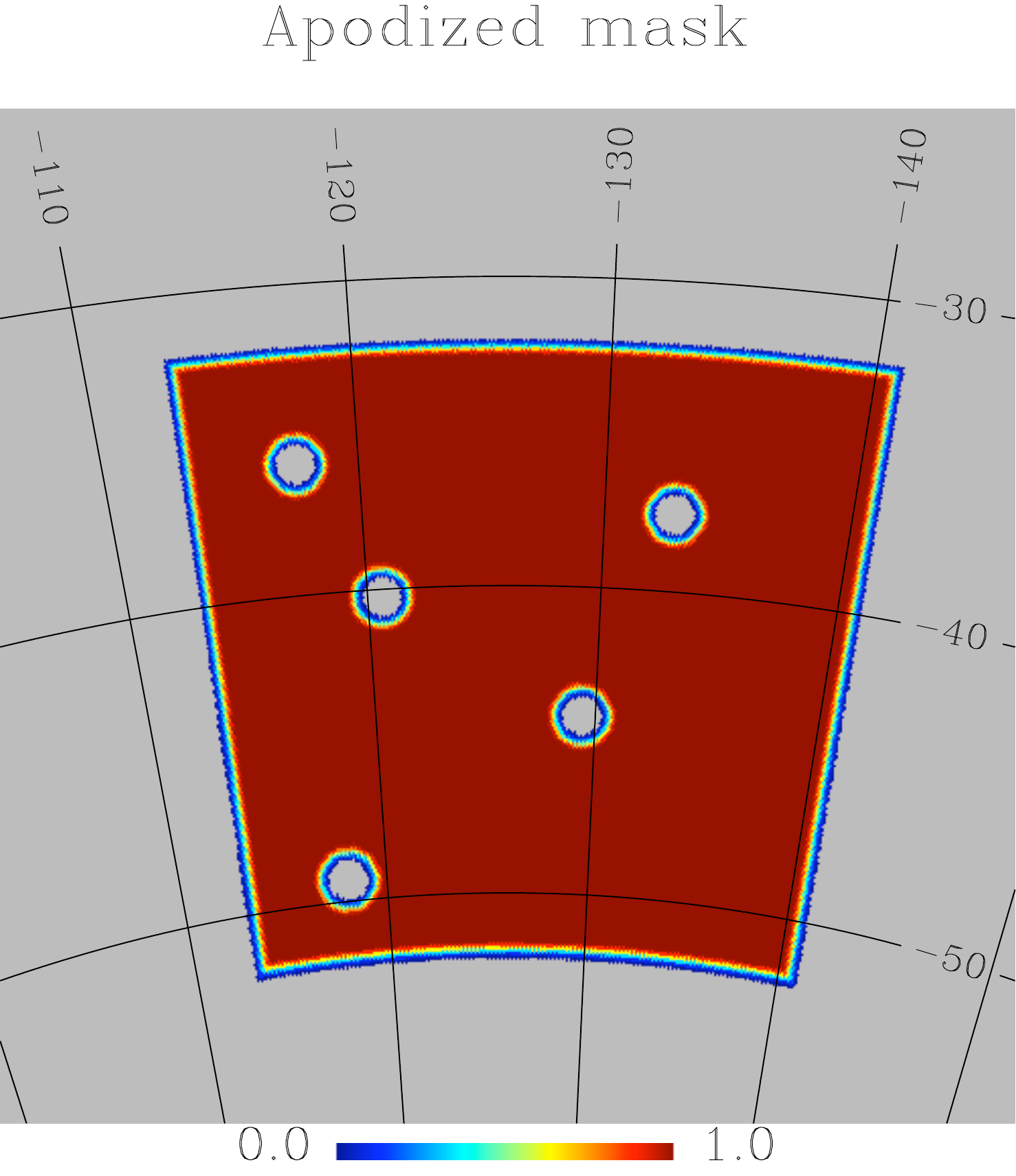} \includegraphics[scale=0.245]{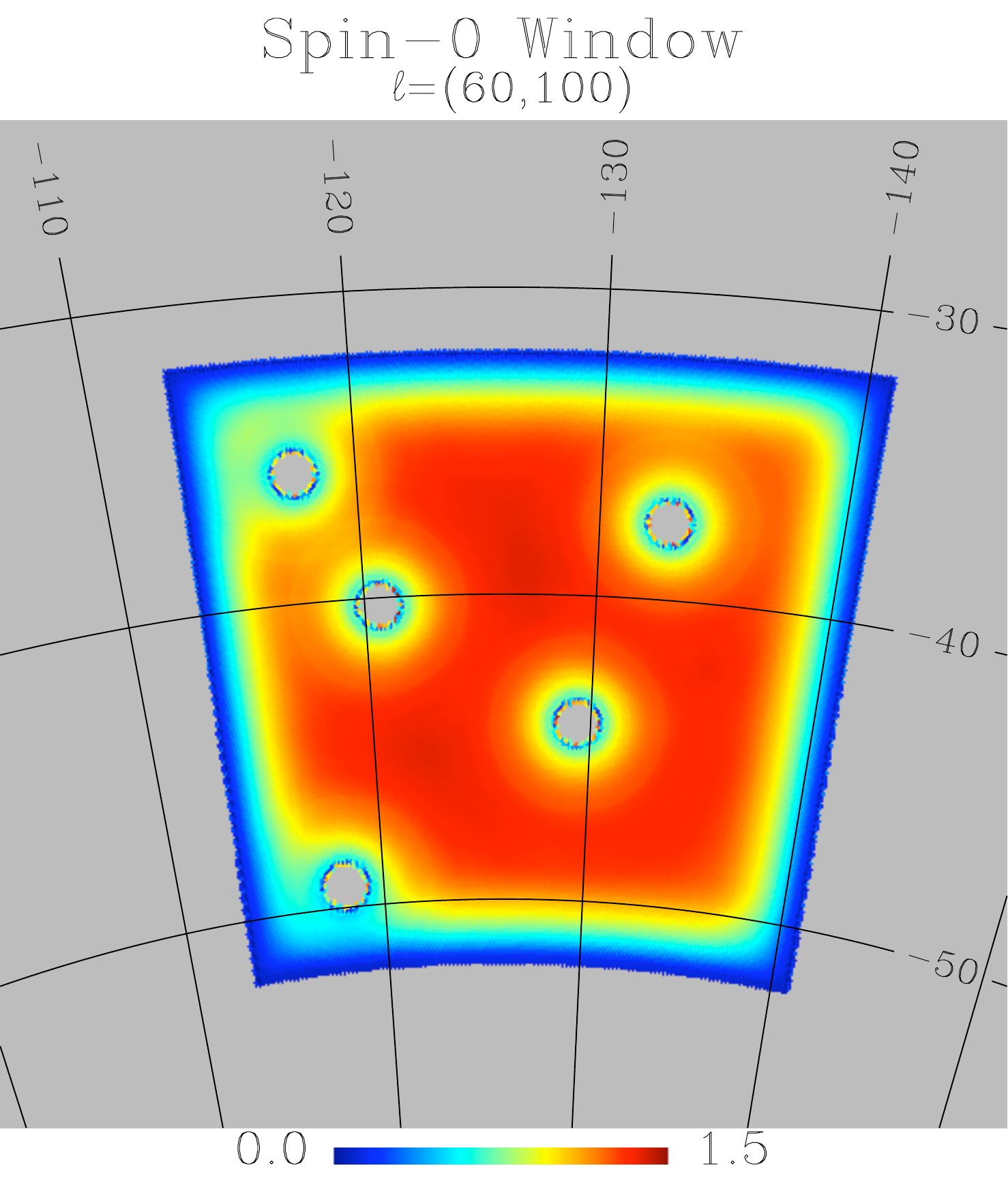} \includegraphics[scale=0.245]{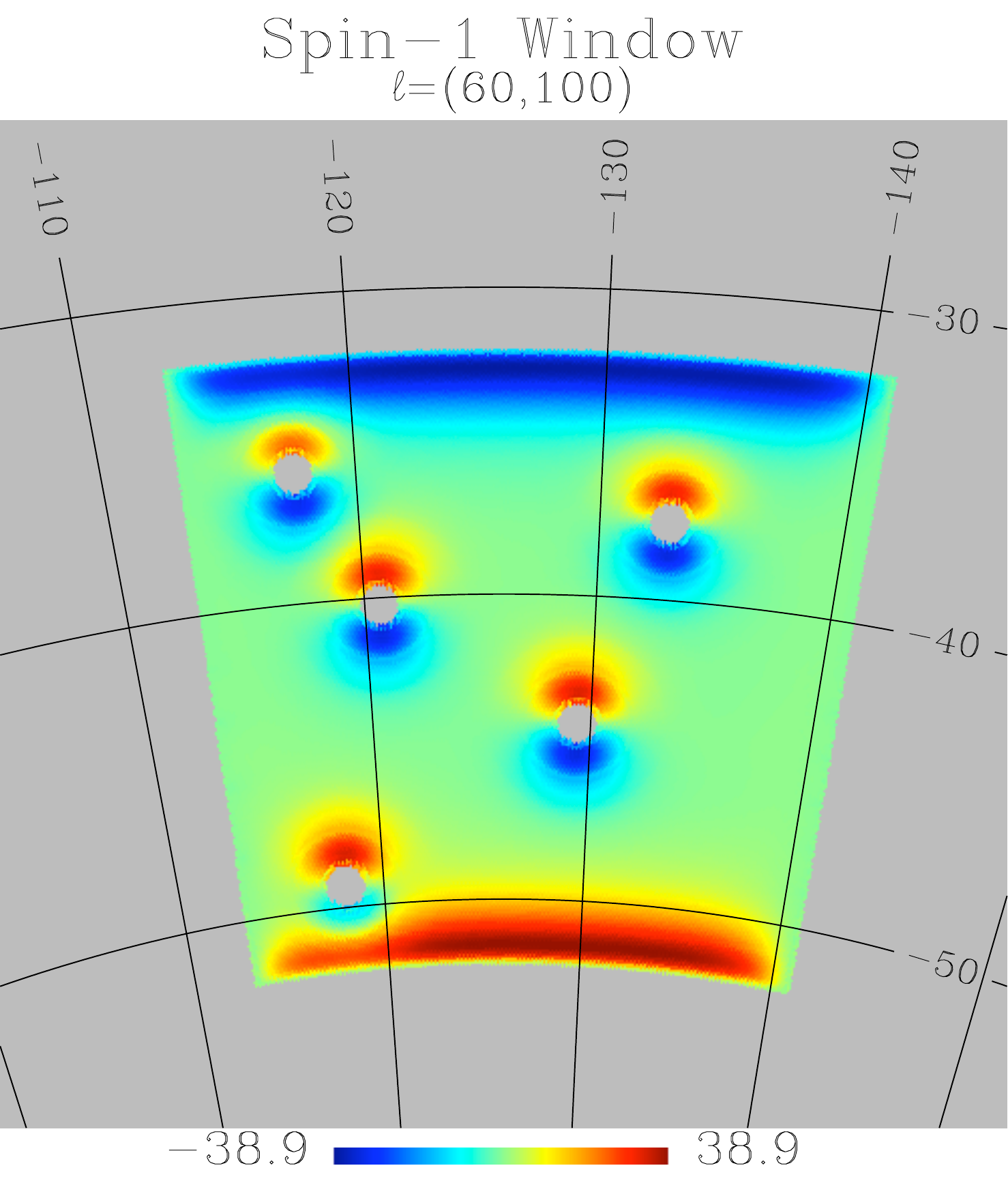} \includegraphics[scale=0.245]{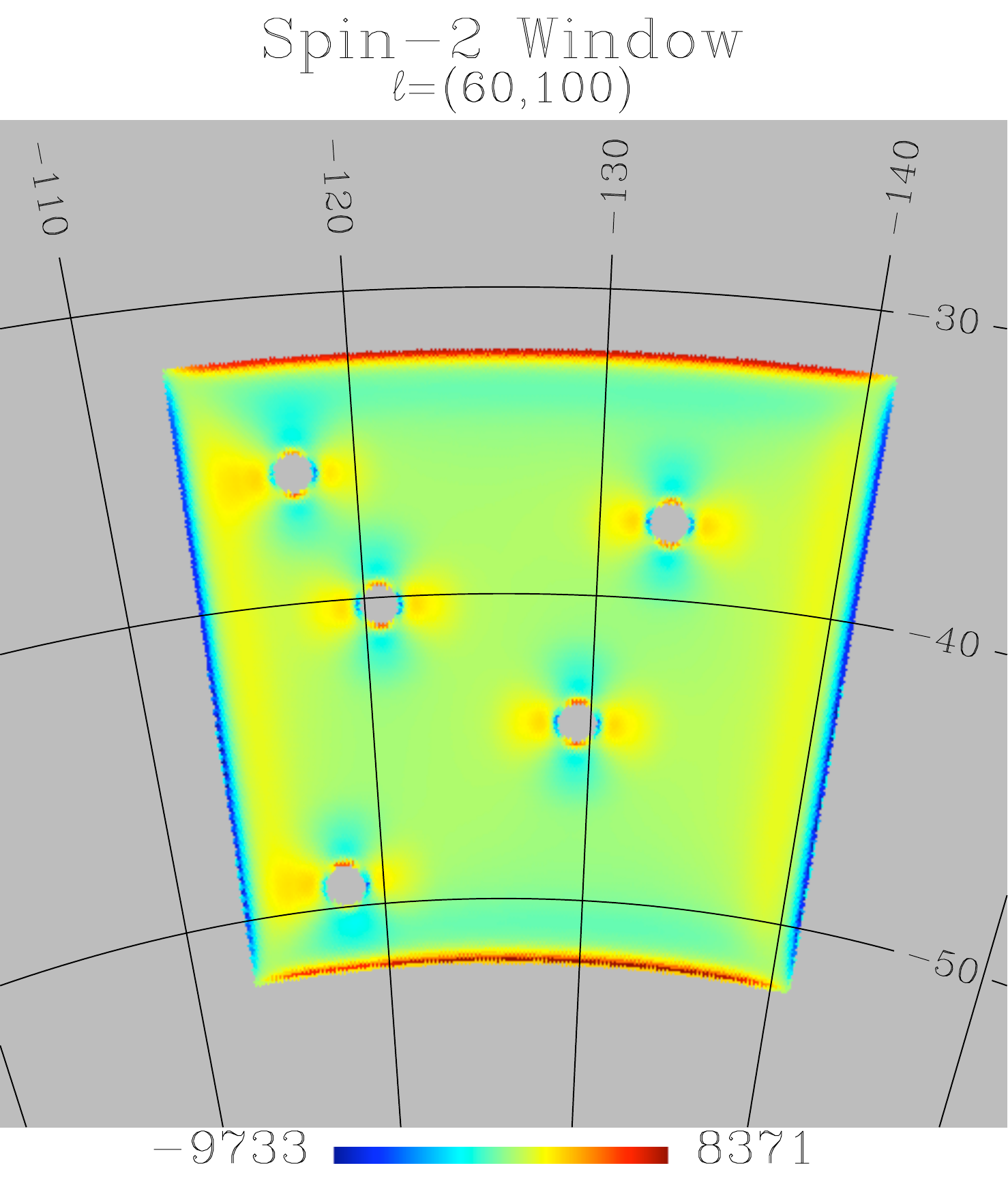}
	\caption{Same as Fig. \ref{fig:SkyapIdeal} for the case of a mock survey with sharpened boundaries and holes.}
	\label{fig:SkyapHoles}
\end{center}
\end{figure}	

\subsection{Toward a realistic survey: Effect of holes and sharp boundaries}
In this section we consider a more intricate observed patch shape consisting of a square patch with internal holes mimicking effects of point source masking~\cite{smith_zaldarriaga_2007}.  The specific patch used hereafter is depicted in Fig. \ref{fig:SkyapHoles}.
\begin{figure*}
\begin{center}
	\includegraphics[scale=0.35]{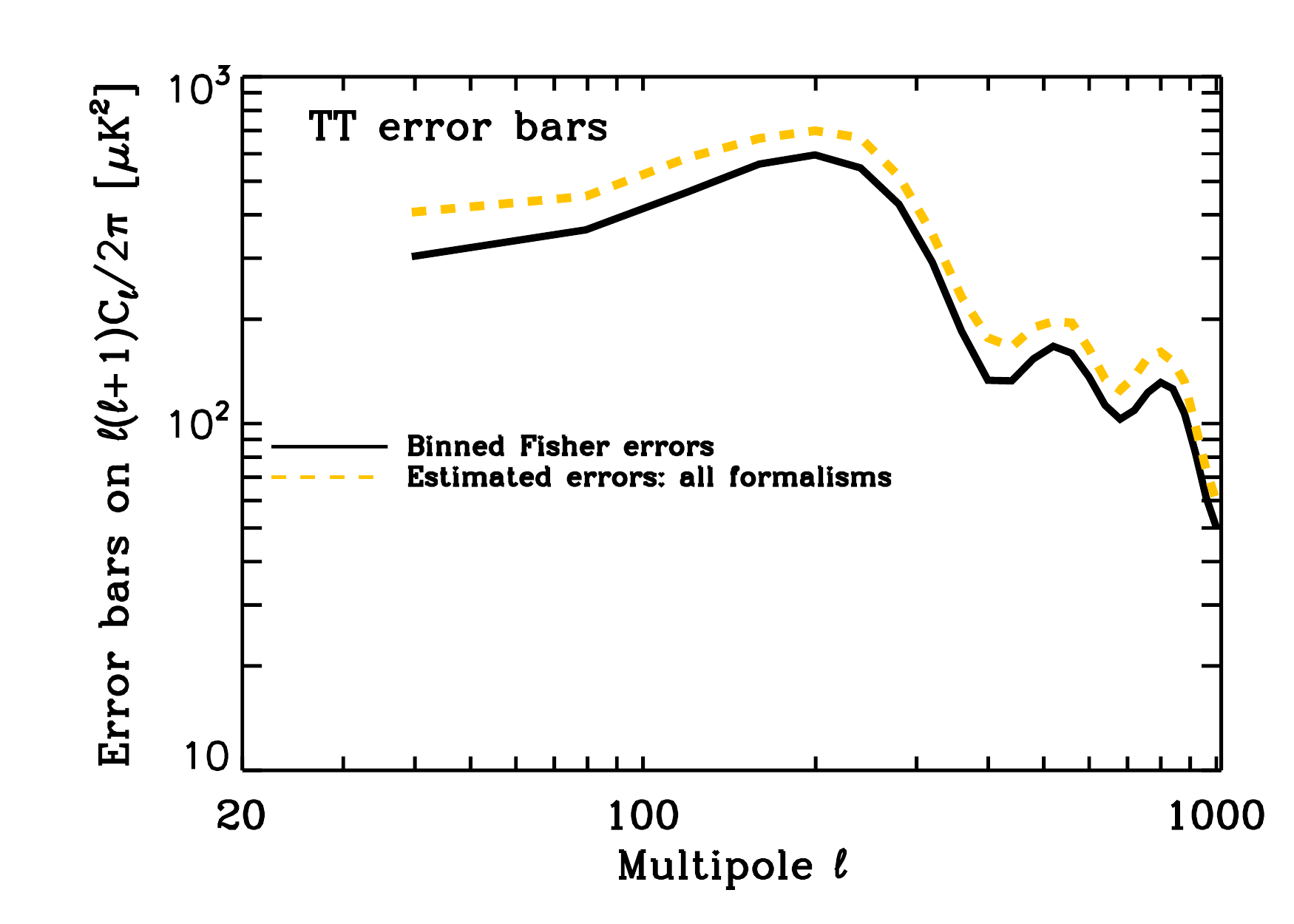}\includegraphics[scale=0.35]{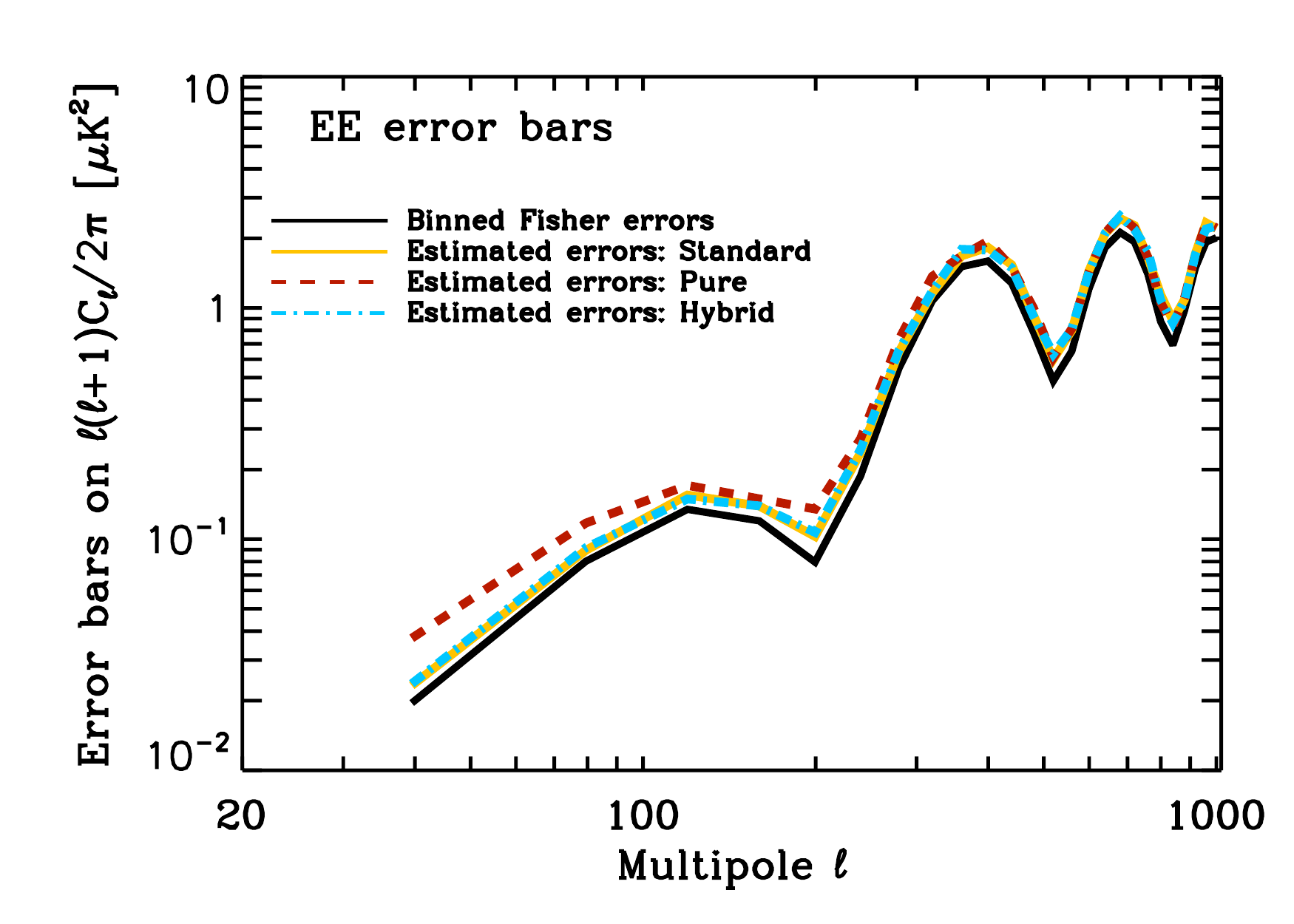}\includegraphics[scale=0.35]{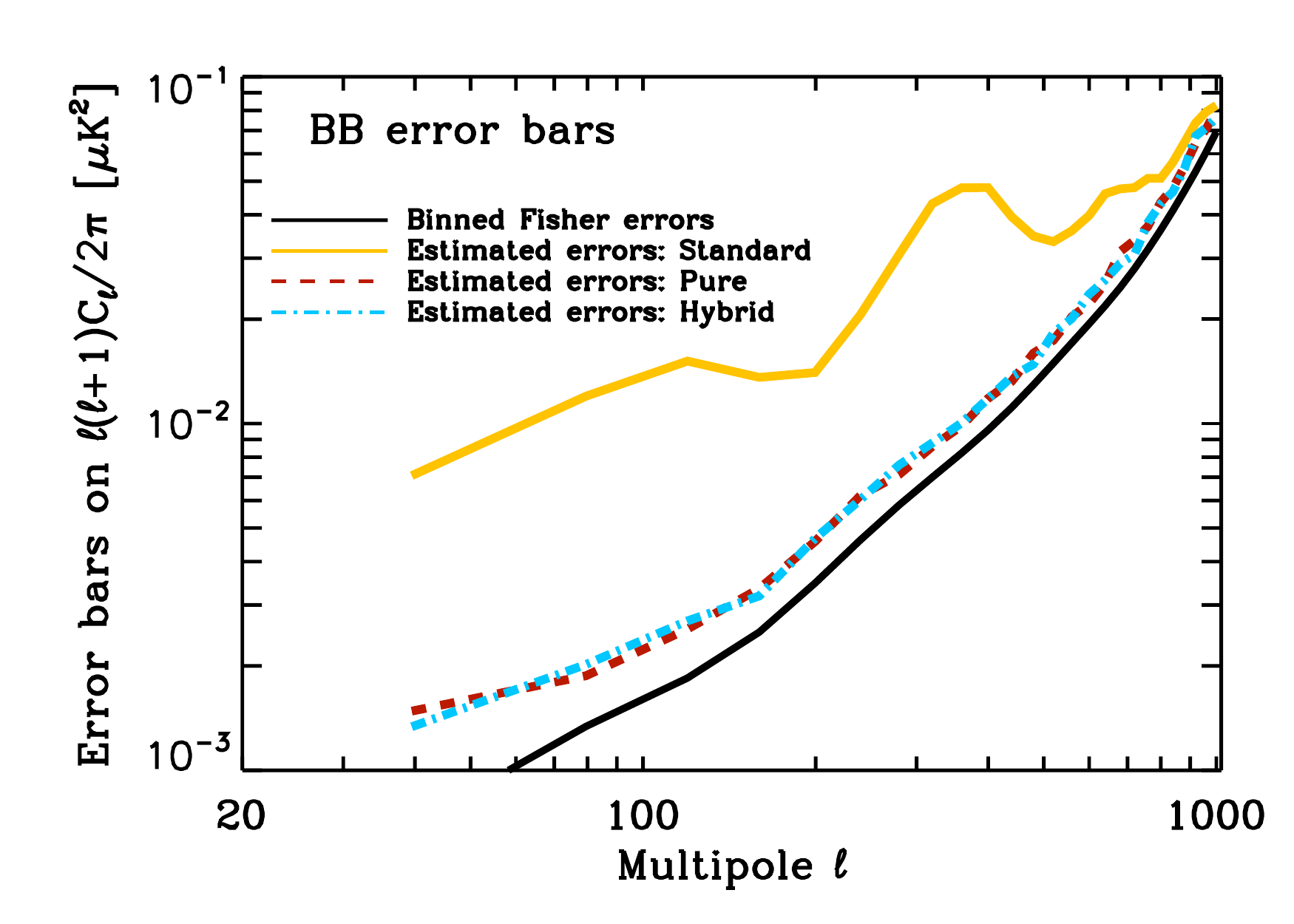}\\
	\includegraphics[scale=0.35]{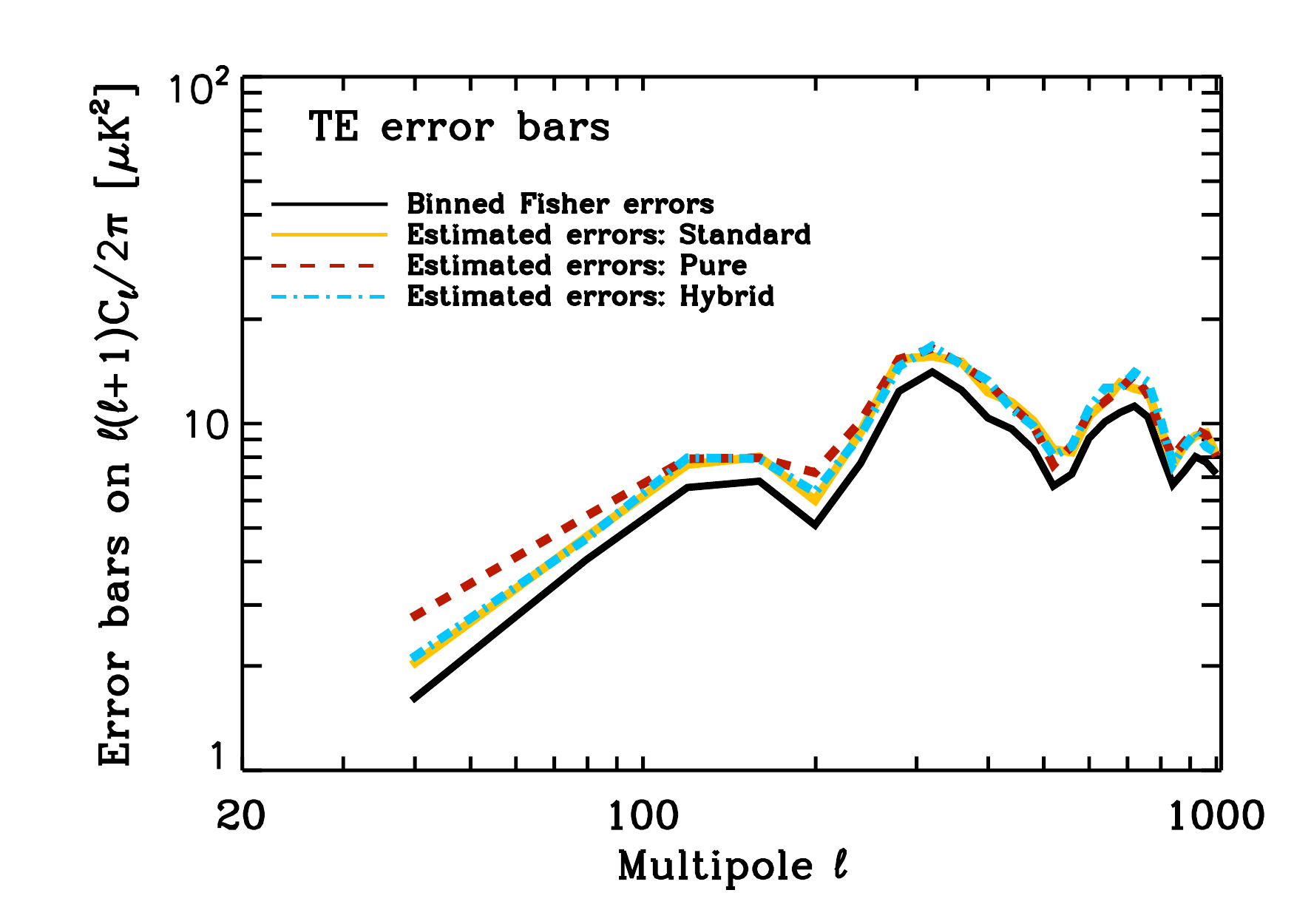}\includegraphics[scale=0.35]{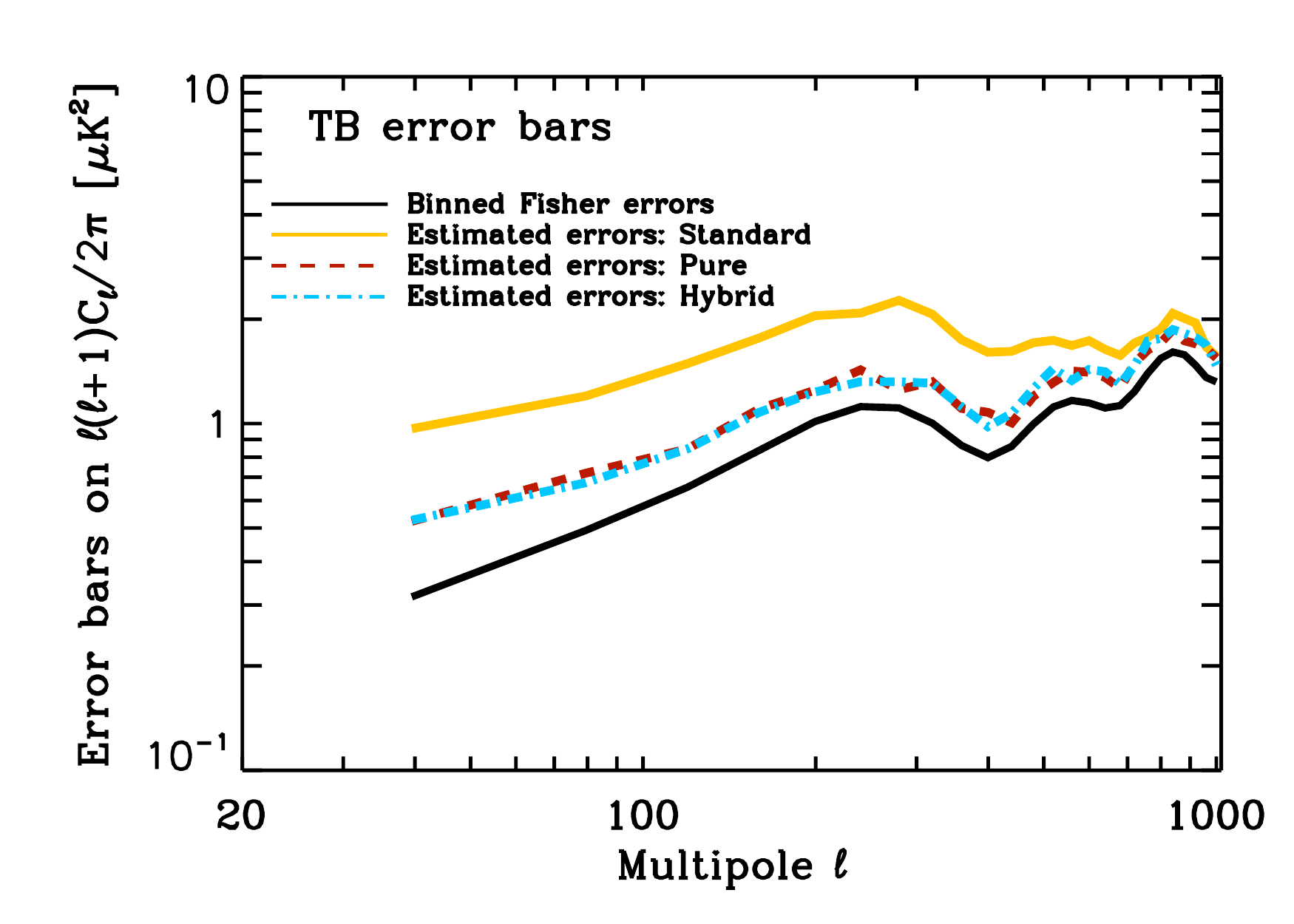}\includegraphics[scale=0.35]{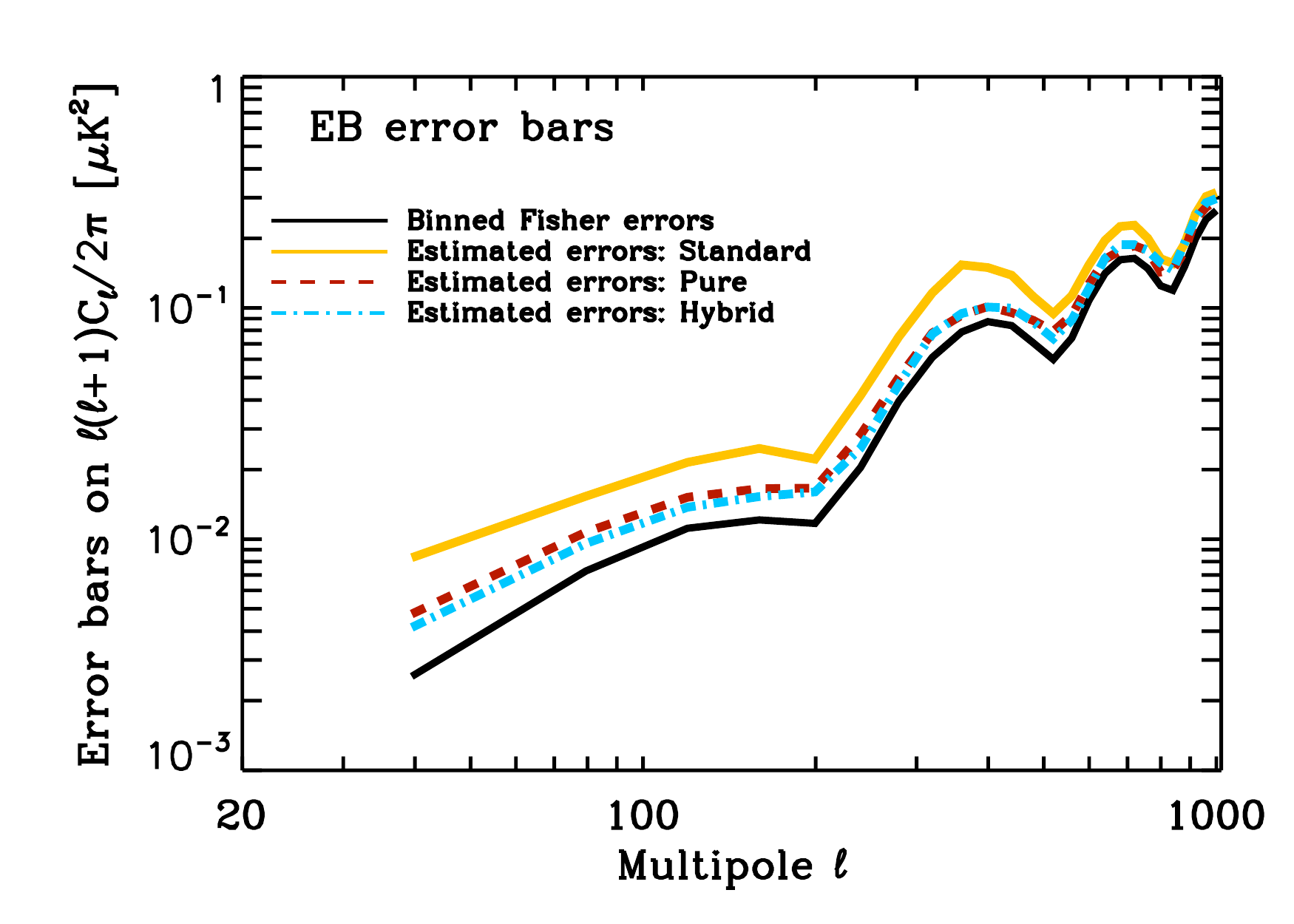}
	\caption{Error bars on the reconstructed angular power spectra for each of the three formalisms (colored curves) alongside the naive (binned) Fisher estimate of such uncertainties. The assumed sky survey is a square patch filled with holes and with homogeneous noise (see Fig. \ref{fig:SkyapHoles}).The error bars are obtained as the standard deviation over 500 MC simulations.}
	\label{fig:varholes}
\end{center}
\end{figure*}

The fraction of the sky covered by this patch is $f_\mathrm{sky}\simeq0.99$\%. For the $TT$ power spectrum (all formalisms),  the $EE$ and $TE$ power spectra (standard and hybrid formalisms), and the $BB$, $TB$, and $EB$ (standard formalism), the effective sky coverage is constant and equal to 0.92\%. For the $BB$ power spectrum (pure and hybrid formalism), and the $EE$ and $EB$ power spectra (pure formalism), it varies from 0.43\% at large angular scales ($20\leq\ell<60$) up to 0.96\% at small angular scales ($980\leq\ell<1020$). Finally, for the $TB$ power spectrum (pure and hybrid formalisms), the $TE$ power spectrum (pure formalism), and the $EB$ power spectrum (hybrid formalism), the sky coverage ranges from 0.78\% at large angular scales ($20\leq\ell<60$) up to 0.93\% at small angular scales ($980\leq\ell<1020$).

We found MC averaged power spectra in perfect agreement with the theoretical input $C_\ell$ showing that power spectra estimators for the three formalisms are unbiased as expected.

The error bars on the estimates of the six angular power spectra are shown in Fig. \ref{fig:varholes}. As already shown for the case of the circular patch, the standard or hybrid estimators gives smaller variance for the $EE$ and $TE$ power spectra, while the pure or hybrid formalisms are to be adopted for the $BB$, $TB$ and $EB$ spectra. One should however expect some differences: since sharpened and internal boundaries enhance the $E$-to-$B$ and $B$-to-$E$ leakages ({\it i.e.} the total power contained in ambiguous modes is increased), the relative merit of each formalism is supposed to be quantitatively strengthened. Though not obvious for $BB$ and $TB$ spectra (for which the gain in using the pure or the hybrid estimators is already dramatic for simple boundaries), it is now clear {\it at large angular scales} ($\ell\leq100$) that the error bars on $EE$ and $TE$ angular power spectra (see top-middle and bottom-left panels of Fig. \ref{fig:varholes}) are lower by using the standard or the hybrid formalism than by using the pure approach. Moreover, a closer inspection of the $EB$ power spectra seen in the  bottom-right of Fig. \ref{fig:varholes} suggests that, at large angular scales, $\ell\leq150$, the hybrid approach performs slightly better than the pure approach, as expected from theoretical considerations.

\subsection{Toward a realistic survey: Effect of inhomogeneous sky sampling}
\begin{figure*} 
\begin{center}
	\includegraphics[scale=0.3]{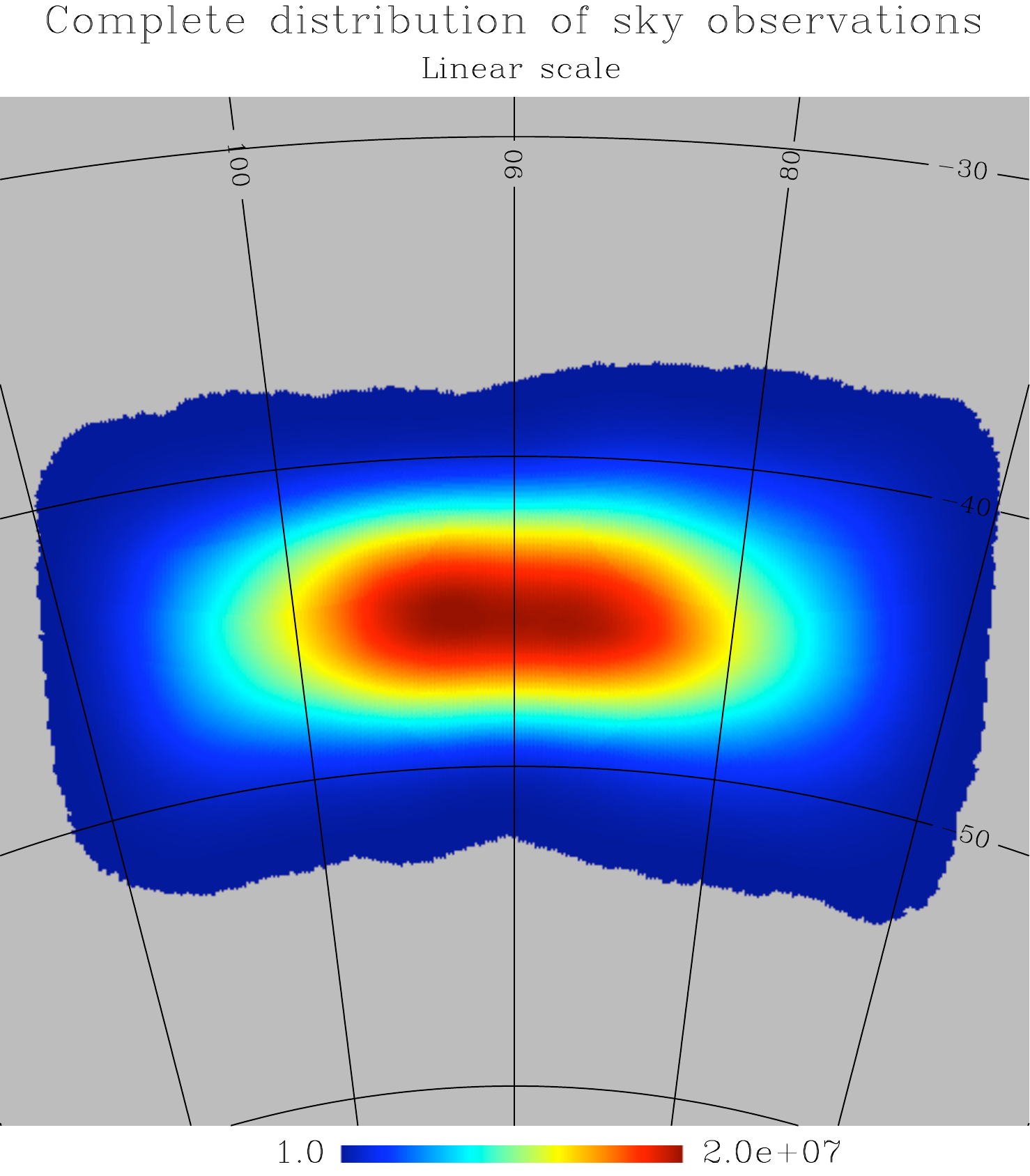} \includegraphics[scale=0.3]{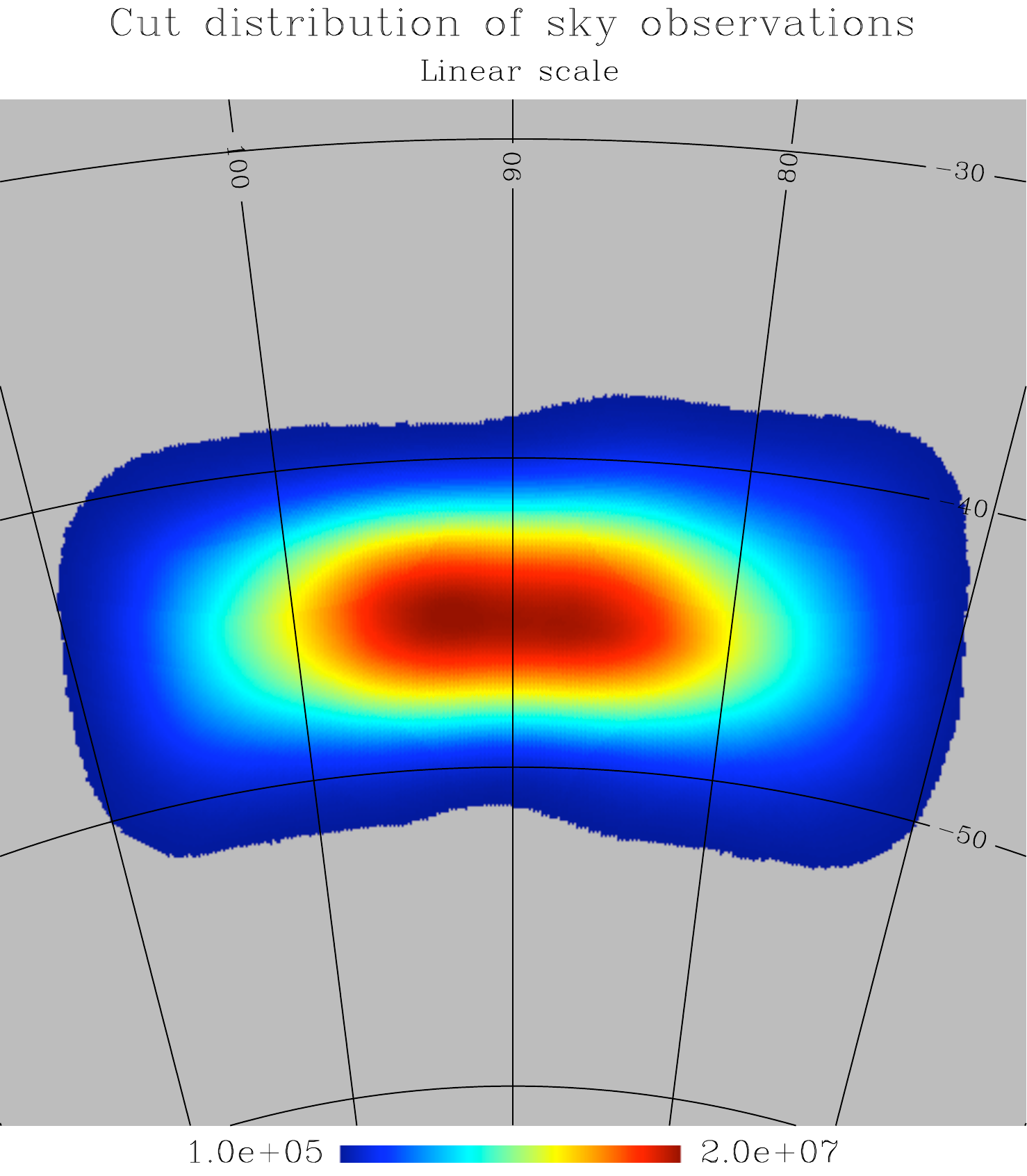} \includegraphics[scale=0.3]{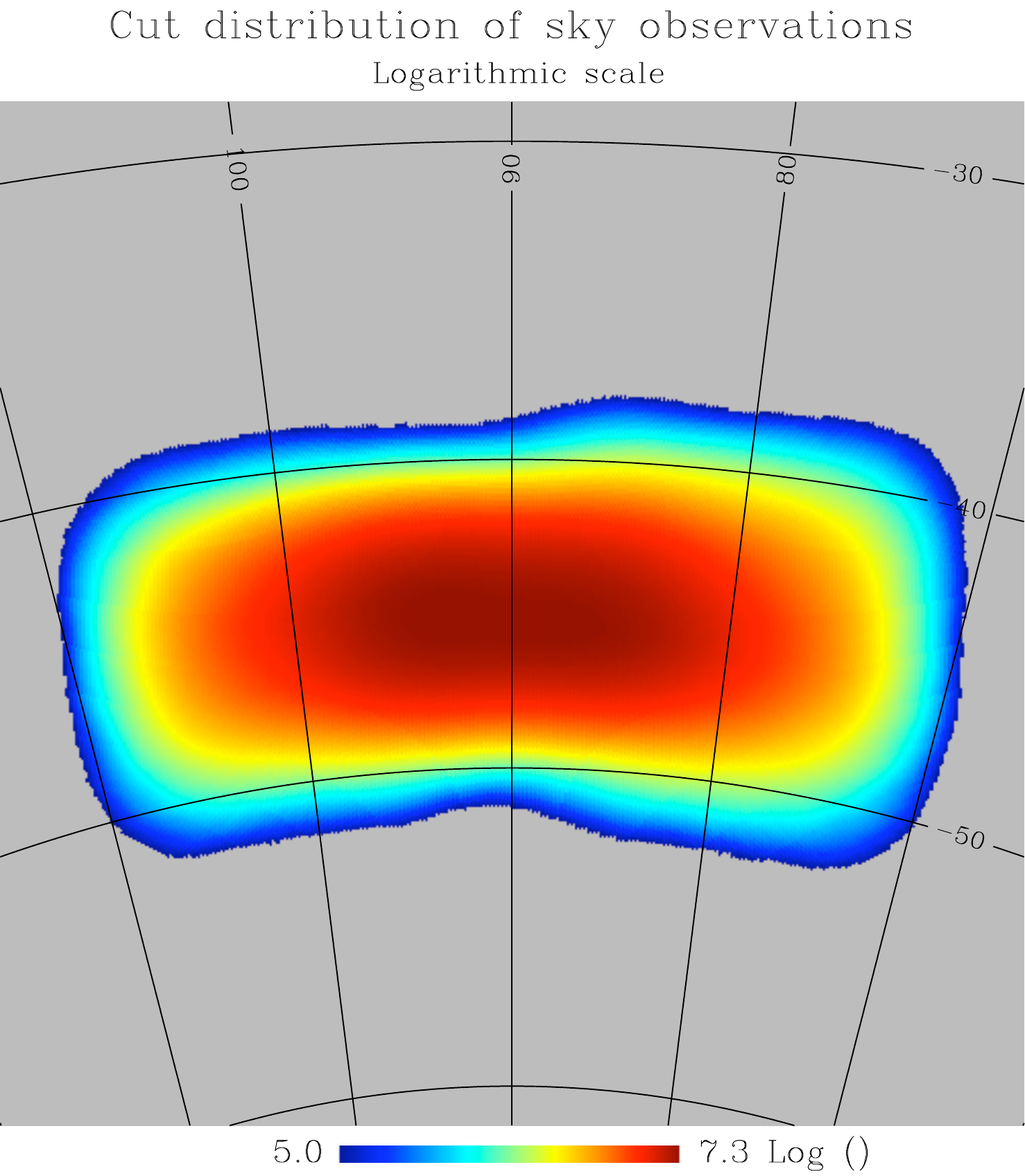}
	\caption{A distribution of sky observation for an observational strategy mimicking the {\sc ebex} experiment. The complete distribution is shown in linear scale (left panel) for which the density of samples ranges from 1 to $\sim2\times10^7$. The noisiest pixels, which lie in the outer part of the observed sky, have been trimmed in order to compute the optimized sky apodizations. This "well-observed" part is displayed in linear scale (middle panel) and in logarithmic scale (right panel). The density of observations for the conserved pixels ranges from $10^5$ to $\sim2\times10^7$.}
	\label{fig:Nhit}
\end{center}
\end{figure*}
For a realistic small-scales experiment, the observed patch is expected to be rather irregular with the density of observations per sky area strongly varying from one pixel to another.
This will unavoidably lead to significant noise inhomogeneity of the reconstructed sky maps. We therefore explore the performance of the $TB$ and $EB$ pseudo-$C_\ell$'s estimators in the case of an observation mimicking a long-duration CMB balloon-borne experiment such as {\sc ebex} \cite{britt_etal_2010}. 
The resulting sky sampling, {\it i.e.} a number of observations per pixel (Fig. \ref{fig:Nhit}), has already been used in \cite{grain_etal_2009} to assess the performance of the pure $B$-mode estimators. It ranges from $10^5$ at the edges of the patch to more than $2\times10^7$ in its core after we cut the noisiest pixels in order to ensure convergency of the optimized sky apodization procedure (see \cite{grain_etal_2009}).
Finally, we set the noise per sample such as the average noise level for the full patch equals the noise level used in the previously-studied homogeneous cases.

\begin{figure}
\begin{center}
	\includegraphics[scale=0.245]{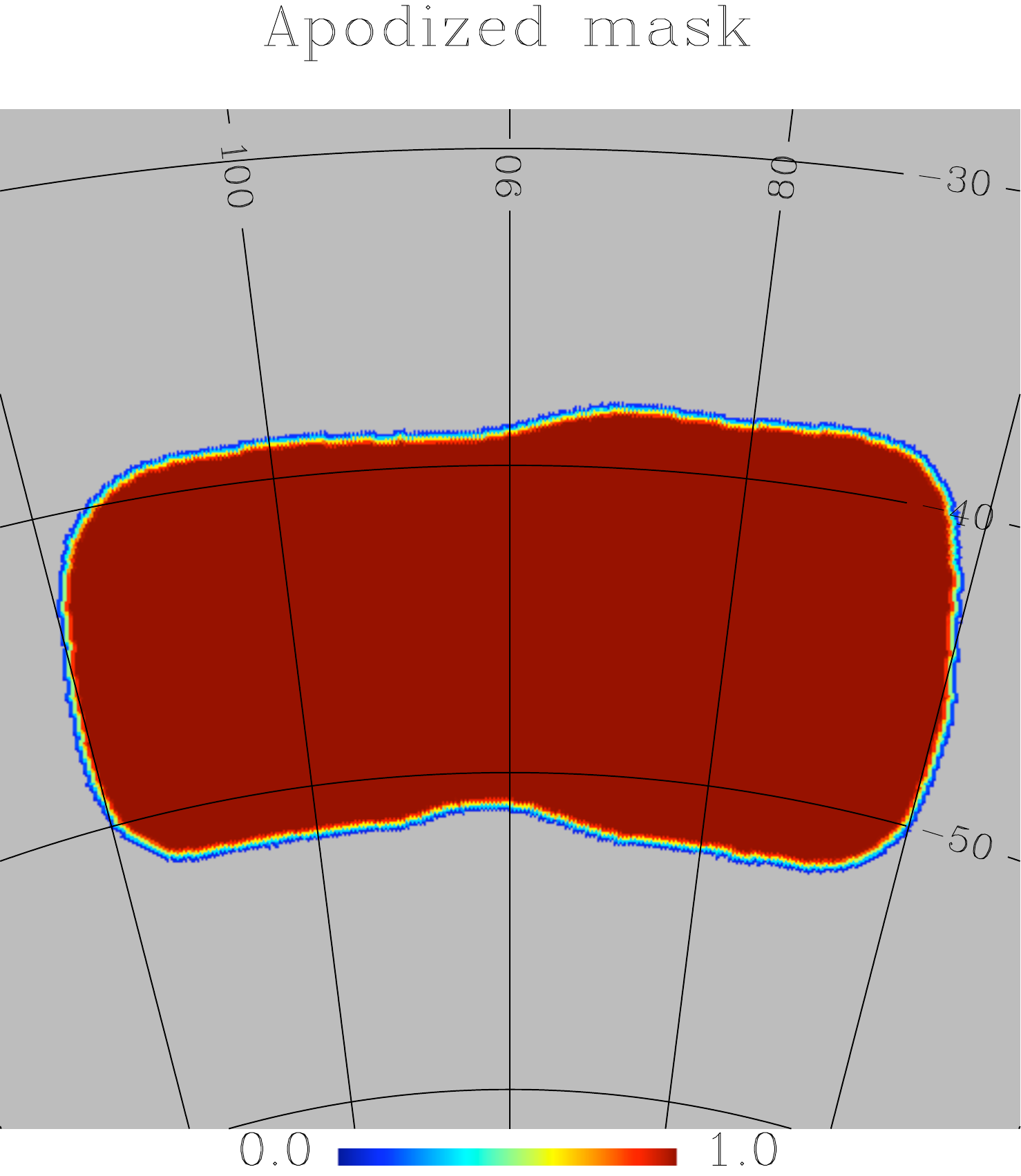} \includegraphics[scale=0.245]{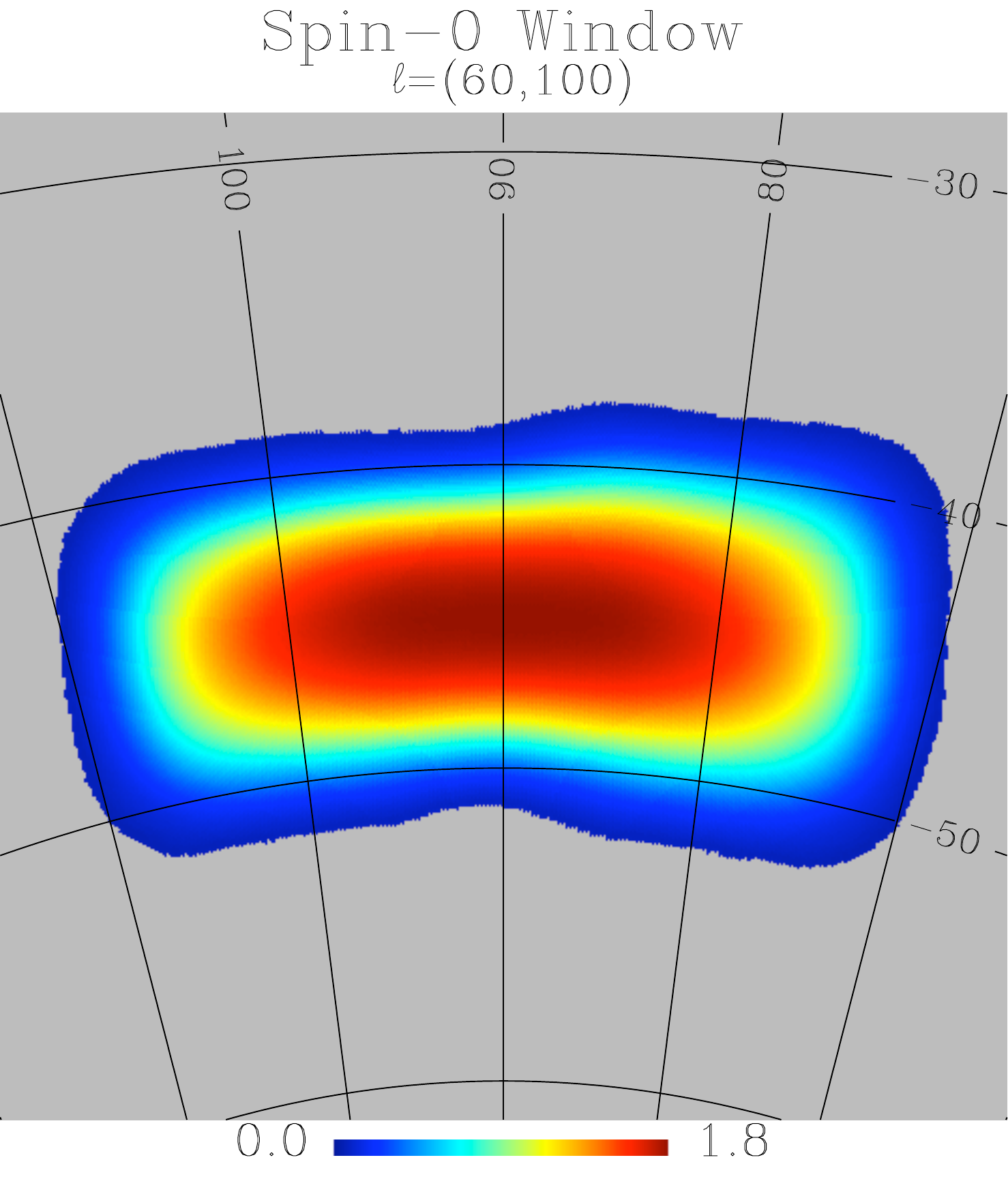} \includegraphics[scale=0.245]{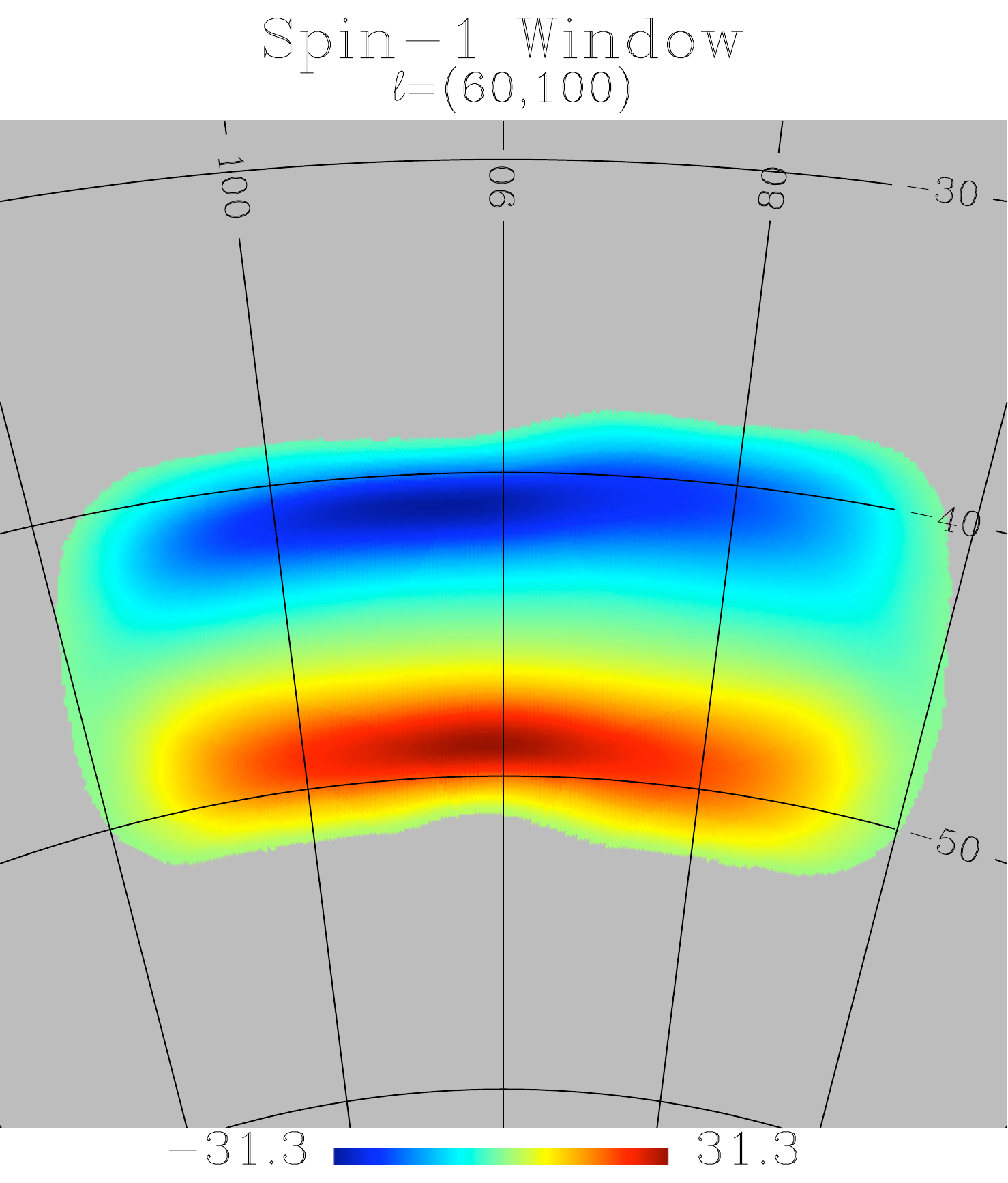} \includegraphics[scale=0.245]{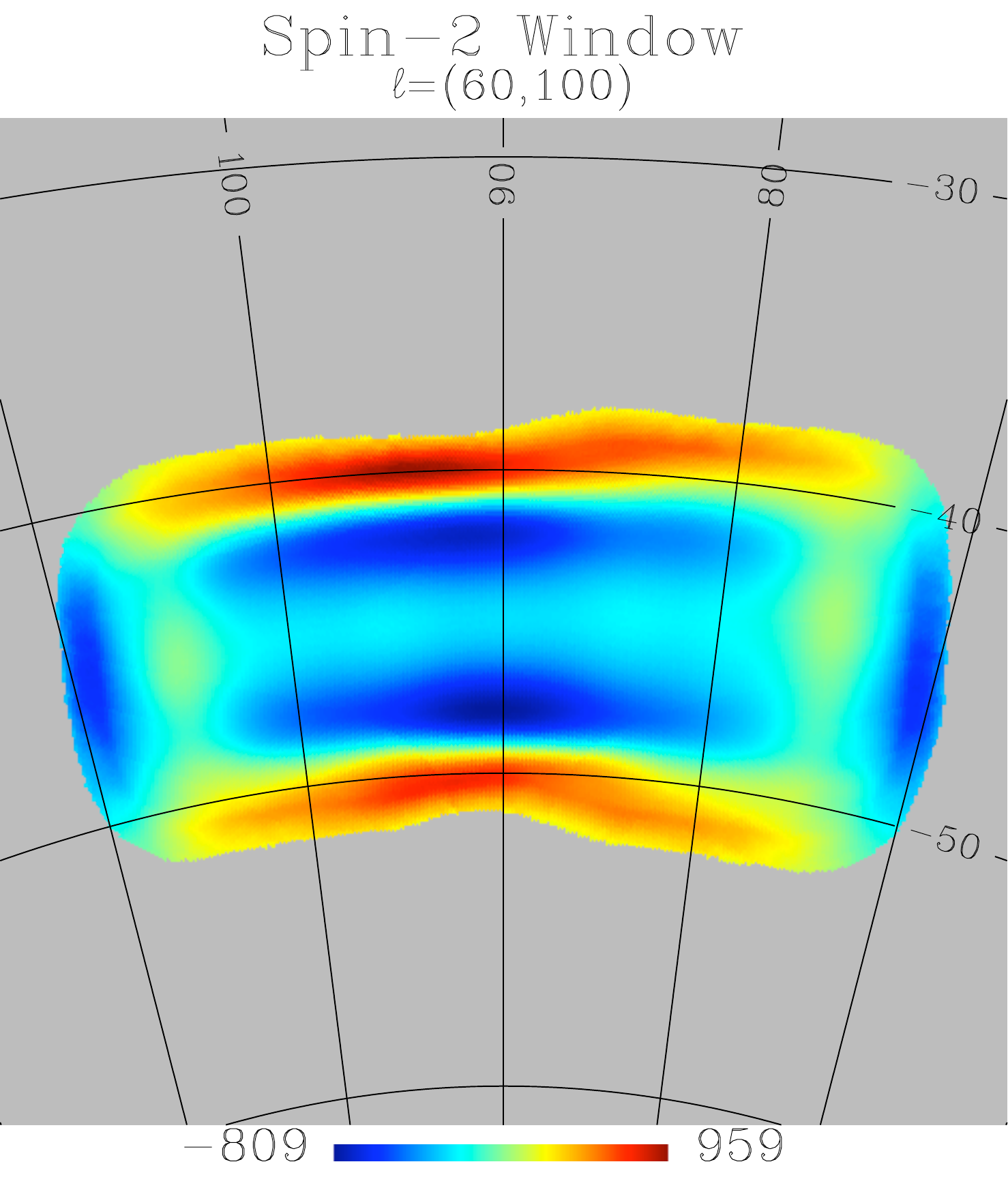}
	\caption{Same as Fig. \ref{fig:SkyapIdeal} for the case of a realistic mock survey with inhomogeneous noise as displayed in Fig. \ref{fig:Nhit}.}
	\label{fig:SkyapEbex}
\end{center}
\end{figure}

The retained sky fraction is equal to $f_\mathrm{sky}\simeq0.89$\%. For the $TT$ power spectrum (all formalisms),  the $EE$ and $TE$ power spectra (standard and hybrid formalisms), and the $BB$, $TB$, and $EB$ (standard formalism, the effective sky coverage is constant and equal to 0.85\%. For the $BB$ power spectrum (pure and hybrid formalism), and the $EE$ and $EB$ power spectra (pure formalism), it varies from 0.42\% at large angular scales ($20\leq\ell<60$) down to 0.29\% at small angular scales ($980\leq\ell<1020$). Finally, for the $TB$ power spectrum (pure and hybrid formalisms), the $TE$ power spectrum (pure formalism), and the $EB$ power spectrum (hybrid formalism), the sky coverage ranges from 0.6\% at large angular scales ($20\leq\ell<60$) down to 0.46\% at small angular scales ($980\leq\ell<1020$). We note that for this {\it inhomogeneous} noise case, the effective sky coverage decreases for smaller scales as the optimized sky apodization converges toward inverse-noise weighting, which is "naturally" apodized as a consequence of the scanning strategy. This difference of the apodization lengths can be easily seen by comparing the left panel of Fig. \ref{fig:Nhit}, showing the inverse noise weighting,  with the right top panel of Fig. \ref{fig:SkyapEbex}, depicting one of the optimized, large scale apodizations.

Once again, we first checked that each formalism leads to unbiased estimates for the six power spectra ($TT$, $EE$, $BB$, $TE$, $TB$, $EB$).

\begin{figure*}
\begin{center}
	\includegraphics[scale=0.35]{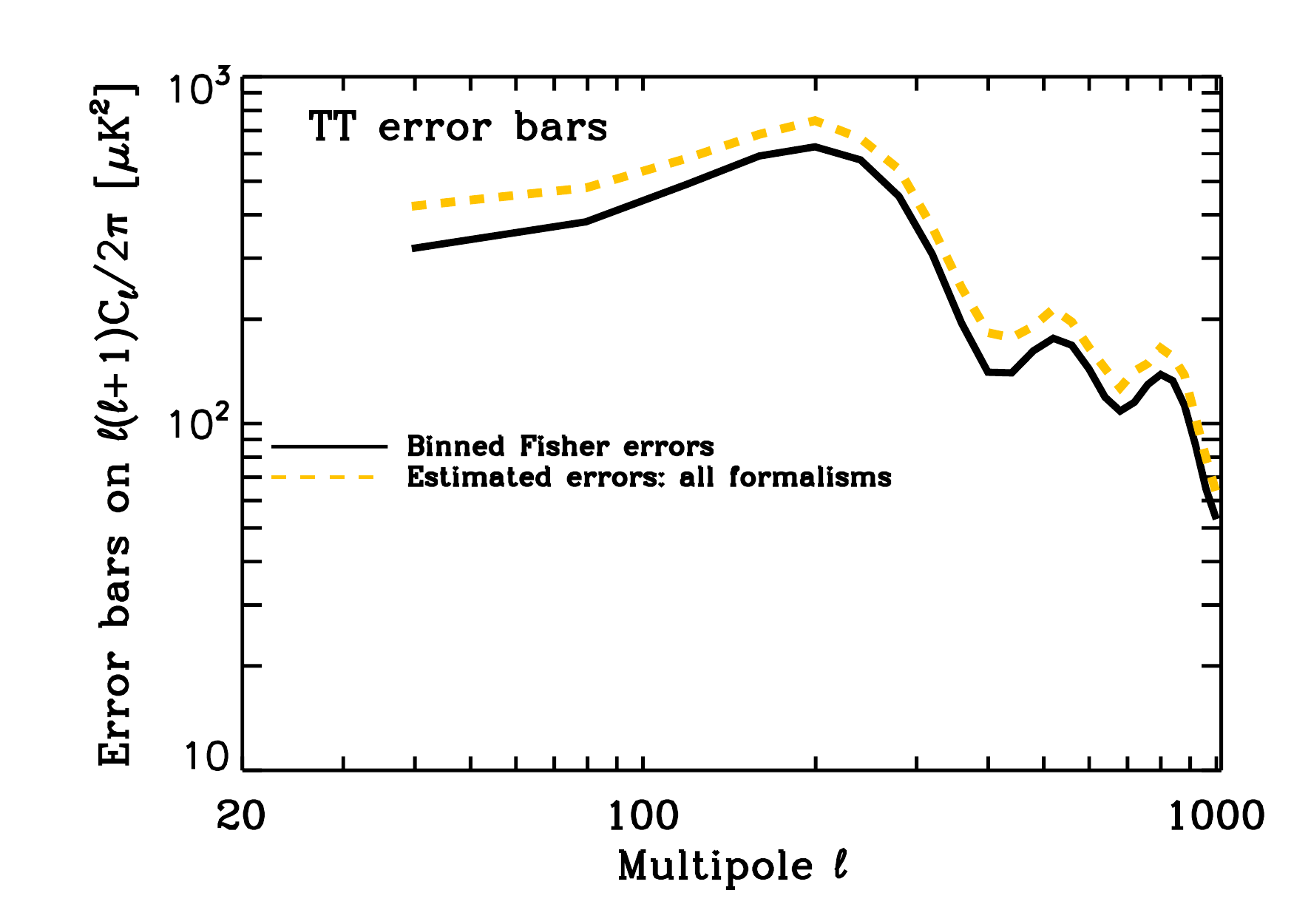}\includegraphics[scale=0.35]{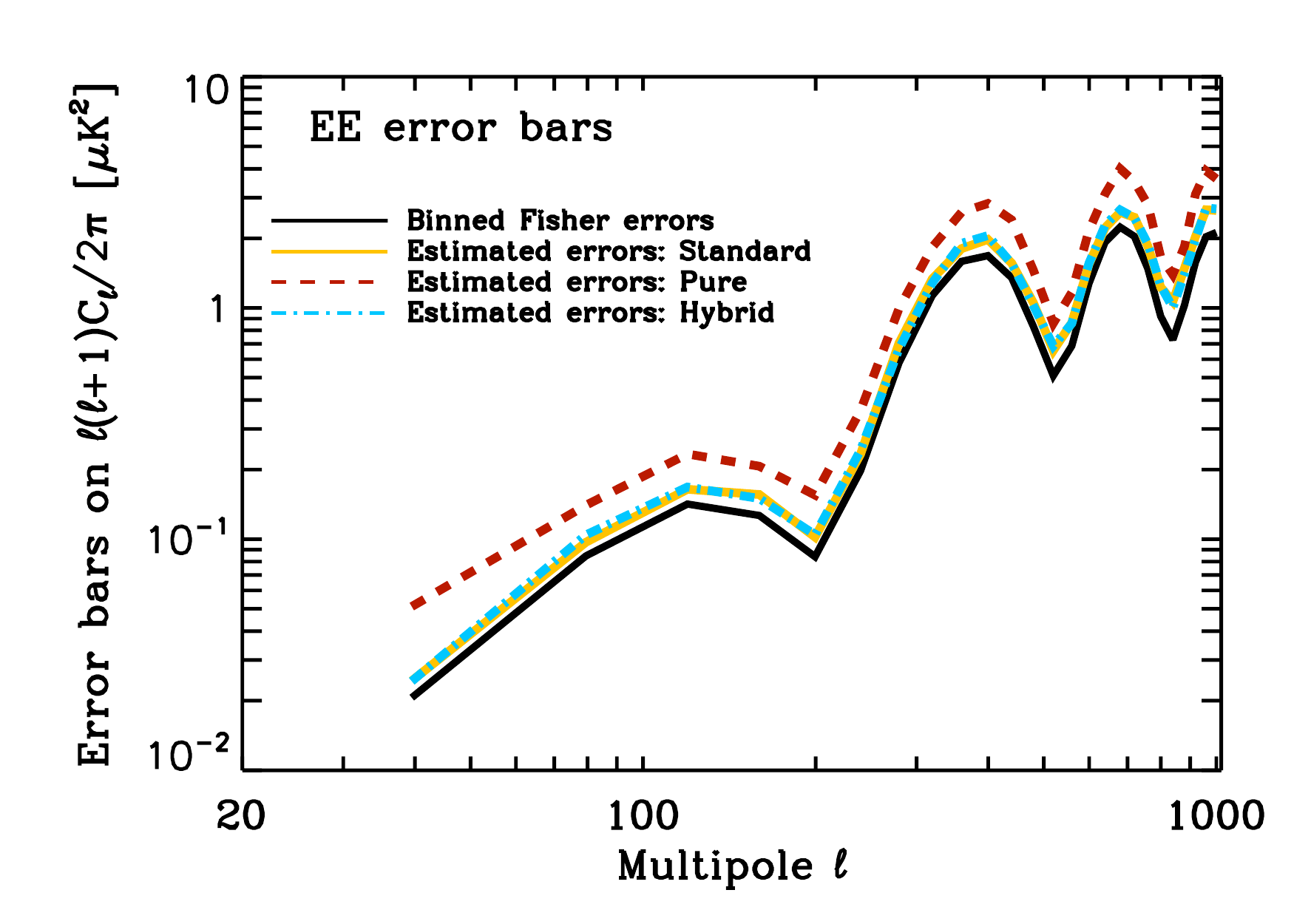}\includegraphics[scale=0.35]{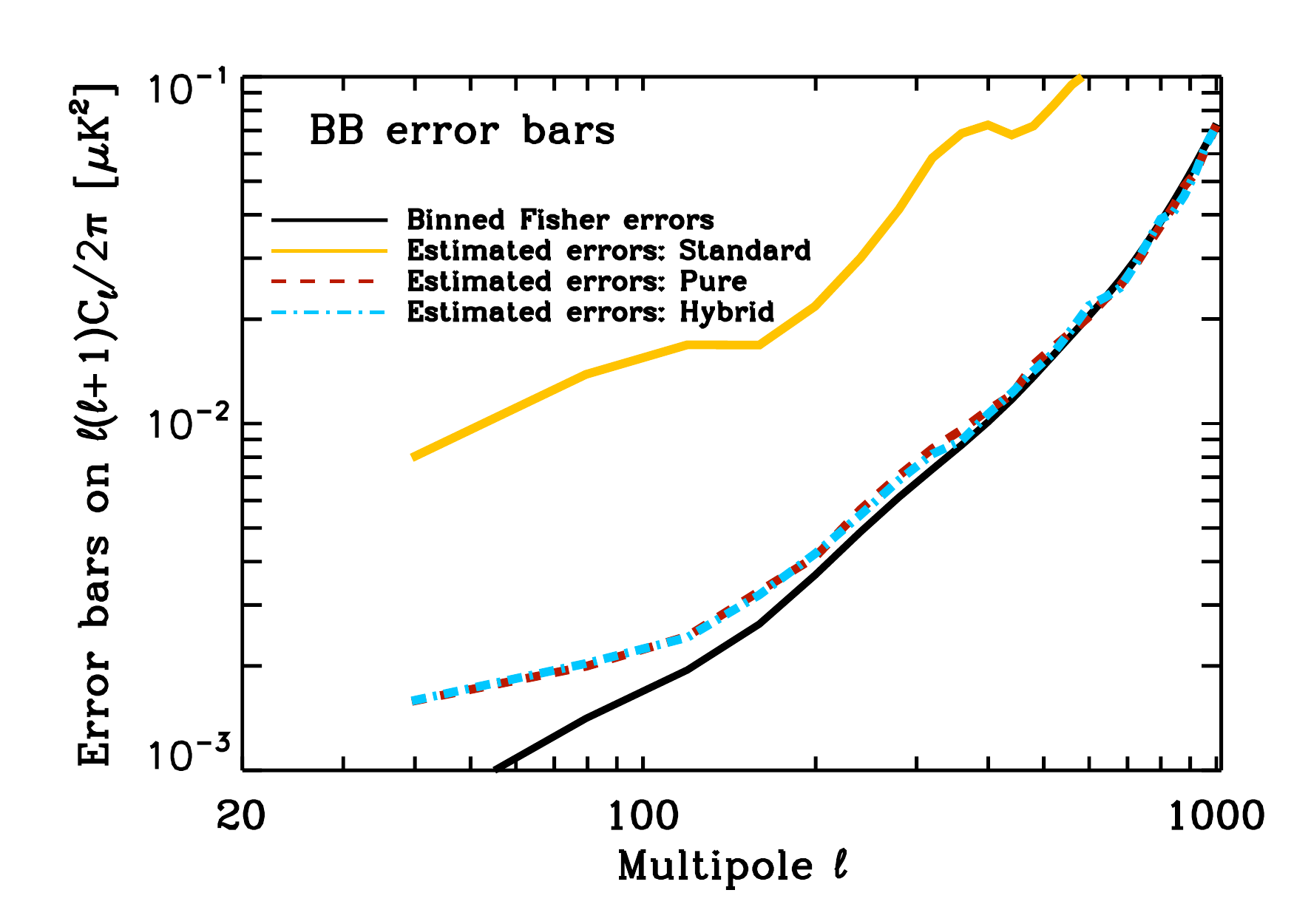} \\
	\includegraphics[scale=0.35]{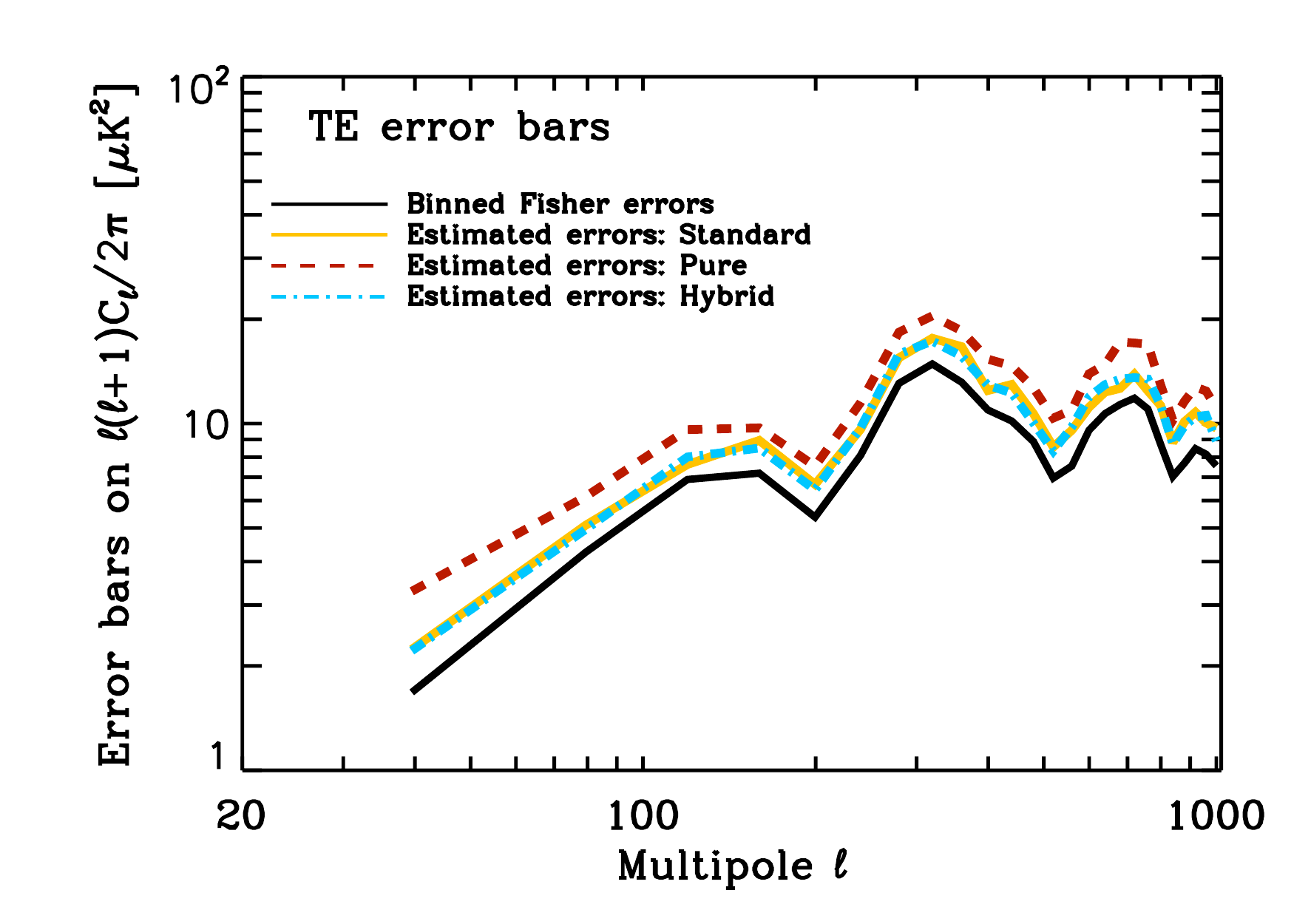}\includegraphics[scale=0.35]{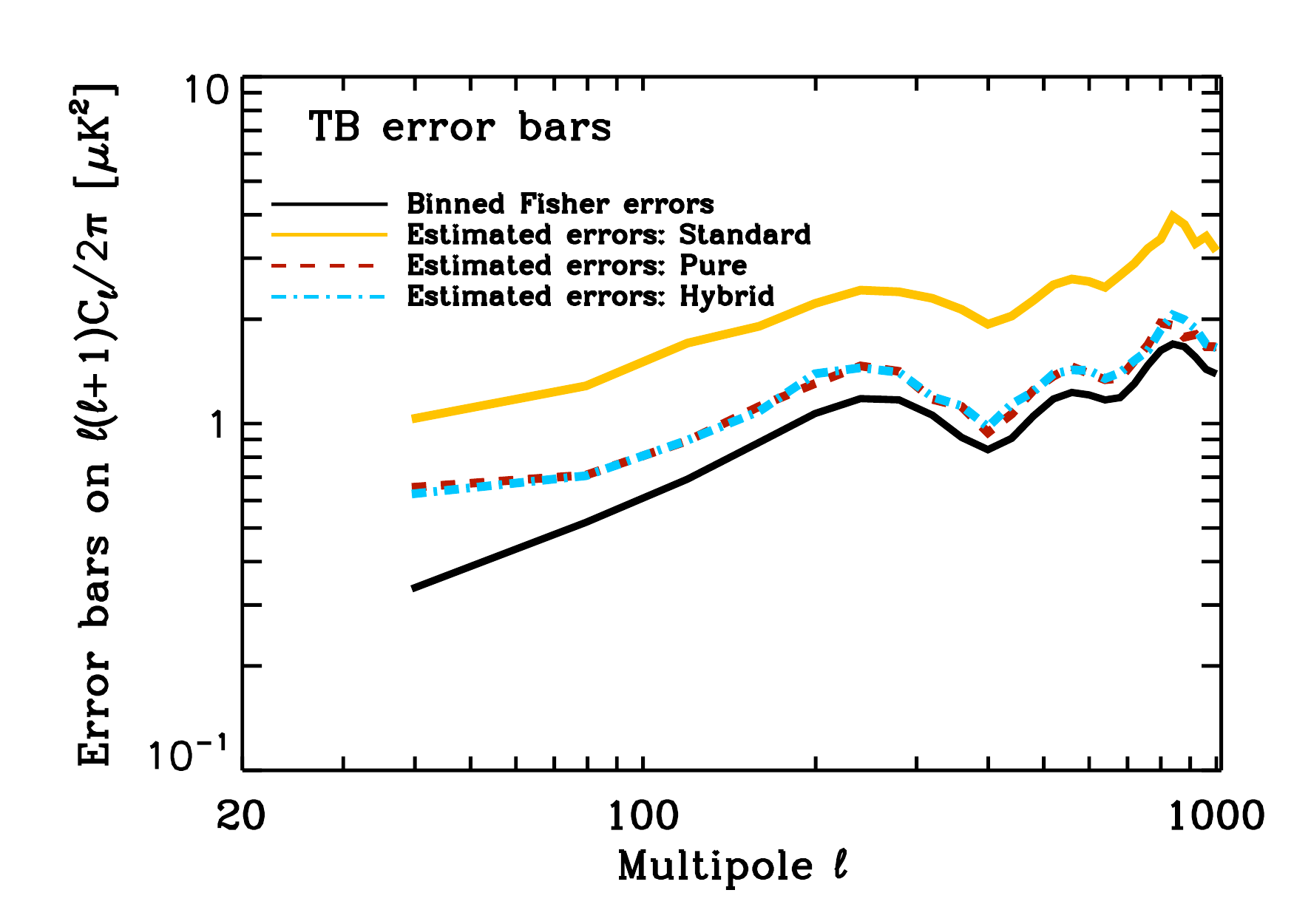}\includegraphics[scale=0.35]{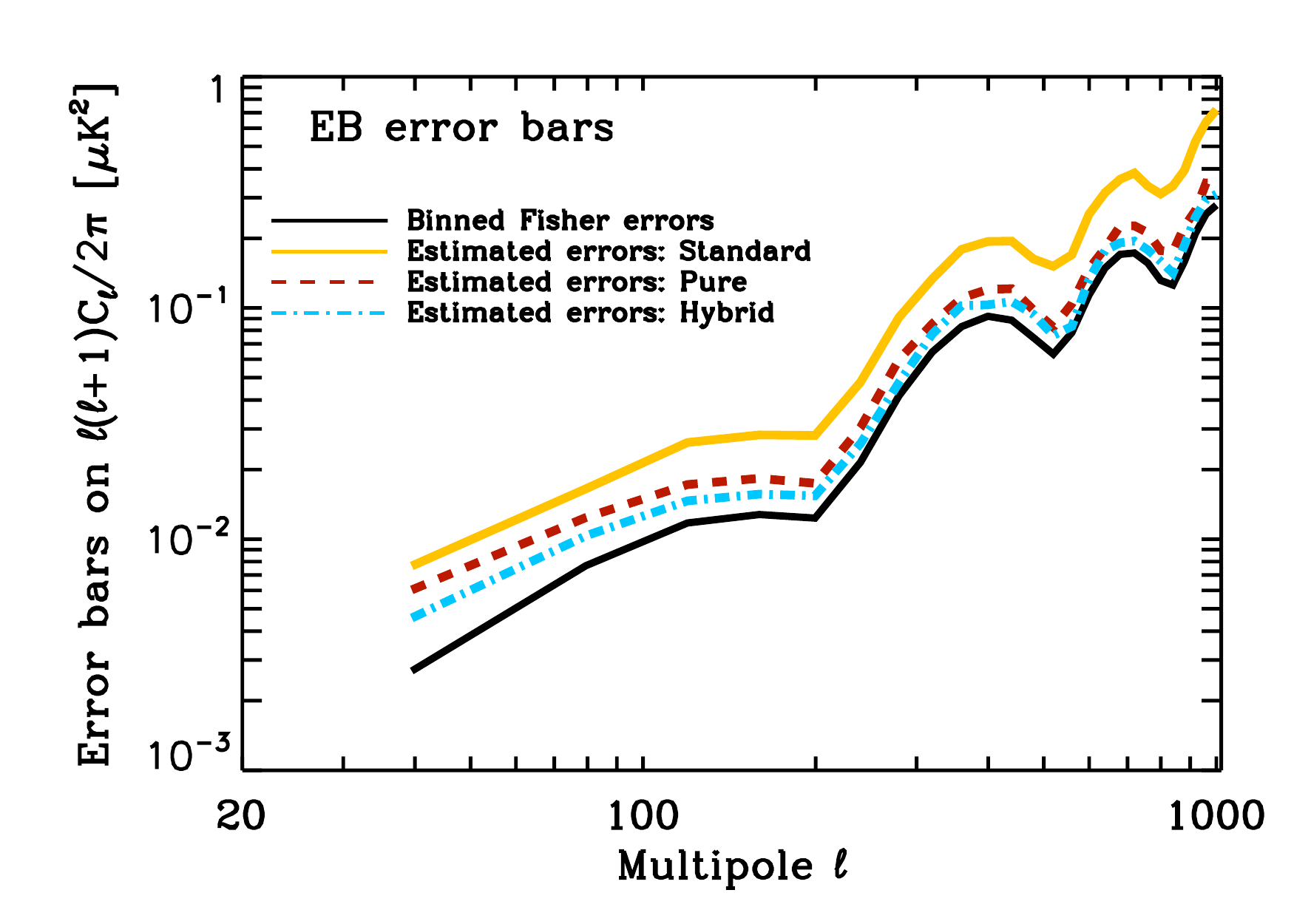}
	\caption{Error bars on the reconstructed angular power spectra for each of the three formalisms (colored curves) alongside the naive (binned) Fisher estimate of such uncertainties. The assumed sky survey is shown in Fig. \ref{fig:SkyapEbex} with inhomogeneous noise as displayed on Fig. \ref{fig:Nhit}).The error bars are obtained as the standard deviation over 500 MC simulations.}
	\label{fig:varebex}
\end{center}
\end{figure*}

The uncertainties  of the estimated angular power spectra are shown in Fig. \ref{fig:varebex} alongside the Fisher estimates~\footnote{We note here in passing that the difference of the $BB$ pure spectrum shown here with that shown in Fig. 24 of~\cite{grain_etal_2009} is due to a different noise level assumed erroneously in this earlier work.}. Overall the hybrid formalism leads to the smallest error bars and allowing for the inhomogeneous noise does not change  at a qualitative level, the main conclusion drawn earlier based on the homogeneous noise cases. A similar conclusion applies to the relative merits of each formalism. On the quantitative level, the noise inhomogeneities seem to enhance the differences between the techniques, as, for example, can be seen clearly  in the case of the $EE$ and $TE$ power spectra in  Fig.~\ref{fig:varebex}.  The larger uncertainties seen in these cases for the pure  estimator are due to the fact that sky apodizations employed in the pure estimation of $E$-modes are {\it a priori} not optimized to
account for the noise variation.

\section{Conclusions}
\label{sec:Conclusion}
We have presented three unbiased alternative approaches to estimate the six angular power spectra associated to CMB temperature and polarization anisotropies. Those techniques are based on the so-called pseudo-$C_\ell$'s estimators and generalize to the full set of angular power spectra the approaches already proposed in \cite{hivon_etal_2002,tristram_etal_2005,smith_2006,smith_zaldarriaga_2007,grain_etal_2009}. Each technique differs from the other in the way $E$- and $B$-pseudo-multipoles are computed from the maps of Stokes parameters. The three approaches are equivalent from the perspective of the $TT$ power spectrum. 
The first approach, dubbed {\it standard}, estimates polarization multipoles by projecting the {\it masked} $Q$ and $U$ maps on the standard spherical harmonics of $E$- and $B$-type. This technique does not lose any information contained in the observed maps but its variance suffers from the $E$-to-$B$ and $B$-to-$E$ leakages.
The second approach, denoted {\it pure}, makes use of the pure-$E$, pure-$B$ and ambiguous modes decomposition \cite{bunn_etal_2003} to project the $Q$ and $U$ maps on a $E$- and $B$-basis free from any leakages. This second technique therefore corrects for $E$-to-$B$ and $B$-to-$E$ leakages on any single realization but loses some part of the information by removing ambiguous modes. 
Finally, in the third approach, called {\it hybrid}, $E$-pseudo-multipoles are numerically computed using the standard approach while $B$-pseudo-multipoles are computed using the pure technique. For the specific case of CMB polarized anisotropies, $E$-modes are much greater than $B$-modes. From theoretical arguments and previous numerical experiments, statistical uncertainties on the $E$-pseudo-multipoles are therefore expected to be mainly increased by the loss of information while statistical uncertainties on the $B$-pseudo-multipoles are expected to be mainly increased by $E$-to-$B$ leakage. As a result, the hybrid technique should provide the smallest error bars at least as long as the $B$-mode power is not larger than that of the $E$-modes.

We have developed and implemented those three approaches by providing the set of expressions to efficiently compute the different mode-mode coupling kernels for each of the formalisms, as well as the noise bias for the specific case of uncorrelated noise from pixel to pixel. The overall consistency of those computations has been demonstrated via numerical experiments.

Finally, we have assessed the relative performances of the three proposed pseudo-$C_\ell$'s estimators for the case of small-scales experiments, 
assuming three different experimental configurations.  In all these cases and for all 6 angular power spectra, the hybrid estimator has been found to result  in the smallest statistical uncertainties, confirming the theoretical expectations, while the other estimators can produce similar uncertainties in some specific cases. For instance, i) the standard and hybrid formalisms performs equally well for estimating the $EE$ and $TE$ power spectra, ii) the pure and hybrid formalisms performs equally well to estimate the $BB$ and $TB$, and iii) the hybrid technique performs the best (at least at large angular scales) for the $EB$ power spectra, though the pure approach is on par with the error bars achieved with the hybrid estimator.

\begin{acknowledgments}
 This research used resources of the National Energy Research Scientific Computing Center, which is supported by the Office of Science of the U.S. Department of Energy under Contract No. DE-AC02-05CH11231. Some of the results in this paper have been derived using {\sc s$^2$hat} \cite{s2hat,pures2hat,hupca_etal_2010,szydlarski_etal_2011}, {\sc HEALPix}~\cite{gorski_etal_2005} and {\sc camb}~\cite{lewis_etal_2000} software packages. We thank Sam Leach, Will Grainger, Chris Cantalupo and Ted Kisner and the {\sc ebex} team fo providing the tools to simulate the realistic experiment study. JG acknowledges financial support from the Groupement d'Int\'er\^et Scientifique (GIS) 'consortium Physique des 2 Infinis~(P2I)'. We also thank the ANR-MIDAS'09 project team for helpful discussions.
\end{acknowledgments}

\begin{widetext}
\appendix

\section{Gaunt integrals and Wigner-$3j$}
\label{app:Gaunt}
We provide in this appendix some general definitions and properties of the Gaunt integrals and Wigner-$3j$ symbols which are useful for the computation of the mode-mode coupling matrices of the pseudo-$C_\ell$ estimators.

The general definition of the Gaunt integrals is 
\begin{equation}
	\mathcal{G}^{\ell m s}_{\ell'm's';\ell''m''s''}\equiv\ds\int_{4\pi}\Ylm{s}{\ell}{m}~\Ylm{s'}{\ell'}{m'}~\Ylm{s''}{\ell''}{m''}d\vec{n},
\end{equation}
and it can be expressed by use of the Wigner-$3j$ symbols
\begin{equation}
	\mathcal{G}^{\ell m s}_{\ell'm's';\ell''m''s''}=\sqrt{\frac{(2\ell+1)(2\ell'+1)(2\ell''+1)}{4\pi}}\left(\begin{array}{ccc}
		\ell&\ell'&\ell'' \\
		s&s'&s''
	\end{array}\right)\left(\begin{array}{ccc}
		\ell&\ell'&\ell'' \\
		m&m'&m''
	\end{array}\right).
\end{equation}
Ths is easily generalized to integrals involving complex conjugates of the spin-weighted spherical harmonics by using $\Ylmcc{s}{\ell}{m}=(-1)^{s+m}\Ylm{-s}{\ell}{(-m)}$. The Wigner-$3j$ symbols verify the following symmetry and orthogonality relations 
\begin{equation}
	\left(\begin{array}{ccc}
		\ell&\ell'&\ell'' \\
		-m&-m'&-m''
	\end{array}\right)=(-1)^{\ell+\ell'+\ell''}\left(\begin{array}{ccc}
		\ell&\ell'&\ell'' \\
		m&m'&m''
	\end{array}\right)
\end{equation}
and
\begin{equation}
	(2\ell''+1)\ds\sum_{mm'}\left(\begin{array}{ccc}
		\ell&\ell'&\ell'' \\
		m&m'&m''
	\end{array}\right)\left(\begin{array}{ccc}
		\ell&\ell'&\ell''' \\
		m&m'&m'''
	\end{array}\right)=\delta_{\ell'',\ell'''}\delta_{m'',m'''}.
\end{equation}

Two linear combinations of Wigner-$3j$ symbols and Gaunt integrals are involved in the expression of the pseudo-multipoles mode-mode coupling matrices. To lighten our expressions, we defined the following quantities:
\begin{equation}
	\displaystyle J^{(\pm)}_{2-s}(\ell,\ell',\ell'')=\displaystyle\left(\begin{array}{ccc}
		\ell&\ell'&\ell'' \\
		-s&2&-2+s
	\end{array}\right)\pm\left(\begin{array}{ccc}
		\ell&\ell'&\ell'' \\
		s&-2&2-s
	\end{array}\right),
\end{equation}
and
\begin{eqnarray}
	\displaystyle G^{(\pm)}_{2-s}(L,L',L'')&\displaystyle=&\displaystyle\int_{4\pi}\bigg[\left(\Ylmcc{s}{\ell}{m}\Ylm{2}{\ell'}{m'}\Ylm{-2+s}{\ell''}{m''}\right)\pm\left(\Ylmcc{-s}{\ell}{m}\Ylm{-2}{\ell'}{m'}\Ylm{2-s}{\ell''}{m''}\right)\bigg]d\vec{n}, \\
	&=&(-1)^{s+m}\sqrt{\frac{(2\ell+1)(2\ell'+1)(2\ell''+1)}{4\pi}}\left(\begin{array}{ccc}
		\ell&\ell'&\ell'' \\
		m&m'&m''
	\end{array}\right)J^{(\pm)}_{2-s}(\ell,\ell',\ell''), \nonumber
\end{eqnarray}
with $L$ labeling the couple $(\ell,m)$ in the Gaunt integrals. From the parity relation of the Wigner-$3j$, it is easily shown that $J^{(+)}_{2-s}=0$ for odd values of $(\ell+\ell'+\ell'')$ and $J^{(-)}_{2-s}=0$ for even values of $(\ell+\ell'+\ell'')$. Finally, we also define the following Wigner-$3j$ symbols
\begin{equation}
	J^{(T)}(\ell,\ell',\ell'')=\displaystyle\left(\begin{array}{ccc}
		\ell&\ell'&\ell'' \\
		0&0&0
	\end{array}\right)
\end{equation}
which is equal to zero for odd values of $(\ell+\ell'+\ell'')$.

\section{Mode-mode coupling matrices}
\label{app:mixtotal}
\subsection{Mode-mode coupling for (pure) pseudo-$a_{\ell m}$}
\label{app:mixingAlm}
In this appendix, we present the explicit expressions of the different mode-mode coupling matrices related the pseudo-multipoles to the CMB multipoles. We remind that standard pseudo-$a_{\ell m}$'s are defined by
\begin{eqnarray}
	E^{(std)}_{\ell m}&\equiv&-\frac{1}{2N_{\ell,2}}\displaystyle\int_{4\pi}W\left[\left(Q+iU\right)\left(\partial^2Y_{\ell m}\right)^\star+\left(Q-iU\right)\left(\bar\partial^2Y_{\ell m}\right)^\star\right]d\vec{n}, \\
	B^{(std)}_{\ell m}&\equiv&\frac{i}{2N_{\ell,2}}\displaystyle\int_{4\pi}W\left[\left(Q+iU\right)\left(\partial^2Y_{\ell m}\right)^\star-\left(Q-iU\right)\left(\bar\partial^2Y_{\ell m}\right)^\star\right]d\vec{n},
\end{eqnarray}
and pure pseudo-$a_{\ell m}$'s are defined by
\begin{eqnarray}
	E^{(pure)}_{\ell m}&\equiv&-\frac{1}{2N_{\ell,2}}\displaystyle\int_{4\pi}\left[\left(Q+iU\right)\left(\partial^2WY_{\ell m}\right)^\star+\left(Q-iU\right)\left(\bar\partial^2WY_{\ell m}\right)^\star\right]d\vec{n}, \\
	B^{(pure)}_{\ell m}&\equiv&\frac{i}{2N_{\ell,2}}\displaystyle\int_{4\pi}\left[\left(Q+iU\right)\left(\partial^2WY_{\ell m}\right)^\star-\left(Q-iU\right)\left(\bar\partial^2WY_{\ell m}\right)^\star\right]d\vec{n}.
\end{eqnarray}
We will first provide the expressions for pure pseudo-multipoles as the standard approach can be easily deduced from the pure one.

\subsubsection{Pure pseudo-multipoles} 
By expanding the spin-raising(spin-lowering) operators acting on the product $W\times Y_{\ell m}$ and using the definition of the spin-weighted spherical harmonics, one obtains
\begin{eqnarray}
	E^{(pure)}_{\ell m}&\equiv&-\frac{1}{2N_{\ell,2}}\displaystyle\int_{4\pi}\displaystyle\sum_{s=0}^2\alpha_{s}N_{\ell,s}\left[\left(Q+iU\right)\left(\partial^{2-s}W\right)^\star{}_sY^\star_{\ell m}+(-1)^s\left(Q-iU\right)\left(\bar\partial^{2-s}W\right)^\star{}_{-s}Y^\star_{\ell m}\right]d\vec{n},
	\label{pureE} 
	\\
	B^{(pure)}_{\ell m}&\equiv&\frac{i}{2N_{\ell,2}}\displaystyle\int_{4\pi}\displaystyle\sum_{s=0}^2\alpha_{s}N_{\ell,s}\left[\left(Q+iU\right)\left(\partial^{2-s}W\right)^\star{}_{s}Y^\star_{\ell m}-(-1)^s\left(Q-iU\right)\left(\bar\partial^{2-s}W\right)^\star{}_{-s}Y_{\ell m}\right]d\vec{n},
	\label{pureB} 
\end{eqnarray}
with $\alpha_{s}=1$ for $s=0,~2$ and $\alpha_{s}=2$ for $s=1$. The $s=2$ term in the above summation over the spin index simply corresponds to the standard multipoles. Therefore, one can first compute the mode-mode coupling matrices for the pure case and then keep only the $s=2$ term in the summation to recover the mode-mode coupling matrices for the standard estimation. This computation is done by plugging the multipolar decomposition of the $(Q\pm iU)$ fields and of the window function $W$, i.e.
$$
\begin{array}{l}
	W=-\ds\sum_{\ell m}w^{(E)}_{0,\ell m}Y_{\ell m}, \\
	Q\pm iU=-\ds\sum_{\ell m}\left(a_{E,\ell m}\pm ia_{B,\ell m}\right)\Ylm{\pm2}{\ell}{m}.
\end{array}
$$
If the derivative relationship between the different spin-weighted window function is completely satisfied, i.e., $W_{s}=\partial^{s}W$, the multipolar decomposition of the spin-0 window function $W$ is sufficient as
$$
w^{(E)}_{s,\ell m}=N_{\ell,s}w_{\ell m}~~~\mathrm{and}~~~w^{(B)}_{s,\ell m}=0.
$$
However, as shown in \cite{grain_etal_2009}, pixelization on the sphere partially breaks this derivative relationship and the multipolar decomposition of each of the spin-weighted window function has to be considered as independent from the spin-0 ones, i.e.
$$
\begin{array}{lcl}
	W_{s}&=&-\ds\sum_{\ell m}\left(w^{(E)}_{s,\ell m}+iw^{(B)}_{s,\ell m}\right)\Ylm{s}{\ell}{m}, \\
	W_{-s}&=&(-1)^{s+1}\ds\sum_{\ell m}\left(w^{(E)}_{s,\ell m}-iw^{(B)}_{s,\ell m}\right)\Ylm{s}{\ell}{m}. \\
\end{array}
$$
By plugging the above multipolar decomposition into Eqs. (\ref{pureE}) and (\ref{pureB}) and using the Gaunt integrals, it is straightforward, though rather long, to show that
$$
\begin{array}{lcl}
	E^{(pure)}_{\ell m}&=&\ds\sum_{\ell'm'}\left[H^{diag}_{\ell m,\ell'm'}a_{E,\ell'm'}+iH^{off}_{\ell m,\ell'm'}a_{B,\ell'm'}\right], \\
	B^{(pure)}_{\ell m}&=&\ds\sum_{\ell'm'}\left[H^{diag}_{\ell m,\ell'm'}a_{B,\ell'm'}-iH^{off}_{\ell m,\ell'm'}a_{E,\ell'm'}\right],
\end{array}
$$
with
\begin{equation}
\begin{array}{lcl}
	\ds H^{diag}_{\ell m,\ell' m'}&=&\ds \frac{-1}{2N_{\ell,2}}\ds\sum_{\ell''m''}\sum_{s=0}^2(-1)^s\alpha_sN_{\ell,s}\left[w^{(E)}_{2-s,\ell''m''}G^{(+)}_{2-s}(L,L',L'')-iw^{(B)}_{2-s,\ell''m''}G^{(-)}_{2-s}(L,L',L'')\right], \nonumber \\
	\ds H^{off}_{\ell m,\ell' m'}&=&\ds \frac{-1}{2N_{\ell,2}}\ds\sum_{\ell''m''}\sum_{s=0}^2(-1)^s\alpha_sN_{\ell,s}\left[w^{(E)}_{2-s,\ell''m''}G^{(-)}_{2-s}(L,L',L'')-iw^{(B)}_{2-s,\ell''m''}G^{(+)}_{2-s}(L,L',L'')\right]. \nonumber
\end{array}
\label{Hll-pure}
\end{equation}

We underline that the spin-dependant pseudo-multipoles, $E_{s,\ell m}$ and $B_{s,\ell m}$, as defined in Eqs. (\ref{aeapp}) and (\ref{abapp}) are also linked to the CMB multipoles via a mode-mode coupling
\begin{eqnarray}
	E_{s,\ell m}&=&\ds\sum_{\ell'm'}\left[K^{s,diag}_{\ell m,\ell'm'}a_{E,\ell'm'}+iK^{s,off}_{\ell m,\ell'm'}a_{B,\ell'm'}\right], \nonumber \\
	B_{s,\ell m}&=&\ds\sum_{\ell'm'}\left[K^{s,diag}_{\ell m,\ell'm'}a_{B,\ell'm'}-iK^{s,off}_{\ell m,\ell'm'}a_{E,\ell'm'}\right], \nonumber 
\end{eqnarray}
with
\begin{equation}
\begin{array}{lcl}
	\ds K^{s,diag}_{\ell m,\ell'm'}&=&\ds -\frac{1}{2}\ds\sum_{\ell''m''}(-1)^s\alpha_s\left[w^{(E)}_{2-s,\ell''m''}G^{(+)}_{2-s}(L,L',L'')-iw^{(B)}_{2-s,\ell''m''}G^{(-)}_{2-s}(L,L',L'')\right], \nonumber  \\
	\ds K^{s,off}_{\ell m,\ell'm'}&=&\ds -\frac{1}{2}\ds\sum_{\ell''m''}(-1)^s\alpha_s\left[w^{(E)}_{2-s,\ell''m''}G^{(-)}_{2-s}(L,L',L'')-iw^{(B)}_{2-s,\ell''m''}G^{(+)}_{2-s}(L,L',L'')\right]. \nonumber 
\end{array}
\label{Kll-pure}
\end{equation}
The full mode-mode coupling matrices are then related to the spin-weighted mode-mode coupling matrices by
\begin{eqnarray}
	H^{diag}_{\ell m,\ell'm'}=\frac{1}{N_{\ell,2}}\ds\sum_{s=0}^2N_{\ell,s}K^{s,diag}_{\ell m,\ell'm'} & ~~~\mathrm{and}~~~ & H^{off}_{\ell m,\ell'm'}=\frac{1}{N_{\ell,2}}\ds\sum_{s=0}^2N_{\ell,s}K^{s,off}_{\ell m,\ell'm'}. \nonumber
\end{eqnarray}
 
\subsubsection{Standard pseudo-multipoles}
From the above result, we can easily deduce the mode-mode coupling matrices for standard multipoles by simply keeping the $s=2$ terms in the summation over the spin index to get
$$
\begin{array}{lcl}
	E^{(std)}_{\ell m}&=&\ds\sum_{\ell'm'}\left[K^{diag}_{\ell m,\ell'm'}a_{E,\ell'm'}+iK^{off}_{\ell m,\ell'm'}a_{B,\ell'm'}\right], \\
	B^{(std)}_{\ell m}&=&\ds\sum_{\ell'm'}\left[K^{diag}_{\ell m,\ell'm'}a_{B,\ell'm'}-iK^{off}_{\ell m,\ell'm'}a_{E,\ell'm'}\right],
\end{array}
$$
with
\begin{equation}
\begin{array}{lcl}
	\ds K^{diag}_{\ell m,\ell' m'}&=&\ds K^{2,diag}_{\ell m,\ell' m'}=-\frac{1}{2}\ds\sum_{\ell''m''}w^{(E)}_{0,\ell''m''}G^{(+)}_{2-s}(L,L',L''), \\
	\ds K^{off}_{\ell m,\ell' m'}&=&\ds K^{2,off}_{\ell m,\ell' m'}=-\frac{1}{2}\ds\sum_{\ell''m''}w^{(E)}_{0,\ell''m''}G^{(-)}_{2}(L,L',L'').
\end{array}
\label{Kll-std}
\end{equation}
We stress out that for standard estimation of the pseudo-multipoles, only the real-valued, spin-0 window function is used which does not contain any $B$-component.


\subsection{Mode-mode coupling for {\it EB} pseudo-$C_\ell$}
\label{app:MixingEB}
We remind that the pseudo-multipoles are related to the CMB multipoles via a mode-mode coupling matrix
$$
\begin{array}{lcl}
	\ds E^{(x)}_{\ell m}&\ds=&\ds\sum_{\ell'm'}\left[\mathcal{K}^{diag}_{\ell m;\ell'm'}a_{E,\ell'm'}+i\mathcal{K}^{off}_{\ell m;\ell'm'}a_{B,\ell'm'}\right], \\
	\ds B^{(y)}_{\ell m}&\ds=&\ds\sum_{\ell'm'}\left[\mathcal{H}^{diag}_{\ell m;\ell'm'}a_{B,\ell'm'}-i\mathcal{H}^{off}_{\ell m;\ell'm'}a_{E,\ell'm'}\right],
\end{array}
$$
where $(x)$ and $(y)$ corresponds to either "standard" or "pure" estimation. We stress out that the above expressions do not formally depend on the formalism. All the formalism dependancy is encoded in the matrices $\mathcal{K}$ and $\mathcal{H}$, which can be set equal to $K$ or $H$, depending on the values adopted for $(x)$ and $(y)$. In the following, we assume that the polarization field is a gaussian and isotropic random fields
$$
\begin{array}{lcl}
	\ds\left<a_{E,\ell m}a^\star_{E,\ell'm'}\right>&\ds=&C^{EE}_{\ell}\delta_{\ell,\ell'}\delta_{m,m'}, \\
	\ds\left<a_{B,\ell m}a^\star_{B,\ell'm'}\right>&\ds=&C^{BB}_{\ell}\delta_{\ell,\ell'}\delta_{m,m'}, \\
	\ds\left<a_{E,\ell m}a^\star_{B,\ell'm'}\right>&\ds=&\ds\left<a_{B,\ell m}a^\star_{E,\ell'm'}\right>=C^{EB}_{\ell}\delta_{\ell,\ell'}\delta_{m,m'}.
\end{array}
$$
The pseudo-power spectra are defined as follows:
$$
\begin{array}{lcl}
	\ds\mathcal{C}^{EE}_{\ell,(xx)}&\ds=&\ds\frac{1}{2\ell+1}\ds\sum_mE^{(x)}_{\ell m}E^{(x)\star}_{\ell m}, \\
	\ds\mathcal{C}^{BB}_{\ell,(yy)}&\ds=&\ds\frac{1}{2\ell+1}\ds\sum_mB^{(y)}_{\ell m}B^{(y)\star}_{\ell m}, \\
	\ds\mathcal{C}^{EB}_{\ell,(xy)}&\ds=&\ds\frac{1}{2(2\ell+1)}\ds\sum_m\left[E^{(x)}_{\ell m}B^{(y)\star}_{\ell m}+B^{(y)}_{\ell m}E^{(x)\star}_{\ell m}\right].
\end{array}
$$
Starting from the definition of the pseudo-multipoles, it is easily shown that
\begin{eqnarray}
	\left<\mathcal{C}^{EE}_{\ell,(xx)}\right>&=&\frac{1}{2\ell+1}\ds\sum_{\ell'}\sum_{mm'}\left\{\left|\mathcal{K}^{diag}_{\ell m;\ell'm'}\right|^2C^{EE}_{\ell'}+\left|\mathcal{K}^{off}_{\ell m;\ell'm'}\right|^2C^{BB}_{\ell'}+2\mathrm{Im}\left[\mathcal{K}^{diag}_{\ell m;\ell'm'}\mathcal{K}^{off\star}_{\ell m;\ell'm'}\right]C^{EB}_{\ell'}\right\}, \nonumber \\
	\left<\mathcal{C}^{BB}_{\ell,(yy)}\right>&=&\frac{1}{2\ell+1}\ds\sum_{\ell'}\sum_{mm'}\left\{\left|\mathcal{H}^{off}_{\ell m;\ell'm'}\right|^2C^{EE}_{\ell'}+\left|\mathcal{H}^{diag}_{\ell m;\ell'm'}\right|^2C^{BB}_{\ell'}-2\mathrm{Im}\left[\mathcal{H}^{diag}_{\ell m;\ell'm'}\mathcal{H}^{off\star}_{\ell m;\ell'm'}\right]C^{EB}_{\ell'}\right\}, \nonumber \\
	\left<\mathcal{C}^{EB}_{\ell,(xy)}\right>&=&\frac{1}{2\ell+1}\ds\sum_{\ell'}\sum_{mm'}\left\{\mathrm{Re}\left[\mathcal{K}^{diag}_{\ell m;\ell'm'}\mathcal{H}^{diag\star}_{\ell m;\ell'm'}-\mathcal{K}^{off}_{\ell m;\ell'm'}\mathcal{H}^{off\star}_{\ell m;\ell'm'}\right]C^{EB}_{\ell'}\right. \\
	&&\left.+\mathrm{Im}\left[\mathcal{H}^{diag}_{\ell m;\ell'm'}\mathcal{K}^{off\star}_{\ell m;\ell'm'}\right]C^{BB}_{\ell'}-\mathrm{Im}\left[\mathcal{K}^{diag}_{\ell m;\ell'm'}\mathcal{H}^{off\star}_{\ell m;\ell'm'}\right]C^{EE}_{\ell'}\right\}. \nonumber
\end{eqnarray}
The form of the above expressions is also formalism-independant as all the specificities of a given formalism is encoded in the {\it peculiar} expressions of the $\mathcal{K}$ and $\mathcal{H}$ matrices. Without any {\it a priori} on the explicit expressions of those matrices, each of the three pseudo-$C_\ell$'s receives contribution from the three angular power spectra. However, due to the parity properties of Wigner-$3j$ symbols, it will be shown in the following of this appendix that the $(EB,PP)$ and $(PP,EB)$ blocks are zero as long as the window functions {\it do not} contain any $B$-type component.

The mode-mode coupling matrices for the pseudo-power spectra are finally computed by taking the expression of the $\mathcal{K}$ and $\mathcal{H}$ matrices valid for the adopted formalism and then performing the summation over the two azimuthal quantum numbers $m$ and $m'$. This explicit computation is very similar to the temperature case \cite{hivon_etal_2002,hauser_peebles_1973,hinshaw_etal_2003}. The standard and pure formalism being a restricted case of the pure formalism\footnote{We remind that the pure pseudo-multipoles are a linear combination of three spin-dependant pseudo-multipoles ($s=0,~1,~2$) and the standard pseudo-multipoles are given by the spin-$2$ term of the pure ones.}, the computation of the mixing kernels is only detailed for this last case.  The forthcoming formulas~ involve the $J^{(\pm)}_{2-s}(\ell,\ell',\ell'')$ quantities and we will remove their $\ell$, $\ell'$, and $\ell''$ dependance to lighten the writings.

\subsubsection{Standard formalism}
In the standard formalism, the mode-mode coupling matrices for the pseudospectra reads \cite{kogut_etal_2003}:
\begin{equation}
\begin{array}{lcl}
	\ds M^{EE,EE}_{\ell\ell'}&\ds =&\ds M^{BB,BB}_{\ell\ell'}=\frac{2\ell'+1}{16\pi}\ds\sum_{\ell''m''}\left|w^{(E)}_{0,\ell''m''}J^{(+)}_0\right|^2, \\
	\ds M^{EE,BB}_{\ell\ell'}&\ds =&\ds M^{BB,EE}_{\ell\ell'}=\frac{2\ell'+1}{16\pi}\ds\sum_{\ell''m''}\left|w^{(E)}_{0,\ell''m''}J^{(-)}_0\right|^2, \\
	\ds M^{EB,EB}_{\ell\ell'}&\ds =&\ds \frac{2\ell'+1}{16\pi}\ds\sum_{\ell''m''}\left[\left|w^{(E)}_{0,\ell''m''}J^{(+)}_0\right|^2-\left|w^{(E)}_{0,\ell''m''}J^{(-)}_0\right|^2\right], \\
	\ds M^{EE,EB}_{\ell\ell'}&\ds =&\ds -M^{BB,EB}_{\ell\ell'}=\frac{2\ell'+1}{8\pi}\ds\sum_{\ell''m''}\left|w^{(E)}_{0,\ell''m''}\right|^2J^{(+)}_0J^{(-)}_0, \\
	\ds M^{EB,EE}_{\ell\ell'}&\ds =&\ds -M^{EB,BB}_{\ell\ell'}=-\frac{2\ell'+1}{16\pi}\ds\sum_{\ell''m''}\left|w^{(E)}_{0,\ell''m''}\right|^2J^{(+)}_0J^{(-)}_0
\end{array}
\label{Mll-std1}
\end{equation}
However, because $J^{(+)}$ vanish for odd values of $(\ell+\ell'+\ell'')$ while $J^{(-)}$ vanish for even values of $(\ell+\ell'+\ell'')$, the off-diagonal blocks relating autospectra to the $EB$ cross-spectrum become null
\begin{equation}
\begin{array}{lcl}
	\ds M^{EE,EB}_{\ell\ell'}&\ds =&\ds -M^{BB,EB}_{\ell\ell'}=0, \\
	\ds M^{EB,EE}_{\ell\ell'}&\ds =&\ds -M^{EB,BB}_{\ell\ell'}=0.
\end{array}
\label{Mll-std2}
\end{equation}

\subsubsection{Pure formalism}
For the $(EE,EE)$ block computation, the expression of Gaunt integrals as functions of Wigner-$3j$ symbols is first plugged in the expression of $H^{diag}_{\ell m,\ell' m'}$ to get
$$
\begin{array}{lcl}
	\ds M^{EE,EE}_{\ell\ell'}&=&\ds\left(\frac{2\ell'+1}{16\pi N^2_{\ell,2}}\right)\ds\sum_{\ell''m''}\sum_{\ell'''m'''}\sum_{s,s'=0}^2\alpha_s\alpha_{s'}N_{\ell,s}N_{\ell,s'}\sqrt{(2\ell''+1)(2\ell'''+1)}\left(\begin{array}{ccc}
			\ell&\ell'&\ell'' \\
			-m&m'&m''
		\end{array}\right)\left(\begin{array}{ccc}
			\ell&\ell'&\ell''' \\
			-m&m'&m'''
		\end{array}\right) \\
	&&\times\left[w^{(E)}_{2-s,\ell''m''}J^{(+)}_{2-s}(\ell,\ell',\ell'')-iw^{(B)}_{2-s,\ell''m''}J^{(-)}_{2-s}(\ell,\ell',\ell'')\right] \\
	&&\times\left[w^{(E)}_{2-s',\ell'''m'''}J^{(+)}_{2-s'}(\ell,\ell',\ell''')-iw^{(B)}_{2-s',\ell'''m'''}J^{(-)}_{2-s'}(\ell,\ell',\ell''')\right]^\star.
\end{array}
$$
We can now first use the orthonormalization properties of the Wigner-$3j$ listed in App. \ref{app:Gaunt} and secondly, the fact that all the terms proportional to $\left(J^{(+)}_{2-s}J^{(-)}_{2-s'}\right)$ are equal to zero to get
\begin{equation}
	M^{EE,EE}_{\ell\ell'}=\frac{2\ell'+1}{16\pi}\ds\sum_{\ell'm''}\left[\left|\sum_{s=0}^2\alpha_s\frac{N_{\ell,s}}{N_{\ell,2}}w^{(E)}_{2-s,\ell''m''}J^{(+)}_{2-s}\right|^2+\left|\sum_{s=0}^1\alpha_s\frac{N_{\ell,s}}{N_{\ell,2}}w^{(B)}_{2-s,\ell''m''}J^{(-)}_{2-s}\right|^2\right].
	\label{Mll-pure1}
\end{equation}
We emphasize that the spin-summation runs from 0 to 2 for the $E$-component of the window functions but from 0 to 1 for its $B$-component as only the spin-1 and spin-2 windows may have a nonvanishing $B$-part. Applying the same reasoning for the $(EE,BB)$ and $(EE,EB)$ blocks leads to
\begin{eqnarray}
	\ds M^{EE,BB}_{\ell\ell'}&=&\ds \frac{2\ell'+1}{16\pi}\ds\sum_{\ell'm''}\left[\left|\sum_{s=0}^2\alpha_s\frac{N_{\ell,s}}{N_{\ell,2}}w^{(E)}_{2-s,\ell''m''}J^{(-)}_{2-s}\right|^2+\left|\sum_{s=0}^1\alpha_s\frac{N_{\ell,s}}{N_{\ell,2}}w^{(B)}_{2-s,\ell''m''}J^{(+)}_{2-s}\right|^2\right], \nonumber \\
	\ds M^{EE,EB}_{\ell\ell'}&=&\ds \frac{2\ell'+1}{8\pi}\ds\sum_{\ell'm''}\sum_{s,s'=0}^2\alpha_s\alpha_{s'}\frac{N_{\ell,s}N_{\ell,s'}}{N_{\ell,2}^2}\left\{J^{(+)}_{2-s}J^{(+)}_{2-s'}\mathrm{Re}\left[w^{(E)}_{2-s,\ell''m''}w^{(B)\star}_{2-s',\ell''m''}\right]\right. 
	\label{Mll-pure2}
	\\
	&&\left.-J^{(-)}_{2-s}J^{(-)}_{2-s'}\left[w^{(B)}_{2-s,\ell''m''}w^{(E)\star}_{2-s',\ell''m''}\right]\right\}. \nonumber
\end{eqnarray}
The other blocks are easily deduced from the above three ones as
$$
\begin{array}{lcl}
	\ds M^{BB,BB}_{\ell\ell'}=M^{EE,EE}_{\ell\ell'}, & \ds M^{EE,BB}_{\ell\ell'}=M^{BB,EE}_{\ell\ell'}, & \ds M^{EB,EB}_{\ell\ell'}=M^{EE,EE}_{\ell\ell'}-M^{EB,EB}_{\ell\ell'} \nonumber \\
	\ds M^{BB,EB}_{\ell\ell'}=-M^{EE,EB}_{\ell\ell'}, & \ds M^{EB,EE}_{\ell\ell'}=-\frac{1}{2}M^{EE,EB}_{\ell\ell'}, & \ds M^{EB,BB}_{\ell\ell'}=\frac{1}{2}M^{EE,EB}_{\ell\ell'}. \nonumber
\end{array}
$$
The expression of the $M^{EE,EB}_{\ell\ell'}$ block shows that in the pure formalism, the $EB$-$EE$ coupling and the $EB$-$BB$ coupling are not zero (unlike the standard formalism). However, such couplings will vanish if the spin-1 and spin-2 window functions have a vanishing component of type $B$.

\subsubsection{Hybrid formalism}
In the hybrid formalism, only the $M^{EB,PP'}_{\ell\ell'}$ blocks needs to be computed as the $M^{EE,PP'}_{\ell\ell'}$ blocks are equal to the $M^{EE,PP'}_{\ell\ell'}$ blocks computed in the standard formalism and the $M^{BB,PP'}_{\ell\ell'}$ blocks equal to the $M^{BB,PP'}_{\ell\ell'}$ blocks derived in the pure formalism. For the remaining $M^{EB,PP'}_{\ell\ell'}$ blocks, we found
\begin{eqnarray}
	M^{EB,EE}_{\ell\ell'}&=&-\frac{2\ell'+1}{16\pi}\ds\sum_{\ell'm''}\frac{1}{N_{\ell,2}}\sum_{s=0}^1\alpha_sN_{\ell,s}J^{(+)}_{0}J^{(+)}_{2-s}\mathrm{Re}\left[w^{(E)}_{0,\ell''m''}w^{(B)\star}_{2-s,\ell''m''}\right], \nonumber \\
	M^{EB,BB}_{\ell\ell'}&=&-\frac{2\ell'+1}{16\pi}\ds\sum_{\ell'm''}\frac{1}{N_{\ell,2}}\sum_{s=0}^1\alpha_sN_{\ell,s}J^{(-)}_{0}J^{(-)}_{2-s}\mathrm{Re}\left[w^{(E)\star}_{0,\ell''m''}w^{(B)}_{2-s,\ell''m''}\right], 
	\label{Mll-hybrid} 
	\\
	M^{EB,EB}_{\ell\ell'}&=&\frac{2\ell'+1}{16\pi}\ds\sum_{\ell'm''}\frac{1}{N_{\ell,2}}\sum_{s=0}^2\alpha_sN_{\ell,s}\left[J^{(+)}_{0}J^{(+)}_{2-s}-J^{(-)}_{0}J^{(-)}_{2-s}\right]\mathrm{Re}\left[w^{(E)}_{0,\ell''m''}w^{(E)\star}_{2-s,\ell''m''}\right], \nonumber
\end{eqnarray}
As is the case for the pure formalism, the $EB$-$EE$ and $EB$-$BB$ couplings are not zero if the spin-1 and spin-2 window functions have a nonvanishing $B$-component.

\subsection{Mode-mode coupling for {\it TB} pseudo-$C_\ell$}
\label{app:MixingTB}
The temperature pseudo-multipoles are related to the CMB multipoles via the following coupling \cite{hivon_etal_2002,hauser_peebles_1973,hinshaw_etal_2003}:
\begin{equation}
	T_{\ell m}=-\ds\sum_{\ell'm'}\sum_{\ell''m''}(-1)^m\sqrt{\frac{(2\ell+1)(2\ell'+1)(2\ell''+1)}{4\pi}}J^{(T)}(\ell,\ell',\ell'')\left(\begin{array}{ccc}
		\ell&\ell'&\ell'' \\
		m&m'&m''
	\end{array}\right)w^{(E)}_{0,\ell''m''}a_{T,\ell'm'}.
	\label{equ:TempMulti}
\end{equation}
Combining the above pseudo-$a^T_{\ell m}$ with pseudo-$a^{E/B}_{\ell m}$ allows us to compute the mode-mode coupling matrices for the pseudo-power spectra in the different formalism. The procedure is similar, though simpler, to the polarization case described in the previous appendix and we only list here our final results.

First, in such the {\it standard} formalism, the mode-mode coupling matrices read
\begin{equation}
\begin{array}{lcl}
	\ds M^{TE,TE}_{\ell\ell'}&=&\ds M^{TB,TB}_{\ell\ell'}=\frac{2\ell'+1}{8\pi}\ds\sum_{\ell''m''}\left|w^{(E)}_{0,\ell''m''}\right|^2J^{(T)}(\ell,\ell',\ell'')J^{(+)}_{0}(\ell,\ell',\ell''), \\
	\ds M^{TE,TB}_{\ell\ell'}&=&\ds -M^{TB,TE}_{\ell\ell'}=0.
\end{array}
\end{equation}

Second, in the {\it pure} formalism, the mode-mode coupling matrices are given by 
\begin{equation}
\begin{array}{lcl}
	\ds M^{TE,TE}_{\ell\ell'}&=&\ds M^{TB,TB}_{\ell\ell'}=\frac{2\ell'+1}{8\pi}\ds\sum_{\ell''m''}\sum_{s}\alpha_s\frac{N_{\ell,s}}{N_{\ell,2}}J^{(T)}J^{(+)}_{2-s}\mathrm{Re}\left[w^{(E)}_{0,\ell''m''}w^{(E)\star}_{2-s,\ell''m''}\right], \\
	\ds M^{TE,TB}_{\ell\ell'}&=&\ds -M^{TB,TE}_{\ell\ell'}=\frac{2\ell'+1}{8\pi}\ds\sum_{\ell''m''}\sum_{s}\alpha_s\frac{N_{\ell,s}}{N_{\ell,2}}J^{(T)}J^{(+)}_{2-s}\mathrm{Re}\left[w^{(E)}_{0,\ell''m''}w^{(B)\star}_{2-s,\ell''m''}\right].
\end{array}
\end{equation}

Finally, in the {\it hybrid} formalism, the mode-mode coupling matrices read
\begin{equation}
\begin{array}{lcl}
	\ds M^{TE,TE}_{\ell\ell'}&=&\ds\frac{2\ell'+1}{8\pi}\ds\sum_{\ell''m''}\left|w^{(E)}_{0,\ell''m''}\right|^2J^{(T)}(\ell,\ell',\ell'')J^{(+)}_{0}(\ell,\ell',\ell''), \\
	\ds M^{TE,TB}_{\ell\ell'}&=&\ds 0,
\end{array}
\end{equation}
and
\begin{equation}
\begin{array}{lcl}
	\ds M^{TB,TB}_{\ell\ell'}&=&\ds \frac{2\ell'+1}{8\pi}\ds\sum_{\ell''m''}\sum_{s}\alpha_s\frac{N_{\ell,s}}{N_{\ell,2}}J^{(T)}J^{(+)}_{2-s}\mathrm{Re}\left[w^{(E)}_{0,\ell''m''}w^{(E)\star}_{2-s,\ell''m''}\right], \\
	\ds M^{TB,TE}_{\ell\ell'}&=&\ds -\frac{2\ell'+1}{8\pi}\ds\sum_{\ell''m''}\sum_{s}\alpha_s\frac{N_{\ell,s}}{N_{\ell,2}}J^{(T)}J^{(+)}_{2-s}\mathrm{Re}\left[w^{(E)}_{0,\ell''m''}w^{(B)\star}_{2-s,\ell''m''}\right].
\end{array}
\end{equation}

\section{Noise bias}
\label{app:Noise}
The noise bias in the pseudo-$C_\ell$ estimators differs from one formalism to another. To illustrate this point, we provide here the explicit formulas~for such biases assuming that noise is uncorrelated from pixel to pixel but allowing correlation between the different measured Stokes parameters, i.e.
$$
\left<N_{S}(\vec{n})N_{S'}(\vec{n}')\right>=\sigma^2_{SS'}(\vec{n})\delta(\vec{n}-\vec{n}')
$$
where $S,~S'$ stands for the Stokes parameters, $T,~Q$, or $U$, and assuming autocorrelation of a given map. To perform such a computation, we recall that \cite{ng_liu_1999}
\begin{equation}
\ds\frac{4\pi}{2\ell+1}\sum_{m=-\ell}^\ell\Ylmcc{s}{\ell}{m}(\vec{n}_i)\Ylm{s'}{\ell}{m}(\vec{n}_j)=(-1)^{s-s'}D^\ell_{s,s'}(\alpha,\beta,\gamma)e^{-2is\gamma},
\end{equation}
with $\alpha,\beta$, and $\gamma$ the three Euler angles defined by the rotation multiplication $R(\alpha,\beta,-\gamma)=R(\theta_i,\varphi_i,0)R^{-1}(\theta_j,\varphi_j,0)$. For $\vec{n}_i=\vec{n}_j$, those three Euler angles are equal to zero. The $D^\ell_{s,s'}$ stands for the Wigner rotation matrices and they satisfy $D^\ell_{s,s'}(0,0,0)=\delta_{s,s'}$ (see \cite{vrashalovich_etal_1988}). With such an hypothesis about the noise properties and making use of the above addition theorem for spin-weighted spherical harmonics, it is straightforward to derive the noise bias in the three formalisms, as presented below.

In the {\it standard} formalism first, all the noise biases vanish except for the $EE$ and $BB$ autospectra. For such spectra, the noise contribution reads:
\begin{equation}
	\mathcal{N}^{EE}_{\ell}=\mathcal{N}^{BB}_{\ell}=\frac{1}{8\pi}\ds\int_{4\pi}\left(\sigma^2_{QQ}+\sigma^2_{UU}\right)W^2~d\vec{n},
\end{equation}
and
\begin{equation}
	\mathcal{N}^{EB}_{\ell}=\mathcal{N}^{TE}_{\ell}=\mathcal{N}^{TB}_{\ell}=0.
\end{equation}

Second, in the {\it pure} formalism, the noise biases for autospectra are given by
\begin{equation}
	\mathcal{N}^{EE}_{\ell}=\mathcal{N}^{BB}_{\ell}=\frac{1}{8\pi}\ds\int_{4\pi}\left(\sigma^2_{QQ}+\sigma^2_{UU}\right)\left(W^2+4\frac{N^2_{\ell,1}}{N^2_{\ell,2}}\left|W_1\right|^2+\frac{1}{N^2_{\ell,2}}\left|W_2\right|^2\right)~d\vec{n},
\end{equation}
while for cross-spectra they reads
\begin{equation}
\begin{array}{lcl}
	\ds \mathcal{N}^{EB}_{\ell}&=&\ds \frac{1}{4\pi}\ds\int_{4\pi}\frac{\sigma^2_{QU}}{N^2_{\ell,2}}\mathrm{Re}\left[W^2_2\right]~d\vec{n}, \\
	\ds \mathcal{N}^{TE}_{\ell}&=&\ds -\frac{1}{4\pi}\ds\int_{4\pi}\frac{W}{N_{\ell,2}}\left(\sigma^2_{TQ}\mathrm{Re}[W_2]+\sigma^2_{TU}\mathrm{Im}[W_2]\right)~d\vec{n}, \\
	\ds \mathcal{N}^{TB}_{\ell}&=&\ds -\frac{1}{4\pi}\ds\int_{4\pi}\frac{W}{N_{\ell,2}}\left(\sigma^2_{TQ}\mathrm{Im}[W_2]-\sigma^2_{TU}\mathrm{Re}[W_2]\right)~d\vec{n}.
\end{array}
\end{equation}

Finally, in the {\it hybrid} formalism, the noise biases are easily deduced from the standard and pure formalism computation to be given by
\begin{equation}
\begin{array}{lcl}
	\ds \mathcal{N}^{EE}_\ell&=&\ds \frac{1}{8\pi}\ds\int_{4\pi}\left(\sigma^2_{QQ}+\sigma^2_{UU}\right)W^2~d\vec{n}, \\
	\ds \mathcal{N}^{BB}_{\ell}&=&\ds \frac{1}{8\pi}\ds\int_{4\pi}\left(\sigma^2_{QQ}+\sigma^2_{UU}\right)\left(W^2+4\frac{N^2_{\ell,1}}{N^2_{\ell,2}}\left|W_1\right|^2+\frac{1}{N^2_{\ell,2}}\left|W_2\right|^2\right)~d\vec{n},
\end{array}
\end{equation}
and
\begin{equation}
\begin{array}{lcl}
	\ds \mathcal{N}^{EB}_{\ell}&=&0, \\
	\ds \mathcal{N}^{TE}_\ell&=&0, \\
	\ds \mathcal{N}^{TB}_\ell&=&\ds -\frac{1}{4\pi}\ds\int_{4\pi}\frac{W}{N_{\ell,2}}\left(\sigma^2_{TQ}\mathrm{Im}[W_2]-\sigma^2_{TU}\mathrm{Re}[W_2]\right)~d\vec{n}.
\end{array}
\end{equation}

Unlike the standard calculation, the pure computation induces some noise bias for the two types of odd-parity power spectra. However, the $EB$ noise bias can be set to zero by switching to a hybrid approach. Moreover, the $TB$ and $EB$ noise bias will also vanish if the noise is uncorrelated from one Stokes parameter to another or by invoking cross-spectra between maps coming from different experiments or different, and thus uncorrelated, detectors.

We also stress out that all the noise biases computed above are indeed invariant under any rotation of the polarization reference basis. This can be straightforwardly checked by making use of the spin properties of the Stokes parameters and of the three spin-weighted window functions under a rotation by an angle $\alpha$:
\begin{eqnarray}
	T&\rightarrow&T'=T, \nonumber \\
	(Q\pm iU)&\rightarrow&(Q'\pm iU')=e^{\pm2i\alpha}(Q\pm iU), \nonumber \\
	W_{\pm s}&\rightarrow&W'_{\pm s}=e^{\pm is\alpha}W_{\pm s}. \nonumber
\end{eqnarray}
	
\end{widetext}


\end{document}